# SWIMMING WITH QUARKS

**M. R. Pennington**

Institute for Particle Physics Phenomenology, University of Durham, Durham DH1 3LE, U.K.

**Abstract.**
These six lectures, given at the XI Mexican School of Particles and Fields held at Xalapa in August 2004, are on the subject of strong coupling QCD. How this colours and shapes the hadron world in terms of (i) the hadron spectrum, (ii) chiral symmetry breaking, (iii) dynamical mass generation and (iv) confinement, are the topics discussed.

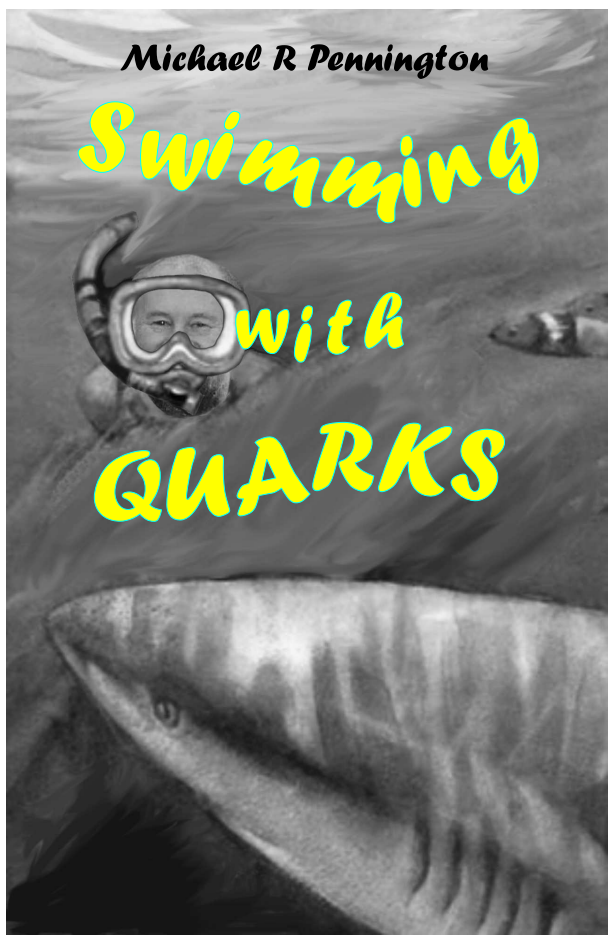

**1. The quark model and beyond**

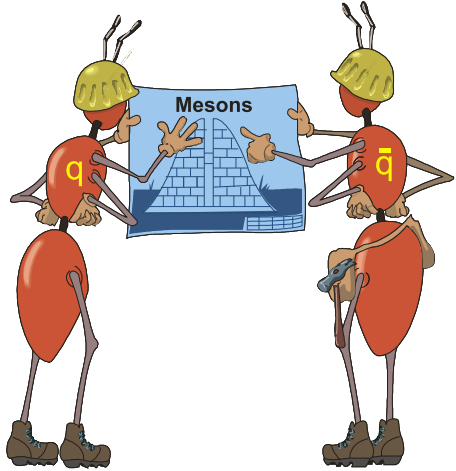

**Hadron spectrum**

**properties of bound states**

QCD is the theory of the strong interactions. It describes the forces between quarks and gluons that make up the hadron world. It may appear that lectures about hadrons describe ancient history, like the world of nuclei before we knew about protons and neutrons. However, there is an essential difference. Hadrons are what we observe in detectors. Though we often like to describe what is going on in terms of quarks and gluons, it is hadrons that either collide or are the result of collisions (or both) and not elementary quarks and gluons. In describing proton collisions at the LHC we imagine quarks and gluons interacting and these may in turn produce an elementary Higgs. However, though quarks and gluons may well interact perturbatively over distance scales of $10^{-18}$ m., we must not forget that as they travel over bigger distances away from the hard scattering process they have to dress themselves into the protons, pions and kaons that trigger detectors. How quarks and gluons are bound into hadrons is a long distance, strong coupling problem. It involves the regime of QCD that makes this theory and the interaction it describes far more interesting than QED. It provides the challenge for our understanding. Strong coupling QCD is responsible for the spectrum of hadrons, particularly those of light flavour. It is responsible for confinement and for the chiral symmetry breaking that shapes the light hadron world reflected in experiment. These are the topics of the present lectures.

QCD has its beginnings in the spectrum of hadrons and the quark model representation of these [1]-[5]. This is our first strong coupling problem. The bound state problem we learnt about at our mother's knee is that of atoms. We know that the spectrum of light, emitted as an excited atom decays, directly reflects the fact that atoms are composed of charged objects, nuclei and electrons, with an electromagnetic force holding them together. Thus the spectrum of the system teaches us about the constituents of the system and the forces that bind them together. This is equally true of the spectrum of hadrons.

Let us concentrate on mesons and recall the simple structure of some of the lightest of these [4]. Quarks have spin 1/2. Consequently a quark and antiquark can only form a system of intrinsic spin $S = 0$ or 1. The quark and antiquark can have relative orbital angular momentum, $L$, which combines with $S$ vectorially as usual in quantum mechanics to give the spin of the meson $J$, where $\mathbf{J} = \mathbf{L} + \mathbf{S}$. This then fixes the parity and charge conjugation of the meson to be

$$P = (-1)^{L+1} \quad , \quad C = (-1)^{L+S} \quad . \tag{1}$$

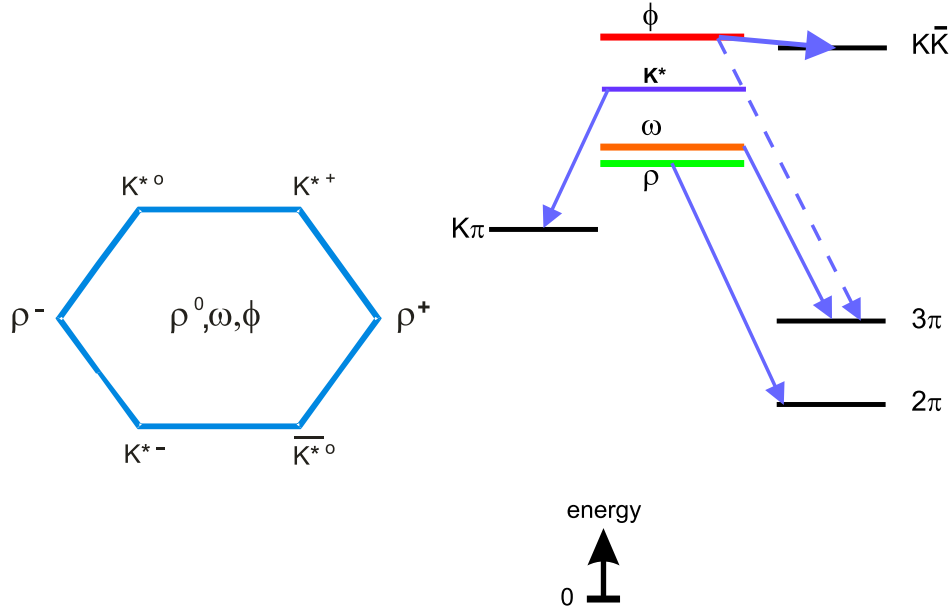

**Figure 1.** The multiplet of the nine lightest vector mesons on the left. The implicit horizontal axis is the $z$-component of isospin and the vertical axis is strangeness. $\rho^0$, $\omega$, $\phi$ are at the origin. The main hadronic decay of each of these vector states is to pseudoscalar mesons as displayed on the right, where the length of the arrow denotes the energy released in that transition.

Let us first consider the quark model states with $S = 1$, $L = 0$. This gives the vector mesons with $J^{PC} = 1^{--}$. With three light flavours, $u, d, s$ we have nine vector mesons. We expect these to be divided into a flavour octet and flavour singlet, since:

$$\mathbf{3} \otimes \mathbf{\bar{3}} = \mathbf{8} \oplus \mathbf{1} \quad . \tag{2}$$

The quantum numbers of the nine observed states are shown in Fig. 1. The two states in the middle of the multiplet with $I = 0$, the $\omega$ and $\phi$, are however not members of an octet and a singlet, respectively, but they magically mix so that they correspond to states of definite quark flavour. We learn this from their decays and from their masses, Fig. 1. The states decay overwhelmingly into pseudoscalar mesons by the creation of $u\bar{u}$ and $d\bar{d}$ pairs from the vacuum. This explains why the strange vector state the $K^*(890)$ decays to $K\pi$ and the $\rho$ to $\pi\pi$, as shown in Fig. 2. The $I = 0$ octet and singlet states would both decay to 3 pions ( $G$-parity forbids the decay to 2 pions ). While the $\omega$ does decay to 3 pions, and has a mass very close to the $\rho$ reflecting their composition of $u$ and $d$ quarks, the $\phi$ overwhelmingly decays to $K\bar{K}$ despite much greater phase space for its $3\pi$ mode, Fig. 1, and an $\bar{s}s$ composition is inferred. This is intimately related to the mass differences

$$m(\phi) - m(K^*) \simeq m(K^*) - m(\omega) \simeq 120\,\text{MeV} \quad ,$$

which reflect the mass difference between the $s$ quark and the $u$ and $d$. For a state made of $\bar{s}s$ quarks to decay to $3\pi$ requires the initial strange quarks to annihilate and for all the quarks of the final state to be created from the vacuum. We learn that such a process is suppressed. This is known as the OZI rule.

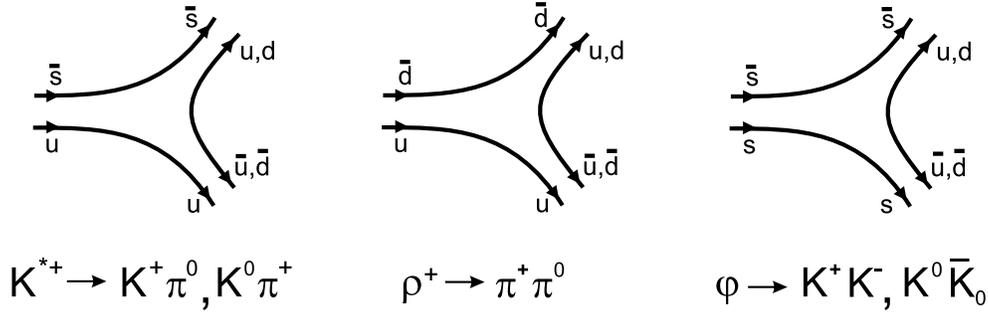

**Figure 2.** Quark line diagrams illustrating how the creation of $\overline{u}u$ and $\overline{d}d$ pairs allows the vector mesons, $K^*$, $\rho$, $\phi$ to decay into two pseudoscalar mesons.

The fact that the $\rho^0$, $\omega$ and $\phi$ have composition

$$\frac{1}{\sqrt{2}}\left(\overline{u}u \pm \overline{d}d\right) \quad , \quad \overline{s}s \quad , \tag{3}$$

is confirmed by their decay rate to $e^+e^-$. As illustrated in Fig. 3, in this decay the quarks annihilate to form a virtual photon that then couples to an $e^+e^-$ pair. Since the coupling of the photon to quarks is proportional to their charge, these decay rates are a measure of the square of the mean charge of the constituent quarks. If all other factors are equal for each

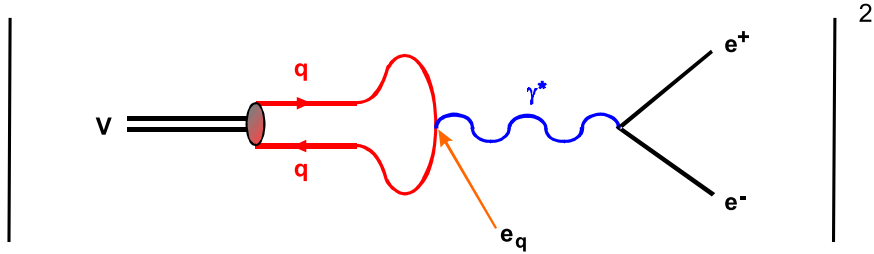

**Figure 3.** Feynman graph of the neutral vector meson, $V$, decaying into $e^+e^-$, illustrating how the matrix element for this process is proportional to the average charge of the constituent quarks in the meson.

meson, then

$$\Gamma(\rho^0 \to e^+e^-) : \Gamma(\phi \to e^+e^-) : \Gamma(\omega \to e^+e^-) =$$

$$\left[\frac{1}{\sqrt{2}}\left(\left(\frac{2}{3}\right) - \left(-\frac{1}{3}\right)\right)\right]^2 : \left(-\frac{1}{3}\right)^2 : \left[\frac{1}{\sqrt{2}}\left(\left(\frac{2}{3}\right) + \left(-\frac{1}{3}\right)\right)\right]^2 = 9 : 2 : 1 \tag{4}$$

Experiment agrees with this very well.

Exactly, the same structure of an ideally mixed nonet shown in Fig. 1 applies to the tensor mesons where the quarks combine with $S = 1$, $L = 1$. There the heaviest, the $f_2'(1525)$ (the spin-2 analogue of the $\phi$) overwhelmingly decays to $K\overline{K}$. The $f_2(1270)$ composed of $u$ and $d$ quarks can decay to $K\overline{K}$ too by the creation of an $\overline{s}s$ pair, but strange quarks being heavier than $u$ and $d$ quarks, this has reduced probability. These assignments can be tested this time in their decay rate to two photons. A spin-2 meson naturally couples to two spin-1 photons. This rate is related to the square of the mean charge squared of the constituent quarks as illustrated in Fig. 4 by

$$\Gamma(T \to \gamma\gamma) \; \sim \; \alpha^2 \, \langle \, e_q^2 \, \rangle^2 \, \Pi_T \quad , \tag{5}$$

where $\Pi_T$ is the probability that the quarks in a tensor meson, $T$, annihilate. In the non-relativistic quark model this is proportional to $\mid \psi(0) \mid^2$, where $\psi(0)$ is the wavefunction at the origin. If all these factors are the same for each neutral tensor meson, then for an ideally mixed multiplet

$$\Gamma(f_2(1270) \to \gamma\gamma) \; : \; \Gamma(a_2(1320) \to \gamma\gamma) \; : \; \Gamma(f_2'(1525) \to \gamma\gamma) \; = \; 25 \; : \; 9 \; : \; 2 \quad .$$

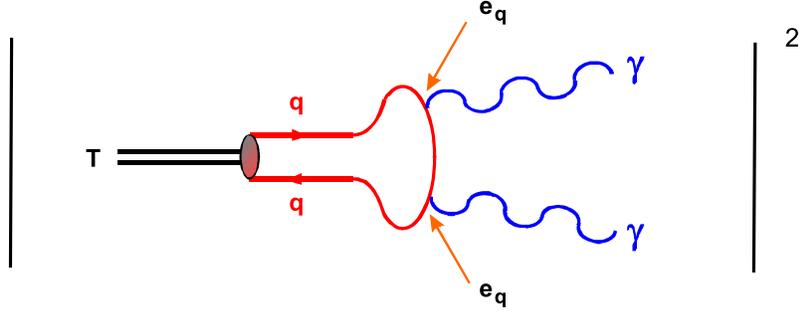

**Figure 4.** Feynman graph for a $\bar{q}q$ tensor meson, $T$, to decay into two photons. The square of its modulus gives the decay rate.

Experiment very approximately reproduces this with $25 : (10 \pm 1) : (1 \pm 1)$. That this is not more exact is a combination of imperfect experiments and the fact that the masses of the states are only roughly equal so there are inevitably corrections to the naive non-relativistic quark model. This is highlighted by the application of the same formula, Eq. (5), to pseudoscalar mesons: $\pi$, $\eta$, $\eta'$, for which experiment gives the ratio $1 : 50 : 600$. But Hayne and Isgur [6] proposed that the naive result of Eq. (5) should be corrected by a factor of $M_P^3$, where $M_P$ is the mass of the pseudoscalar meson. For the $\pi$, $\eta$, $\eta'$ this contributes factors of $1 : 67 : 357$ multiplying the quark charges, bringing far better agreement with experiment.

What does QCD add to this picture of $\bar{q}q$ mesons? We believe confinement requires that hadronic states should be colour singlets. This means their wavefunctions must be symmetric under permutations of the colours. Then the $\rho^+$ instead of being simply $u\bar{d}$ has a colour wavefunction:

$$\mid \rho^+ \rangle \; = \; \frac{1}{\sqrt{N_c}} \, \mid u_R \, \bar{d}_R \, + \, u_G \, \bar{d}_G \, + \, u_B \, \bar{d}_B \, + \, \cdots \rangle \quad , \tag{6}$$

where $R$ (red), $G$ (green), $B$ (blue), $\cdots$ label the $N_c$ colours. While the number of colours matters little for mesons, for baryons it is a different matter. A colour singlet wavefunction for a state of three quarks demands that each comes in a different colour. So a proton, made of $uud$ quarks, has a singlet wavefunction

$$\mid p \, \rangle \; = \; \frac{1}{\sqrt{6}} \, \mid u_R \, u_G \, d_B \, + \, u_G \, u_B \, d_R \, + \, u_B \, u_R \, d_G \, - \, u_R \, u_B \, d_G \, - \, u_B \, u_G \, d_R \, - \, u_G \, u_R \, d_B \, \rangle \, . \tag{7}$$

While the wavefunction is symmetric under permutations of the colours, it is odd under the interchange of any pair. This explains a longstanding problem of the statistics obeyed by quarks. Quarks were known to have spin-1/2 in order to build the observed hadrons and their quantum numbers, but are they fermions? If we consider the wavefunction of the $J = 3/2$ $\Delta(1232)^{++}$ made of three $u$ quarks. It is clearly symmetric in quark flavour and quark spin. The introduction of colour means that the colour singlet wavefunction, like Eq. (6) with three $u$ quarks, is antisymmetric under the interchange of any pair. This tells us that quarks obey Fermi-Dirac statistics, and not some strange *parastatistics* that was needed before the discovery of *colour*. The $\Delta(1232)$ belongs to a decuplet of $qqq$ states, for which the existence of the $sss$ state, the $\Omega^-$, provided historic confirmation. Each decays by creating $\overline{u}u$ or $\overline{d}d$ pairs, as for the mesons of Fig. 2. Thus

$$\Delta^{++} \to \pi^+ p \quad , \qquad u\,u\,u \to (u\,\overline{d})(d\,u\,u) \quad .$$

in terms of quarks. Experiment has more recently hinted that there may be hadrons that decay strongly in a different way, as we will discuss.

If the criterion for the existence of observable isolated hadrons is that they should be colour singlets, then QCD does explain why we see mesons made of $\overline{q}q$ and baryons of $qqq$, but it predicts there could/should be hadrons beyond the quark model, for instance mesons [7], made of

- $q\overline{q}g$, states called hybrids, where gluons $g$ contribute to the intrinsic quantum numbers,
- multiquark states like $qq\overline{qq}$, and
- $gg$ or $ggg$, called glueballs.

or baryons with more than 3 quarks. For thirty years experiment has sought to establish whether any of these exist in nature. Here we will quickly summarise what we have learnt and refer to the experimental and phenomenological analyses [8, 9, 10] for the details.

The development of crystal technology allowing photons to be accurately detected has opened up a new era in spectroscopic experiments, pioneered by the GAMS collaboration at both Protvino and CERN [11], and more recently followed by the high statistics BNL-E852 experiment [12]. Detecting photons allows the decays of neutral particles like the $\pi^0$, $\eta$ and $\eta'$ to be observed in their decays to 2 photons. This has enabled the measurement of $\pi\eta$ final states produced by meson exchange in $\pi^- p$ collisions. Now the simple quark model tells us that any resonance that decays into $\pi\eta$ with odd spin must be a state beyond the quark model, and so be a candidate for a hybrid or a tetraquark state. Long ago in the mid '80's GAMS [11] observed a strong forward-backward asymmetry that indicated the presence of a sizeable odd partial wave. If such a wave were resonant, its quantum numbers would be $1^{-+}$, which is not possible in the simple quark model, Eq. (1). The GAMS group claimed that this state was indeed resonant. In the PDG tables [13] it is listed as the $\pi_1(1400)$, and is an important example of a state beyond the quark model whether hybrid or tetraquark meson. The BNL-E852 experiment [12] has been able to study this in more detail with cleaner partial wave separation. The $\pi\eta$ channel is dominated by the $D$-wave $a_2(1320)$, the standard $\overline{q}q$ tensor meson discussed earlier. As seen in Fig. 5 its signal has $10^4$ events in a 40 MeV bin. The $P$-wave underneath it is a factor 20 or more smaller with sizeable error bars. Consequently any small misassignment from the large $D$-wave signal feeds a big uncertainty into the $P$-wave. There is a $D - P$ wave phase variation in the 1200-1500 MeV region, seen in Fig. 5, but is this what is expected of both waves resonating is a question of fierce debate, not least within the BNL-E852 Collaboration itself [14].

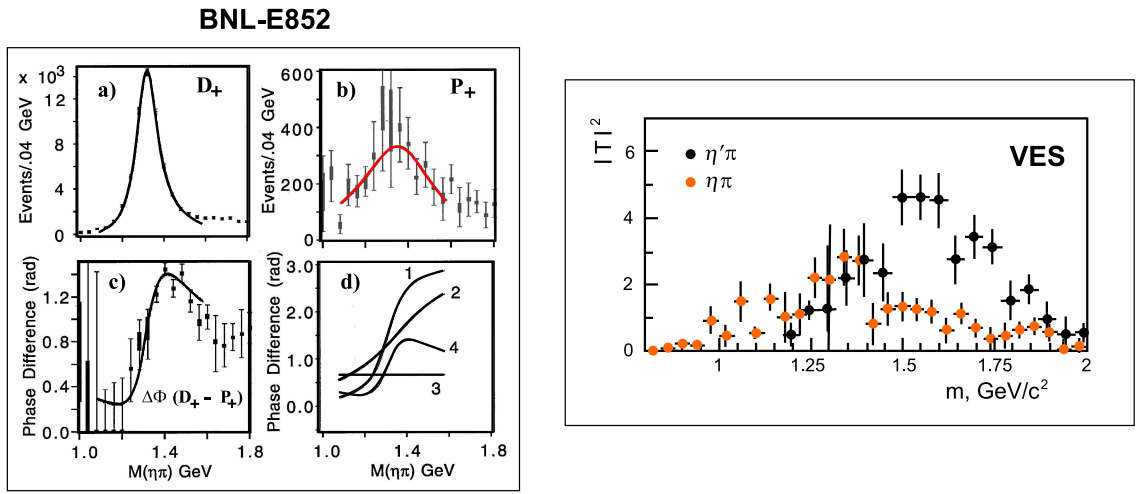

**Figure 5.** Results of the partial wave separation of BNL-E852 results [14] on essentially $\pi^-\eta$ scattering from $\pi^- p \to (\pi^-\eta)p$, showing the modulus of (a) the $D_+$ wave dominated by the well-known tensor meson, $a_2(1320)$ and (b) the $P_+$ wave, as well as (c) their relative phases, while (d) indicates (1) the fitted $D_+$ phase, (2) the fitted $P_+$ phase, (3) the relative production phase and (4) the overall phase difference. The graph on the right displays the VES results [15] on $\eta\pi$ and $\eta'\pi$ production as a function of the di-meson mass in $\pi^- Be$ collsions at 28 GeV/$c$.

The VES experiment [15] at Serpukhov has also studied what is effectively $\pi\eta$ and $\pi\eta'$ scattering and find, as seen in Fig. 5, a rather small enhancement in the $\pi\eta$ channel at 1.4 GeV, but a much more pronounced signal in the $\pi\eta'$ channel at 1.5-1.6 GeV. A $\pi_1(1600)$ is also consistent with the analysis of the BNL-E852 results [14]. To be certain that a hybrid or tetraquark meson has been observed, we await future results from the range of experiments planned for Hall D at Jefferson Laboratory [16], which will allow the photoproduction of such states with high statistics.

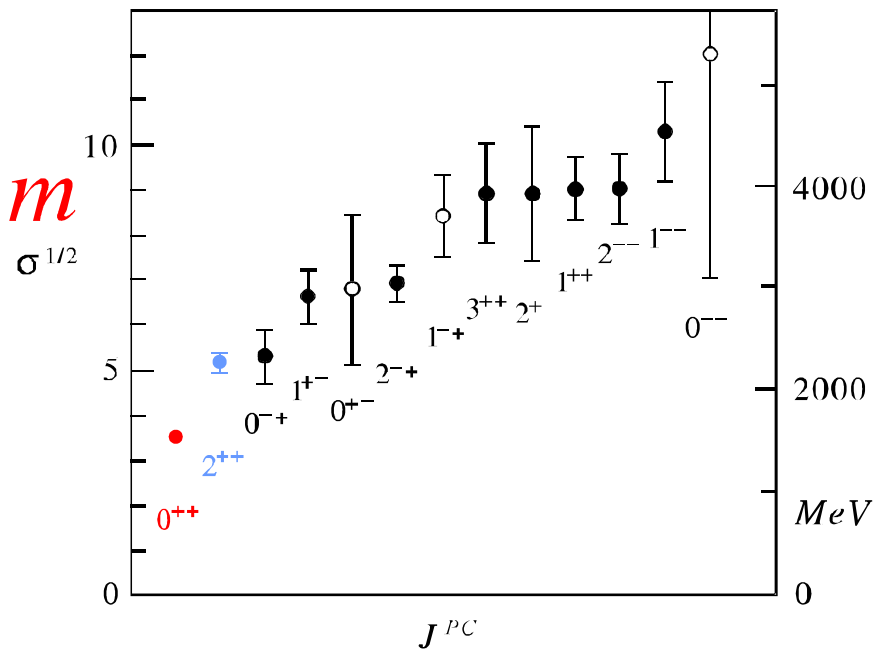

**Figure 6.** The spectrum of the lightest states of pure glue with different $J^{PC}$ quantum numbers as calculated in computations of pure gluonic QCD on the lattice [17], with masses on the left in terms of the 'string tension' and on the right in energy units.

Glueballs should exist in a world without quarks. A world that is difficult to study in experiment, but is readily computable in lattice gauge theories. Remarkably such calculations have long shown (see Fig. 6 from [17]) that in a world of pure glue there is a spectrum of colour singlet hadrons. The lightest of these is a scalar (with $J^{PC} = 0^{++}$) with mass of 1500-1700 MeV. Though these calculations must be corrected by quark loops to have physical meaning, experiment reveals possible candidates with masses in exactly this same region. The different protagonists (see [18]) being certain that the candidate closest to their *quenched* lattice calculation must be the scalar glueball.

The folklore for 25 years has been that glueballs should appear in "glue-rich" processes, like those illustrated in Fig. 7: in $J/\psi$-decay, or in $\bar{p}p$ annihilation. $J/\psi$ decay was first extensively studied at SLAC, and now comprehensively investigated by the BEPC, the Beijing electron-positron collider, while the biggest series of $\bar{p}p$ annihilation experiments [19] were performed last century at LEAR, the low energy antiproton ring, at CERN. These experiments have thrown up candidates for glueballs: $f_J(1710)$, $f_0(1510)$ in particular, very close to the quenched lattice predictions of Fig. 6. As seen in the $J/\psi \to \gamma K\bar{K}$ mass spectrum there are a series of peaks that look like resonances highlighted in Fig. 8 in the results from DM2 [20] and Mark III [21].

Which of these is a glueball, if any? Of course, conventional tensor resonances like the $f_2'(1525)$ also appear in such a spectrum, as shown in Fig. 8, since once again tensors naturally couple to two spin-1 objects, here the $J/\psi$ and photon. So how do we tell what is a glueball, with the gluons resonating as pictured in Fig. 7? Or once the gluons have created a quark and an antiquark perhaps they resonate, as with the $f_2'(1525)$? This can be answered in the framework of perturbative QCD by a method suggested by Farrar and Çakir [22], as applied together with Close and Li [23]. While the result that the $f_2'(1525)$ is a $q\bar{q}$ state is hardly surprising, for all of the $f_0(1500)$, $f_0(1700)$, $f_2(1710)$ the conclusions on their glue content is far from unambiguous [18], not helped by inconsistency between datasets. More data from BES should clarify the situation. Interestingly, Mark III [21] and BES [24] both observe a higher mass state, the $\xi(2230)$, arrowed in Fig. 8, but not seen in DM2. The new increased statistics from BES should tell us if this is the signal for the tensor glueball of Fig. 6, or not. In the case of the scalar glueball, it is inevitable that its decays to light hadrons mean it must mix with $\bar{q}q$ states and so no longer remain "pure glue" (see [18] for some further references). Again more analysis of more data is needed and this is part of the Hall D programme at Jefferson Laboratory (TJNAL), as well as a focus of the ongoing projects at BES and CLEO-c.

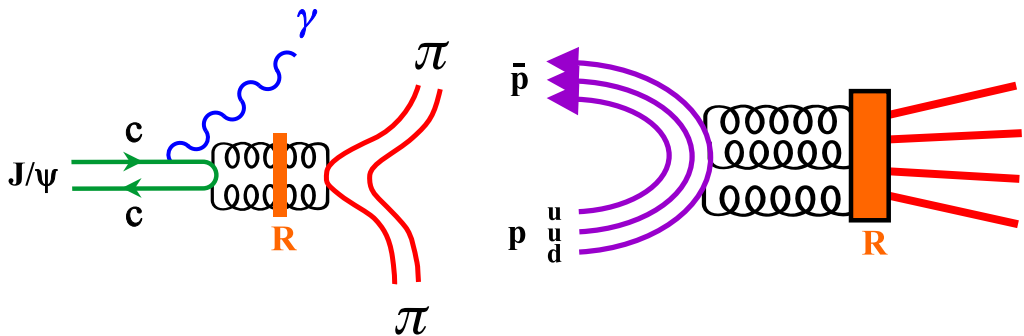

**Figure 7.** Perturbative Feynman graphs for the processes $J/\psi$-radiative decay to $\pi\pi$, and $\bar{p}p$ annihilation into hadrons, illustrating why these reactions are often claimed to be glue-rich and provide ways in which largely gluonic resonances, $R$, might be formed.

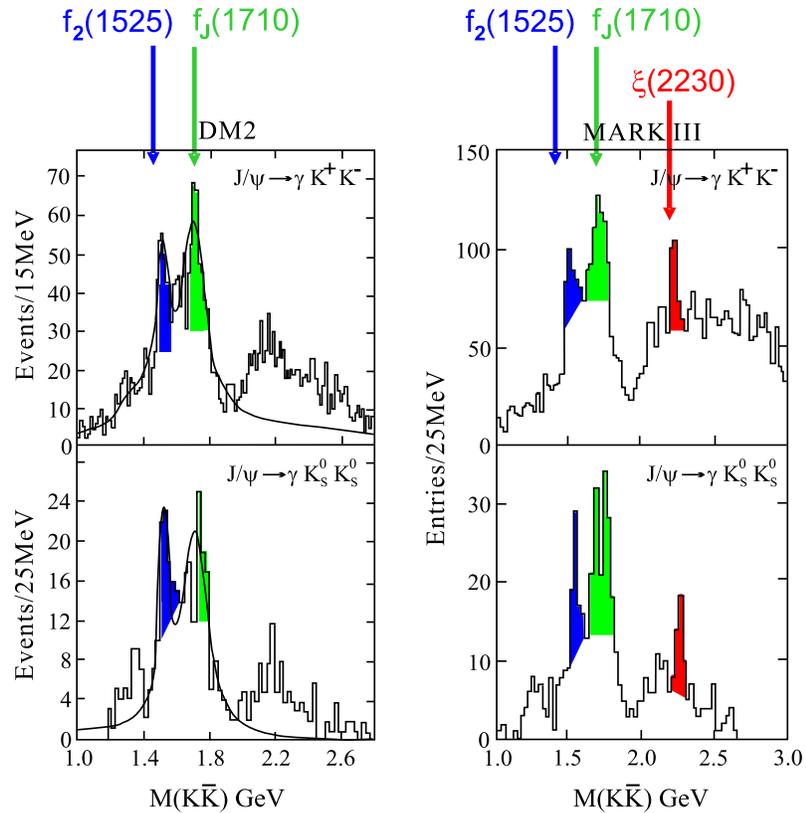

**Figure 8.** $\overline{K}K$ mass spectrum for charged and neutral kaon pairs resulting from $J/\psi$ radiative decay as measured by DM2 [20] and Mark III [21] experiments. These both show enhancements consistent with the formation of the $f_2'(1525)$ and $f_J(1710)$. Mark III see the highlighted $\xi(2230)$ at a statistically significant level, as do BES [24].

## 2. Pentaquarks and tetraquarks

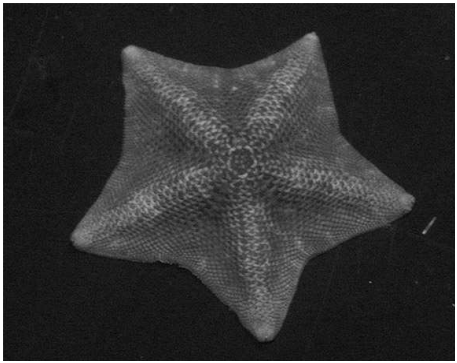

**Figure 9.** Pentaquark state: each apex represents either one of 4 $q$ or 1 $\overline{q}$

A remarkable development in the past year has been evidence for a baryon that does not fit the pattern of decays we described earlier. The $\Theta(1530)$ is seen in photoproduction at SPring-8 in Japan [25] and in the CLAS experiment at TJNAL [26], Fig. 10. What makes this state unusual is not just its longevity (having a width consistent with the resolution of the experiments at less than 20 MeV) but that it decays to $K^+ n$. Since this has strangeness +1, it must contain at least one $\bar{s}$ quark. To have baryon number $+1$ it must then have more quarks than the minimum three. Thus its simplest composition is $\bar{s}uudd$. To decay to $K^+n$ does not require the creation of additional quark-antiquark pairs (as in Fig. 2), but rather a rearrangement of the 4 quarks and an antiquark into two colour singlets. The possibility of a baryon with positive strangeness is a prediction of the Skyrme model of Diakonov, Petrov and Polyakov [27]. These models with their non-trivial topology are complicated and require background beyond the scope of these lectures [5]. An alternative modelling is in terms of diquarks as key entities [28, 29], which is simpler to discuss.

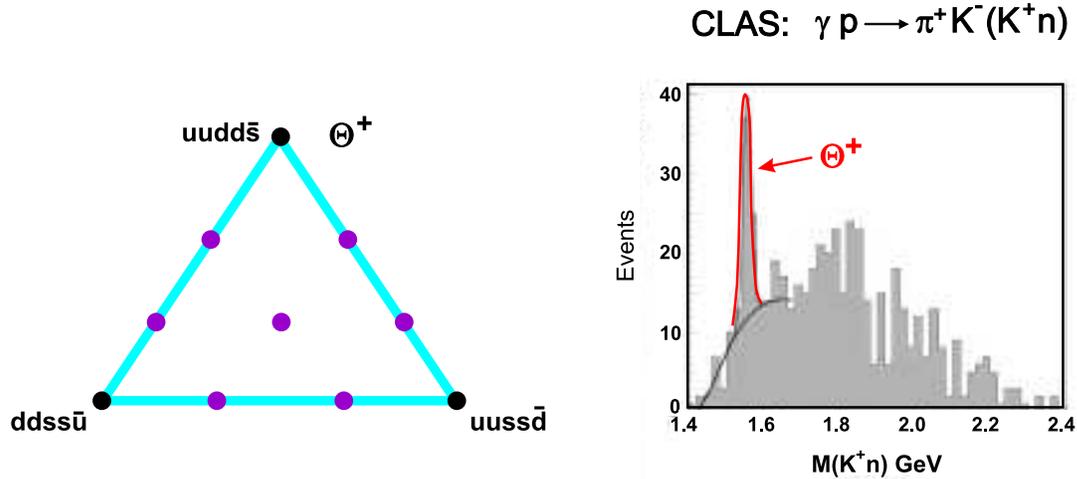

**Figure 10.** A five quark decuplet that could contain the claimed $\Theta^+(1530)$ state seen, for instance, in the missing $K^+n$ mass spectrum found by CLAS [26] in $\gamma p \to \pi^+ K^- (K^+n)$. Only the three states at each apex of the decuplet have flavour quantum numbers distinct from those of a nucleon octet.

To understand this idea let us return to something more basic. Let us consider the creation of hadrons in $e^+e^-$ annihilation — a process that actually counts the number of colours. Consider the electron and positron approaching each other head-on at energies above 3 GeV or so, as first performed at SLAC. The $e^+e^-$ deposit all their energy in a virtual photon of size of a tenth of a fermi, much smaller than the size of the hadrons the photon is eventually going to produce. By the uncertainty principle this photon must turn into matter in $10^{-25}$-$10^{-24}$ of a second, and there is no time to create something big like a pion or a proton, so it creates a quark-antiquark pair. These move apart radiating gluons that in turn create more $q\bar{q}$ pairs, which eventually arrange themselves into colour singlet hadrons. The most energetic of these are still moving in the direction of the initial quark and antiquark that are created. The distribution of hadron jets remembers the spin of these quarks and so the distribution of jets follows an average $(1+\cos^2\theta)$ form in the centre-of-mass frame, just as when $\mu^+\mu^-$ pairs are produced. Moreover, the ratio of the cross-section for $e^+e^- \to hadrons$ to that for $e^+e^- \to \mu^+\mu^-$ measures the average charge squared of the quarks that are produced. Experiment is explained if each quark flavour comes in 3 colours.

The mechanism of how the quarks turn into hadrons is a key topic of strong coupling physics to which we will return, but here is a mystery. At low energy to produce a $\pi^+\pi^-$ pair the primordial $q\bar{q}$ pair radiate gluons, one of which produces another back-to-back $q\bar{q}$ pair. But how do we form a $p\bar{p}$? Do additional $q\bar{q}$ pairs have to be created, which then have to get themselves in just the right kinematic configuration to form a proton-antiproton pair. A simpler possibility is that there exists a system of diquarks, $qq$. In terms of colour, $qq$ can be just like $\overline{q}$:
$$\mathbf{3} \otimes \mathbf{3} = \mathbf{6} \oplus \overline{\mathbf{3}} \tag{8}$$
making constructing both a meson and a baryon the colour equivalent of the multiplication rule of Eq. (2) (there for flavour) to form a colour singlet. Thus in $e^+e^-$ annihilation, we can produce a $p\overline{p}$ final state by the initial $q\overline{q}$ radiating just one energetic gluon that creates a $qq$-$\overline{qq}$ pair. Then we can think of baryons as having an intrinsic $q$-$qq$ composition. One can see that the Pauli exclusion principle would inhibit two quarks of the same flavour forming a closely bound system. Consequently a diquark must involve quarks of different flavours. In a low spin state, like a nucleon, the wavefunctions of the 3 quarks overlap sufficiently for this not to be so apparent. However, we can see this diquark pattern when the single quark carries most of the baryon's momentum in deep inelastic scattering. There the cross-section determines the structure function, $F_2(x)$ [4, 1], which is a measure of the probability of any particular constituent carrying fraction $x$ of the nucleon's longitudinal momentum weighted by the square of the charge of this constituent times the momentum fraction $x$. It has long been known that the ratio of $F_2$ for a neutron to that for a proton tends to $1/4$ [30] as $x \to 1$, Fig. 11. Since $e_d^2/e_u^2 = 1/4$, this ratio is natural if it is the $d$ quark in a neutron and the $u$ quark in a proton that carries all the momentum. But since each contains additional $u$ and $d$ quarks how can this be? If one constituent carries all the nucleon's momentum, then the other constituents carry *no* momentum. The exclusion principle then tells us that those carrying *no* momentum cannot be in the same quantum mechanical state. Thus if we consider each nucleon to be composed of a $ud$ di-quark, which when $x \to 1$ carries no momentum, plus the quark carrying all the momentum, then this naturally leads to the ratio of structure functions tending to $1/4$.

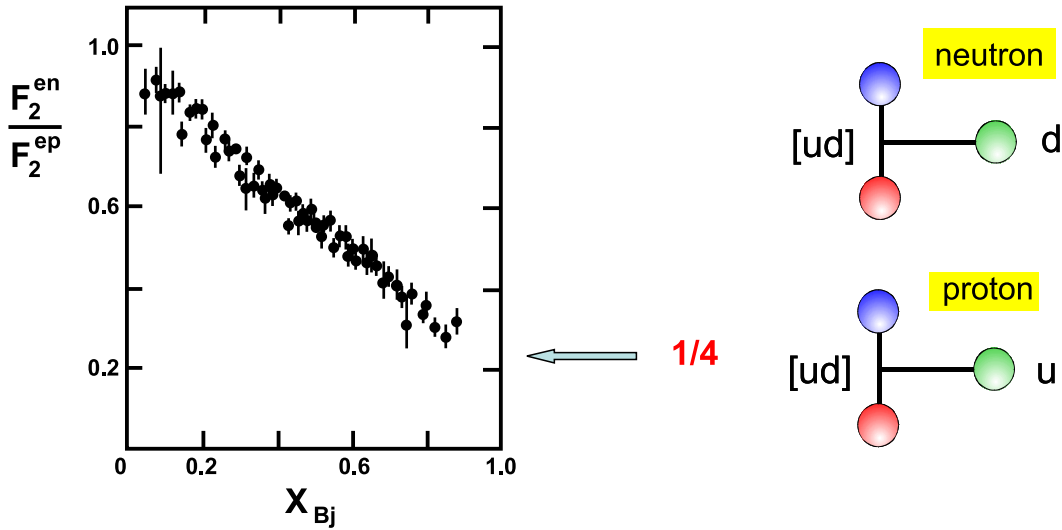

**Figure 11.** Ratio of the structure function $F_2$ determined in deep inelastic $en$ scattering to that in the $ep$ process as a function of Bjorken $x_{Bj}$ [30]. On the right is a picture of each nucleon with a $[ud]$ diquark.

Diquarks that bind are scalars and the quarks have different flavours. Thus we expect $[ud]$, $[su]$, $[sd]$ systems each in $\overline{\mathbf{3}}$ of colour [28]. A pentaquark state would then be in a diquark-diquark-antiquark configuration. Among the flavour multiplets of $\overline{\mathbf{3}} \otimes \overline{\mathbf{3}} \otimes \mathbf{3}$, there would be a decuplet, where the state at the apex (Fig. 10) with configuration $[ud][ud]\overline{s}$ would be the

| EXPERIMENT | REACTION | MASS (MeV) | Width (MeV) | Statistical sig. |
|---|---|---|---|---|
| LEPS | $\gamma n(C^{12}) \to \Theta^+ K^-$ | $1540 \pm 10$ | $< 25$ | $4.6\sigma$ |
| DIANA | $K^+ Xe \to \Theta^+ X$ | $1539 \pm 2$ | $< 9$ | $4.4\sigma$ |
| CLAS(d) | $\gamma d \to \Theta^+ K^- p$ | $1542 \pm 5$ | $< 21$ | $5.2\sigma$ |
| CLAS(p) | $\gamma p \to \Theta^+ K^- \pi^+$ | $1555 \pm 10$ | $< 26$ | $7.8\sigma$ |
| SAPHIR | $\gamma p \to \Theta^+ K^0$ | $1540 \pm 4$ | $< 25$ | $4.8\sigma$ |
| HERMES | $\gamma^* d \to \Theta^+ X$ | $1528 \pm 4$ | $17 \pm 9$ | $4.6\sigma$ |
| ITEP | $\nu A \to \Theta^+ X$ | $1533 \pm 5$ | $< 20$ | $6.7\sigma$ |
| SVD-2 | $pA \to \Theta^+ X$ | $1526 \pm 3$ | $< 24$ | $5.6\sigma$ |
| COSY | $pp \to \Theta^+ \Sigma^+$ | $1530 \pm 4$ | $< 18$ | $4.6\sigma$ |
| ZEUS | $\gamma^* p \to \Theta^+ X$ | $1522 \pm 3$ | $8 \pm 4$ | $4.6\sigma$ |

**Table 1.** Summary table of experiments claiming a signal for the putative pentaquark state, the $\Theta^+(1530)$, see [31] for the detailed references.

$\Theta^+$. At the other two corners of Fig. 10 there would be $[su][su]\overline{d}$ and $[sd][sd]\overline{u}$, the $\Xi^+$, $\Xi^{--}$, respectively. These three states at the corners of the pentaquark decuplet have quantum numbers distinct from that of a conventional excited nucleon octet and so most readily distinguishable from conventional baryons. Such states may have been observed. Presuming that the $N^*(1710)$ is a non-strange member of the same family, Diakonov *et al.* [27] within their Skyrme model predict the "exotic" $\Theta^+$ to have a mass of 1530 MeV. Remarkably SPring-8 [25] and CLAS [26] find such a state (Fig. 10). Other experiments confirm signals that agree with this as listed in the summary table 1 [31]. In no case are the statistics overwhelmingly conclusive. They all require rather specific cuts to observe identifiable signals even with such limited statistical significance. It could be that such cuts are indeed related to the production mechanism needed to create a multiquark state that is not seen in years of $KN$ phase-shift analyses. As seen from the summary table, many of the reactions that claim to find the $\Theta^+$ involve photoproduction. This must be a clue to the composition [32, 33, 29] and dynamics of this strange state.

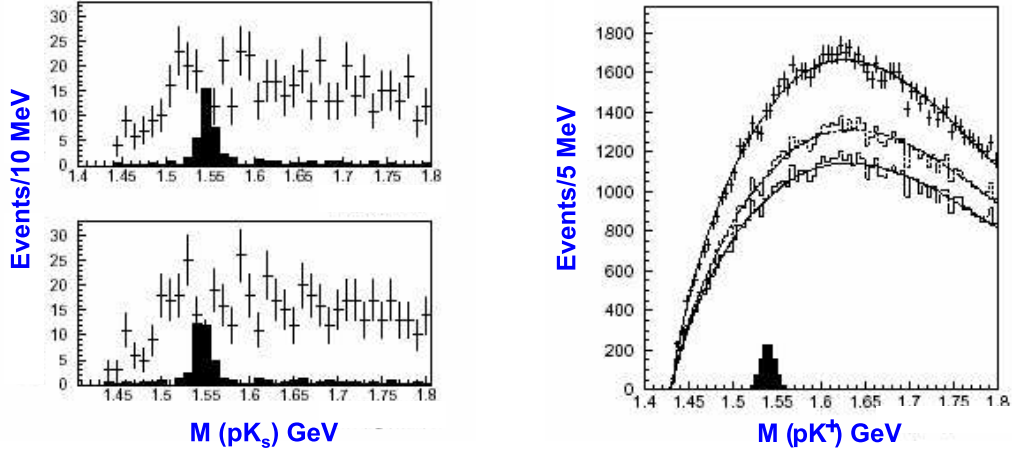

**Figure 12.** On the left are the effective $pK_s$ mass spectra for the reaction $pN \to p\,K_s\,K_s\,N$ from the SPHINX Collaboration of Antipov *et al.* [37]. The upper and lower plots have a bin shift of 5 MeV. The black histogram displays the Monte Carlo signal assuming a cross-section for $\Theta^+$ production to be 10% that for producing the $\Lambda(1520)$. The right hand plot shows the $pK^+$ mass distribution searching for a $\Theta^{++}$ baryon in the reaction $pN \to (pK^+)\,K^-\,N$. The three histograms with curves correspond to the results with no cuts, with $\Lambda(1520)$ and with $\Lambda(1520)\phi$ cuts. The black histogram depicts the Monte Carlo signal assuming the production cross-section for a $\Theta^{++}$ being 1% of that for the $\Lambda(1520)$.

Many experiments, not listed in Table 1, have scoured their data for such states and have not found them, most notably BaBar [34]. What are needed are higher statistics experiments and a common analysis approach. This is highlighted by the fact that H1 at HERA [35] claims to see a possible charmed companion of the $\Theta$ (with a $\bar{c}$ rather than an $\bar{s}$) the $\Theta_c^+$, but ZEUS [36] does not. While we wait for the new run at CLAS at Jefferson Lab. we do just have a new analysis of the older data from the SPHINX Collaboration [37] at Protvino of 70 GeV/$c$ collisions of protons on both carbon and nucleon targets. They look for the production of the $\Theta^+(1540)$ in conjunction with a $\overline{K}^0$, where the $\Theta$ is sought in $pK_s$ with the $nK_s$ and $pK^+$ channels as controls, Fig. 12. No signal of the $\Theta$ is seen, either a $\Theta^+$ of the decuplet of Fig. 10 or the $\Theta^{++}$ of a possible **27** pentaquark multiplet. The exact limits deduced can be found in Ref. [37] by Antipov *et al.* Hopefully more data from dedicated experiments will clarify whether the putative pentaquark states are *dead or alive*.

The existence of baryons with an essential diquark component, makes it highly likely that mesons with four quarks $\overline{qq}qq$ should occur. We have for long had hints [8] that the scalars $f_0(980)$ and $a_0(980)$, which we will discuss in more detail in the final lecture, have a close affinity with $K\overline{K}$. The $a_0^+$, for example, might be a tetraquark $[us][\overline{ds}]$. However, their dynamics is complicated as we will explain. In the charm sector there have also been surprises in the past year, several narrow states like the $X(3872)$ [38] and the $D_{sJ}(2317),(2460),(2632)$ [39]. The $X(3872)$ is a state of hidden charm, that does not fit in with the known structure of charmonium [40]. It sits very close to $D^*\overline{D}$ threshold, and could very well be a tetraquark state made of $[cn][\overline{cn}]$, with $n = u,d$. Similarly, the very narrow $D_{sJ}$ states have affinity with $D\overline{K}$-like channels and may be $[cn][\overline{sn}]$ tetraquarks. At last we may well be uncovering evidence that QCD has a far richer spectrum than the simple quark model we previously reviewed. Something that has been long expected [7], but never uncovered with much certitude till now.

## 3. Low energy hadron dynamics

Having discussed the hadron spectrum in sketch form, we now focus on some details that are a key to the structure of the QCD vacuum. If we look at the spectrum of hadrons, in particular mesons, we are struck by the fact that the pseudoscalars are so much lighter than all other mesons, Fig. 13. Indeed, in the quark model these have $L = 0$, like the vector mesons of Fig. 1. They only differ by having $S = 0$ rather than $S = 1$, so why is their mass splitting so great with the pion especially light.

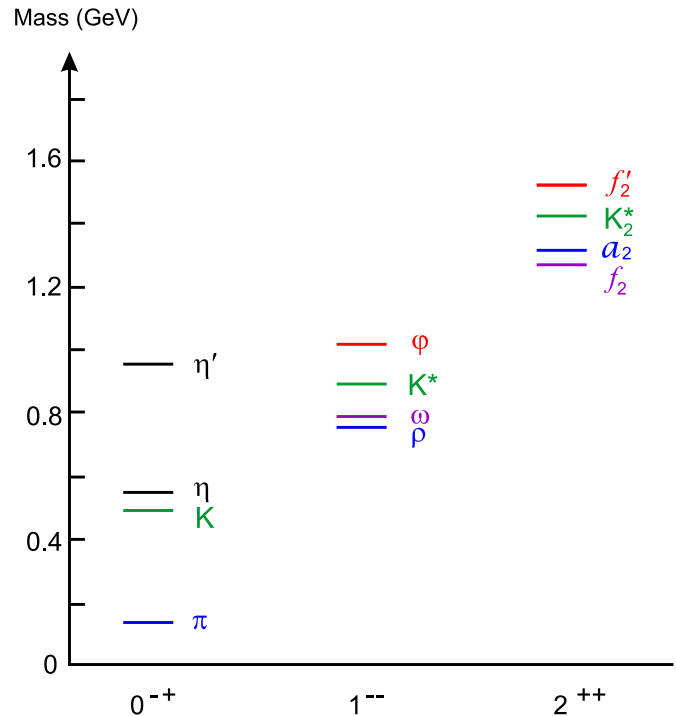

**Figure 13.** The masses of typical hadrons with quantum numbers labelled by $J^{PC}$, where $J$ is spin, $P$ parity and $C$ charge conjugation.

Pions were first predicted by Yukawa to be the mediators of the strong nuclear force that binds protons and neutrons to make nuclei. Yukawa made the prediction based on the maximum range of nuclear forces of $10^{-14}$ m and so pions are the lightest of all hadrons. What is surprising is that they are so much lighter than any other hadron, Fig. 13. A typical meson like the $\rho$ has a mass of 770 MeV, while the nucleon has a mass of 940 MeV. This accords with idea that the mass of a constituent $u$ or $d$ quark is $\sim 300$ MeV. In contrast the pion only weighs in at 140 MeV — a fifth as heavy as the $\rho$, yet composed of the very same quarks. Indeed, since it is the square of the mass on which dynamics depends, the pion has only 3% of the mass$^2$ of other typical mesons, Fig. 13. This fact has, of course, been known for a very long time. It is nevertheless remarkable that on the hadron scale of 1 GeV$^2$, pions are so very light, almost massless. This surely cannot be an accident. This lecture will describe what this teaches us about the QCD vacuum. There is nothing factual here that is original, only the telling of the story [41].

The world of quarks and gluons is described by QCD with a small number of basic interactions. In contrast the hadron world can be represented by an effective theory incorporating every possible hadron interaction. We do know from experiment that protons and neutrons interact in the same way: they have the same strong interactions, similarly for pions whatever their charge.

This we explain by requiring hadronic interactions respect isospin symmetry. The masses of the proton and neutron are almost equal. This symmetry we expect to be true at the level of QCD. All the quarks have the same interaction with gluons. Only their masses are different. If the near equality of the neutron and proton masses means that the *up* and *down* quarks have the same mass, then QCD will also have this isospin or $SU(2)$ symmetry. We can interchange *up* and *down* quarks and QCD is unchanged, if their masses are equal. In as much as *strange* quarks are not much heavier, the QCD Lagrangian has an approximate $SU(3)$ symmetry too.

Experiment tells us, for instance from the distribution of hadronic jets in $e^+e^-$, that quarks have spin $1/2$, as discussed in sect. 2. The construction of hadronic wavefunctions tells us they obey the Pauli exclusion principle and hence follow Fermi-Dirac statistics. To proceed let us recall some rather particular properties of fermions that will be crucial for understanding low energy QCD. Let us begin with some basics. Fermions can have helicity $\pm 1/2$. They can have their spin either pointing along the direction of their motion or in the opposite direction, Fig. 14. Consider a fermion, which for some observer has helicity $+1/2$. If we now run faster than this

## left-right symmetry

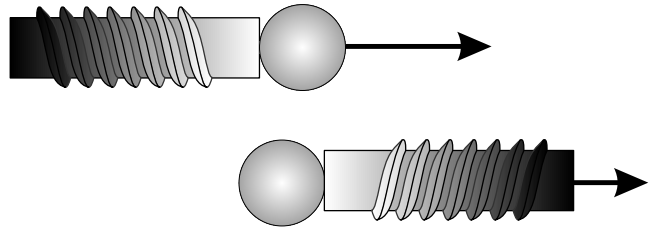

**Figure 14.** Illustration of the helicity of a spin $1/2$ particle as being left or right-handed.

fermion then we would see it moving backwards and its helicity would now be $-1/2$. Helicity is relative to the observer. But if the fermion is massless, it moves at the speed of light. Then if its helicity is along the direction of motion for one observer it is like that for all inertial observers. The world with *plus* helicity, and that with *minus* become decoupled, Fig 14. The fermions now have definite handedness. They are then said to be *chiral*.

To formalise this, let us consider the Dirac equation. If $\psi(x)$ is the solution of this equation with no mass then

$$i\,\not{\partial}\,\psi \;=\; 0 \quad , \tag{9}$$

where $\not{\partial} \equiv \gamma_\mu \partial^\mu$. Recall that the $\gamma_5$ operator, defined by $\gamma_5 = i\gamma_0\gamma_1\gamma_2\gamma_3$, anticommutes with the other four $\gamma_\mu$'s ($\mu = 0,1,2,3$). Consequently, if we multiply Eq. (9) by $\gamma_5$, we then have

$$i\,\gamma_5\,\not{\partial}\,\psi \;=\; -i\,\not{\partial}\,\gamma_5\,\psi \;=\; 0 \tag{10}$$

and we see that $\gamma_5\,\psi(x)$ is also a solution of the massless Dirac equation. With $\psi(x)$ and $\gamma_5\,\psi(x)$ both solutions, clearly any linear combination of them is too. It is useful to form the combinations:

$$\psi_L(x) \;=\; \frac{1}{2}\,(\mathbf{1} - \gamma_5)\,\psi \quad , \quad \psi_R(x) \;=\; \frac{1}{2}\,(\mathbf{1} + \gamma_5)\,\psi \quad . \tag{11}$$

where $\mathbf{1}$ is a unit $4 \times 4$ matrix. The subscripts $L, R$ refer to left and right-handed in the sense of a right-handed corkscrew, Fig. 14. So right-handed means the massless fermion has helicity $+1/2$ and left-handed has helicity $-1/2$.

Now the Lagrangian for a massless non-interacting fermion is

$$\mathcal{L} = i\overline{\psi}\,\displaystyle{\not{\partial}}\,\psi \quad. \tag{12}$$

from which the Euler-Lagrange equation is just the Dirac equation, Eq. (9). We can rewrite this Lagrangian, Eq. (12), as

$$\mathcal{L} = i\overline{\psi}_L\,\displaystyle{\not{\partial}}\,\psi_L + i\overline{\psi}_R\,\displaystyle{\not{\partial}}\,\psi_R \quad. \tag{13}$$

and see that the *chiral* fields $\psi_L$ and $\psi_R$ do not couple to each other. Consequently, the Lagrangian is seen to be invariant under a global chiral phase change

$$\psi_L(x) \to e^{i\alpha_L}\,\psi_L(x) \quad, \quad \psi_R(x) \to e^{i\alpha_R}\,\psi_R(x) \tag{14}$$

with $\alpha_L$, $\alpha_R$ real constant phases.

Now let us embody these ideas in a theory of hadron interactions. Though this will be very simple, it will incorporate some of the key aspects of the real world.

## 4. The Sigma Model

We introduce *Isospin symmetry* into a Lagrangian with the nucleon field $\psi$ and the pion field $\underline{\pi}$. The nucleon is a two component isospinor $\psi = \begin{pmatrix} p \\ n \end{pmatrix}$, while the pion is a 3-component vector. The Lagrangian of nucleons and pions and their interactions is

$$\mathcal{L} = \overline{\psi}\,(i\,\displaystyle{\not{\partial}} - m)\,\psi + \frac{1}{2}\left[\partial^\mu\underline{\pi}\cdot\partial_\mu\underline{\pi} - m_\pi^2\,\underline{\pi}\cdot\underline{\pi}\right] + ig\overline{\psi}\,\underline{\tau}\cdot\underline{\pi}\gamma_5\psi - \frac{\lambda}{4}\left(\underline{\pi}\cdot\underline{\pi}\right)^2 \quad, \tag{15}$$

where $m$ is the $2\times 2$ diagonal nucleon mass matrix and $\underline{\tau}$ is the 3-component vector of $2\times 2$ Pauli matrices, $g$ is the $\pi NN$ coupling and $\lambda$ the self-coupling of pions. This Lagrangian is invariant under global $SU(2)$ rotations

$$U = \exp\left(\frac{i}{2}\underline{\tau}\cdot\underline{\alpha}\right) \tag{16}$$

for any vector $\underline{\alpha}$ of constant phases, if as

$$\psi \to \psi' = U\psi \quad, \quad \underline{\tau}\cdot\underline{\pi} \to \underline{\tau}\cdot\underline{\pi}' = U\,\underline{\tau}\cdot\underline{\pi}\,U^\dagger \tag{17}$$

as then $\underline{\pi}\cdot\underline{\pi} = \underline{\pi}'\cdot\underline{\pi}'$.

We now make this theory of nucleon-meson interactions chirally symmetric. This means that not only must we have the pseudoscalar pions, but also their chiral partners — the scalar field, we call $\sigma$. This field has $I=0$ so only comes in an electrically neutral state. We introduce interactions between the nucleon field and the $\sigma$ with the same coupling as the nucleon has to the pion and we have an interaction potential, $V(\sigma,\underline{\pi})$, for the $\sigma$, $\pi$ fields that we will specify shortly. Thus

$$\mathcal{L} = i\overline{\psi}\,\displaystyle{\not{\partial}}\,\psi - g\overline{\psi}\,(\sigma - i\underline{\tau}\cdot\underline{\pi}\gamma_5)\,\psi$$

$$+ \frac{1}{2}\partial_\mu\underline{\pi}\cdot\partial^\mu\underline{\pi} + \frac{1}{2}\partial_\mu\sigma\,\partial^\mu\sigma - V(\sigma,\underline{\pi}) \quad. \tag{18}$$

Notice that as yet there is no explicit mass term for the nucleon. Nambu [42] proposed a very simple form for the interaction potential which involves quadratic and quartic terms as:

$$V(\sigma,\underline{\pi}) = \frac{\mu^2}{2}\left(\sigma^2 + \underline{\pi}^2\right) + \frac{\lambda}{4}\left(\sigma^2 + \underline{\pi}^2\right)^2 \quad. \tag{19}$$

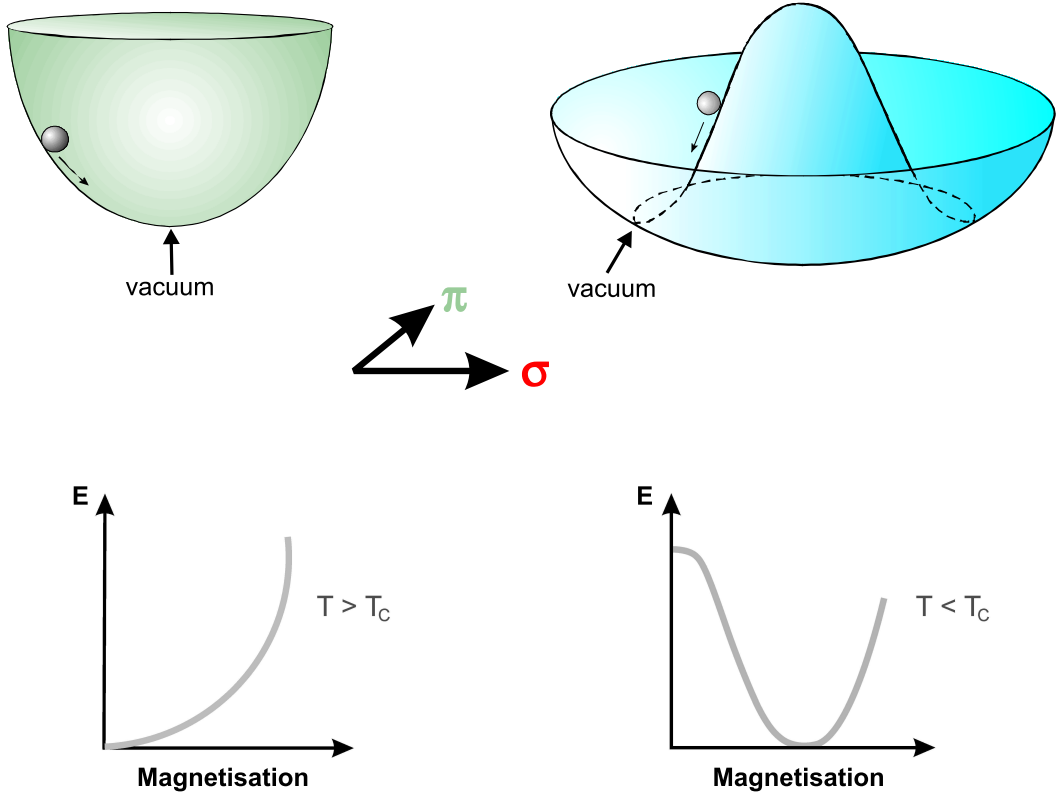

**Figure 15.** The upper pictures illustrate two possible shapes for the potential for the interaction of the $\sigma$ and $\underline{\pi}$ fields. While both are chirally symmetric, the minimum of the potential on the left occurs when both fields are zero, that on the right occurs with a non-zero expectation value for $\sigma^2 + \underline{\pi}^2$. Nature chooses the ground state close to the point labelled *vacuum* as discussed in the text. A slice through each potential is displayed in the lower pictures. These are the analogues of the Gibbs free energy as a function of magnetization for a ferromagnet above and below the critical temperature.

If $\mu^2$ is positive then $\mu$ is the mass of the $\sigma$ and $\pi$ fields, and the interaction potential looks like the picture on the left of Fig. 15. The ground state, which is defined by the lowest energy state, lies at the bottom of the potential, where both the $\pi$ and $\sigma$ fields are zero. The quantum fluctuations about this minimum that correspond to the physical states, rise up the potential well in all directions and so the pion and $\sigma$ have related masses. However, as Nambu noticed, if $\mu^2 < 0$, then the potential has the shape on the right: the famous Mexican hat. The minimum of the potential is round the rim where

$$\sigma^2 + \underline{\pi}^2 = -\frac{\mu^2}{\lambda} \equiv v^2 \quad . \tag{20}$$

Nature then spontaneously chooses one vacuum state. The ground state is at some point on this circle. Wherever it is, it breaks the symmetry between pion and scalar fields. The chiral symmetry is broken. In fact nature chooses the ground state close to the point

$$\sigma_0 + \sqrt{\frac{-\mu^2}{\lambda}} = 0 \quad , \qquad \underline{\pi}_0 = 0 \quad , \tag{21}$$

labelled "vacuum" on the right in Fig. 15. About this point the quantum fluctuations, the physical particles, are no longer symmetric. The fluctuations to left and right in the $\sigma$-direction rise up and down the sides of the Mexican hat and so are massive, while those in the pion direction are round the rim and feel no resistance. Consequently, the pions are massless. They are the (pseudo-)Goldstone bosons [43] of chiral symmetry breaking. We can see this formally by defining the fluctuations of the scalar field about the minimum of the potential at $\sigma_0$, so $\hat{\sigma} = \sigma - \sigma_0$, then

$$\mathcal{L} = \overline{\psi}\,(i\,\slashed{\partial} - g\,v)\,\psi - g\,\overline{\psi}\,(\hat{\sigma} - i\,\underline{\tau}\cdot\underline{\pi}\,\gamma_5)\,\psi$$

$$+ \frac{1}{2}\left[\partial_\mu \hat{\sigma}\,\partial^\mu \hat{\sigma} - \lambda\,(3\sigma_0^2 - v^2)\,\hat{\sigma}^2\right] + \frac{1}{2}\left[\partial_\mu \underline{\pi}\cdot\partial^\mu \underline{\pi} - \lambda\,(\sigma_0^2 - v^2)\,\underline{\pi}\cdot\underline{\pi}\right]$$

$$-\lambda\,\sigma_0\,(\sigma_0^2 - v^2)\,\hat{\sigma} - \lambda\,\sigma_0\,\left(\hat{\sigma}^2 + \underline{\pi}^2\right) - \frac{\lambda}{4}\left[\left(\hat{\sigma}^2 + \underline{\pi}^2\right)^2 + \sigma_0^2(\sigma_0^2 - 2v^2)\right]. \quad (22)$$

From this Lagrangian, we can read off the square of the pion mass from the coefficient of the $\underline{\pi}\cdot\underline{\pi}$ term giving

$$m_\pi^2 = \lambda\,\left(\sigma_0^2 - v^2\right)\quad. \quad (23)$$

Now with the minimum of the $\sigma$ field, $\sigma_0$, having an expectation value of magnitude $v$ as in Eqs. (20,21), the pion mass vanishes as anticipated by Goldstone's theorem [43]. While the pion mass is zero, chiral symmetry breaking gives the $\hat{\sigma}$ a mass of $\sqrt{2\lambda}\,v = \sqrt{-2\mu^2}$ and the nucleon a mass of $gv$.

It is helpful in constructing a general Lagrangian with chiral symmetry, which we do later, to rewrite the pion and sigma fields in a non-linear form, $\mathcal{U}$, by defining the combination

$$\sigma + \underline{\tau}\cdot\underline{\pi} = -(v + S)\mathcal{U} \quad (24)$$

where

$$\mathcal{U} = \exp\left(\frac{i\,\underline{\tau}\cdot\underline{\pi}'}{v}\right)\quad, \quad (25)$$

from which we see that $\underline{\pi}'$ and $\underline{\pi}$, and $S$ and $\hat{\sigma}$, are related. In terms of $\mathcal{U}$ the Lagrangian of Eq. (22) becomes on setting $\sigma_0 = -v$

$$\mathcal{L} = i\overline{\psi}_L\,\slashed{\partial}\psi_L + i\overline{\psi}_R\,\slashed{\partial}\psi_R + g\,(v + S)\left(\overline{\psi}_L\,\mathcal{U}\,\psi_R + \overline{\psi}_R\,\mathcal{U}^\dagger\,\psi_L\right)$$

$$+\frac{1}{2}\left[(\partial_\mu S)^2 - 2\mu^2\,S^2\right] + \frac{1}{4}\,(v+S)^2\,\mathrm{Tr}\left(\partial_\mu \mathcal{U}\,\partial^\mu \mathcal{U}^\dagger\right) + \lambda\,v\,S^3 - \frac{1}{4}\lambda\,S^4\,, \quad (26)$$

where we have dropped irrelevant constant terms and where we have decomposed the nucleon field, $\psi$, into its right and left components as in Eq. (11). We see that it is only through $\mathcal{U}$ (the chiral fields) that $\psi_L$ and $\psi_R$ couple. This (non-linear) representation of the sigma model contains identical physics to the original theory, but is in a form well suited to extracting the general predictions of chiral dynamics. It is invariant under $SU(2)$ rotations $L$ in the left hand space and $R$ in the right hand space, like those of Eq. (14), where

$$\mathcal{L} \Longrightarrow \mathcal{L}\quad, \qquad \mathrm{as}\quad \mathcal{U} \Longrightarrow L\,\mathcal{U}\,R^\dagger\quad. \quad (27)$$

The interaction of pions alone is embodied in just one term of the Lagrangian of Eq. (26), viz.

$$\mathcal{L}_{\mathrm{eff}} = \frac{1}{4}v^2\,\mathrm{Tr}\left(\partial_\mu \mathcal{U}\,\partial^\mu \mathcal{U}^\dagger\right)\quad. \quad (28)$$

With $v$ related to the the pion decay constant, we will see Eq. (28) is the basis of the most general pion interactions in any chiral theory, in particular QCD, to which we will turn in the next section.

Since pions, of course, have mass, chiral symmetry is explicitly broken. It is helpful to consider this first in the context of the sigma model, before we study QCD. Explicit breaking in the sigma model is typically given by

$$\mathcal{L}_{breaking} = c\sigma = \gamma \mu^2 v \sigma \quad, \tag{29}$$

where $c$ is a dimensionful constant, which can be suitably rewritten in terms of the dimensionless constant $\gamma$. This potential with its broken symmetry still has three turning points as in Fig. 15. These are determined by

$$\frac{\partial}{\partial \sigma} V = 0 \quad, \qquad \frac{\partial}{\partial \underline{\pi}} V = 0 \quad. \tag{30}$$

These conditions give

$$\left(-\frac{\mu^2}{\lambda} + \sigma^2 + \underline{\pi}^2\right)\sigma + \frac{c}{\lambda} = 0 \quad, \qquad \left(-\frac{\mu^2}{\lambda} + \sigma^2 + \underline{\pi}^2\right)\underline{\pi} = 0 \quad. \tag{31}$$

All three turning points have $\underline{\pi} = 0$ but instead of being at $\sigma = -v, 0, v$, the potential in Fig. 15 is tilted and the extrema satisfy

$$\sigma^3 - v^2 \sigma + \gamma v^3 = 0 \quad, \tag{32}$$

i.e. at

$$\begin{aligned}
\sigma_1 &= -v\left(\cos\beta + \frac{1}{\sqrt{3}}\sin\beta\right) \quad, \\
\sigma_2 &= \frac{2v}{\sqrt{3}}\sin\beta \quad, \\
\sigma_3 &= v\left(\cos\beta - \frac{1}{\sqrt{3}}\sin\beta\right) \quad,
\end{aligned} \tag{33}$$

where

$$\sin 3\beta = \frac{3\sqrt{3}}{2}\gamma \quad. \tag{34}$$

The solution, $\sigma_1$, corresponds to the absolute minimum labelled *vacuum* on the right in Fig. 15. Consequently, substituting $\sigma_0 = \sigma_1$ into the formula for the pion mass given in Eq. (23)

$$m_\pi^2 = -\frac{2}{\sqrt{3}}\mu^2 \sin\beta \left(\cos\beta - \frac{1}{\sqrt{3}}\sin\beta\right) \quad. \tag{35}$$

If the symmetry breaking in Eq. (29) is small, i.e. $\gamma \ll 1$, then from Eqs. (35,34) we see

$$m_\pi^2 \simeq -\mu^2 \gamma \left(1 - \frac{\gamma}{2}\cdots\right) \quad, \tag{36}$$

where recall $\mu^2 < 0$. The origin of this explicit breaking is crucially related to the structure of the vacuum.

We now turn to what we believe is the underlying theory of the strong interaction and study how these fundamental ideas emerge from QCD.

## 5. QCD

We first recall the QCD Lagrangian with 2 flavours of quark with $\psi = \begin{pmatrix} u \\ d \end{pmatrix}$ and mass $M = \begin{pmatrix} m_u \\ m_d \end{pmatrix}$ so that

$$\begin{aligned}
\mathcal{L} &= \overline{\psi}\,(i\gamma_\mu \mathcal{D}^\mu - M)\,\psi - \frac{1}{4}\mathcal{G}_{\mu\nu}\mathcal{G}^{\mu\nu} \\
&= \overline{u}\,(i\gamma_\mu \mathcal{D}^\mu - m_u)\,u + \overline{d}\,(i\gamma_\mu \mathcal{D}^\mu - m_d)\,d - \frac{1}{4}\mathcal{G}_{\mu\nu}\mathcal{G}^{\mu\nu} \quad,
\end{aligned} \quad (37)$$

$\mathcal{G}^{\mu\nu}$ is the gluon field strength tensor, the non-Abelian generalization of the electromagnetic $\mathcal{F}^{\mu\nu}$. That the hadron world has the well known isospin symmetry is a consequence of symmetry at the quark level. The fact that the neutron and proton have almost the same mass means within the quark model that the masses of the $d$ and $u$ quarks should be equal. Indeed their constituent masses are both about 300 MeV. This means QCD has an $SU(2)$ symmetry

$$\mathcal{L} \to \mathcal{L} \qquad \text{when} \qquad \psi \to \psi' = \exp(i\underline{\tau}\cdot\theta)\,\psi \qquad (38)$$

provided $m_u = m_d$. However, the mass that enters the Lagrangian of Eq. (37) is the mass of the quark when it travels over very short distances, not across a whole hadron. Then its mass is the so called *current* mass. For $u$ and $d$ quarks these are a few MeV, compared to the constituent quark masses of 300-350 MeV. Now, the scale of QCD is fixed on renormalization by the momentum scale at which the strong interaction is strong. This scale $\Lambda_{QCD} \sim 100-200$ MeV. The current masses of the $u$ and $d$ quarks are thus very much smaller than $\Lambda_{QCD}$ and so almost massless. If we separate the quark wavefunction into left-handed and right-handed components, it is only the mass term in the Lagrangian that couples left and right:

$$m_u\,u\overline{u} + m_d\,d\overline{d} = m_u\,(\overline{u_L}u_R + \overline{u_R}u_L) + m_d\,(\overline{d_L}d_R + \overline{d_R}d_L)\,. \qquad (39)$$

To a very good approximation

$$m_u\,,\ m_d \ll \Lambda_{QCD} \qquad \Longrightarrow \qquad m_u = m_d = 0 \quad.$$

As we have already seen when fermions are massless the left and right-handed components decouple, and so the Lagrangian has a bigger symmetry. It has an $SU(2)_L \otimes SU(2)_R$ chiral symmetry. Thus

$$\mathcal{L} \to \mathcal{L}$$

under the transformation

$$\psi_L \to \exp(i\theta_L \cdot \underline{\tau})\,\psi_L \quad;\quad \psi_R \to \exp(i\theta_R \cdot \underline{\tau})\,\psi_R \quad, \qquad (40)$$

that define the transformations $L$ and $R$ of Eq. (27). This chiral symmetry can equally well be regarded as a vector−axial-vector symmetry. Defining

$$\theta_V = \frac{1}{2}\,(\theta_L + \theta_R) \quad,\qquad \theta_A = \frac{1}{2}\,(\theta_L - \theta_R) \quad, \qquad (41)$$

then we can see the Lagrangian is invariant under the transformation

$$\psi \to \exp(i\theta_V \cdot \underline{\tau})\,\psi \quad,\qquad \psi \to \exp(i\theta_A \cdot \underline{\tau}\gamma_5)\,\psi \quad. \qquad (42)$$

A rotation through an equal angle to left and right corresponds to a vector transformation, while a rotation by equal but opposite angles to right and left gives an axial transformation. The QCD Lagrangian with 2 massless quark flavours has the symmetry

$$SU(2)_L \otimes SU(2)_R \quad \Longrightarrow \quad SU(2)_V \otimes SU(2)_A \quad . \tag{43}$$

While the basic symmetry between up and down quarks is reflected in the hadron world by its isospin symmetry, the bigger chiral symmetry does not appear in the world of hadrons. Scalars and pseudoscalars, vectors and axial-vectors are not simply related in masses and interactions. This as we have seen is because of the way nature chooses the ground state or *vacuum*. So let us consider some simple ideas about the vacuum to begin with and then see how we can characterise its complex nature.

In an atom, like hydrogen, the electron moves through a vacuum as it orbits. Once upon a time we thought this vacuum was empty. The discovery of the uncertainty principle and the quantum mechanical nature of forces means it is not quite so simple. The electron continually emits and absorbs virtual photons, as does the positively charged nucleus. Indeed, photons bounce backwards and forwards between the electron(s) and the nucleus in some sort of cosmic dance that holds them bound. The discovery by Dirac of the consequences of a relativistic quantum theory that these photons could borrow enough energy to form $e^+e^-$ pairs for $10^{-21}$ s means that our single electron in its orbit is sometimes two electrons and a positron and even three electrons and two positrons for short lengths of time. Processes that Feynman depicted in his famous loop diagrams, Fig. 16. We can then think of the electron as *swimming* in the Dirac sea past shoals of electron-positron pairs and occasionally teaming up with these in its flight round the nucleus. Thus the vacuum is not empty, but filled with particle-antiparticle pairs, here generated by electromagnetic interactions.

The vacuum of the the strong interaction is not surprisingly still more complicated. We have discovered that the force that binds quarks together to make hadrons has very interesting properties. Experiments on deep inelastic scattering teach us that "yes" there are quarks inside a proton just as the simple quark model picture would lead us to expect, but these quarks move inside the proton over short distances ($\leq 10^{-17}$ m) as though they are free; yet they cannot get out. They are confined within the femto-universe. The effective coupling between quarks and gluons behaves as shown in Fig. 17. This behaviour is automatically embodied in the non-Abelian gauge theory we know as QCD.

We have seen that the fact that the up and down quarks are very nearly massless means that QCD has a chiral symmetry. A symmetry that is self-evidently broken at the hadron level. We have seen that Nambu proposed, long before QCD was discovered, how this symmetry can be spontaneously broken, resulting in the $\sigma$ field having a non-zero vacuum expectation value, while the pseudoscalars are massless. It is natural to identify these Goldstone modes with pions, or in a world with 3 massless quark flavours with the $\pi$, $K$ and $\eta_8$ (the octet $I=0$ member of the pseudoscalar $\overline{q}q$ multiplet). This picture is intuitively similar to the way a ferromagnet breaks rotational symmetry. Consider a lump of iron. For simplicity imagine it is spherical. Then above some critical temperature the magnetons associated with each atom point in random directions. The ball can be oriented in any direction and will look the same. There is rotational symmetry and the minimum energy state is one with no magnetization, cf. on the lower left in Fig. 15. However, as the ball cools, suddenly all the little Bohr magnetons point in one direction (on Earth usually aligned with the ambient magnetic field of the planet), resulting in the state of lowest energy having non-zero magnetization. The ground state no longer has full rotational symmetry, cf. on the lower right in Fig. 15. This illustrates the Goldstone theorem [43] within Nambu's ferromagnetic analogy. This is only a model of the hadron world in which the chiral symmetry is spontaneously broken. But what is happening at the quark and gluon level? Condensates are formed in the vacuum: condensates of quarks, antiquarks and gluons. Which specific condensate drives this dynamical breakdown of chiral symmetry is an issue we will discuss in sect. 6.

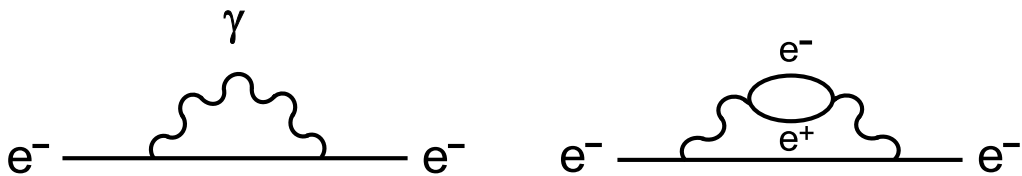

**Figure 16.** Feynman diagrams showing an electron propagating as an electron and a virtual photon, and as an electron and a virtual $e^+e^-$ pair.

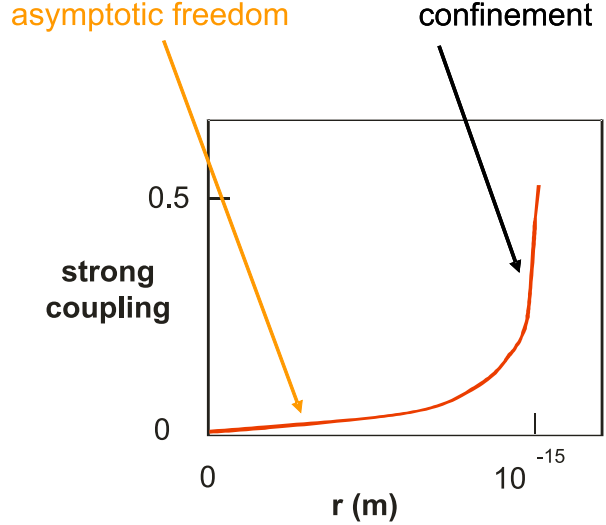

**Figure 17.** How the QCD coupling of interacting quarks and gluons behaves with distance. At short distances the coupling is weak and described as *asymptotic freedom*. Over a distance of a fermi, the coupling becomes strong and this is believed to responsible for *confinement*.

The world of hadrons is described by an effective Lagrangian embodying all the possible hadronic interactions each with its own coupling. We know that this Lagrangian must not only respect isospin symmetry, but the underlying chiral symmetry of $SU(N_f) \otimes SU(N_f)$, with at least $N_f = 2$, but could be 3 if we regard the strange quark as nearly massless too. This invariance is all that is needed to determine the interactions of the corresponding Goldstone bosons. Thus $\pi\pi \to \pi\pi$ scattering has its structure specified by chiral symmetry alone, just as we discussed earlier in sect. 5. The most elegant way to embody this symmetry into an effective Lagrangian of pion interactions is to express this in terms of the exponential representation $\mathcal{U}$ introduced in Eq. (25). We then write down the most general Lagrangian invariant under the chiral transformation of Eq. (27). Since $\mathrm{Tr}\left(\mathcal{U}\,\mathcal{U}^\dagger\right) = 2$ is a constant, the Lagrangian for massless pions must involve derivatives of $\mathcal{U}$ in Lorentz invariant combinations. These can be ordered according to the number of derivatives, those with $n$ derivatives contributing to the sub-Lagrangian, $\mathcal{L}_n$, where $n$ is even. Thus quite generally we can write

$$\begin{aligned}
\mathcal{L} &= \mathcal{L}_2 + \mathcal{L}_4 + \cdots \\
&= \frac{v^2}{4} \mathrm{Tr}\left(\partial_\mu \mathcal{U}\, \partial^\mu \mathcal{U}^\dagger\right) \\
&\quad + \alpha_1 \left[\mathrm{Tr}\left(\partial_\mu \mathcal{U}\, \partial^\mu \mathcal{U}^\dagger\right)\right]^2 + \alpha_2\, \mathrm{Tr}\left(\partial_\mu \mathcal{U}\, \partial_\nu \mathcal{U}^\dagger\right) \cdot \mathrm{Tr}\left(\partial^\mu \mathcal{U}\, \partial^\nu \mathcal{U}^\dagger\right) + \cdots,
\end{aligned} \quad (44)$$

where $v, \alpha_1, \alpha_2 \cdots$ are a series of constants determinable from experiment. Since the vacuum expectation value $v$ has dimensions of mass (indeed we will see that it is related to the pion decay constant $f_\pi$), the $\pi\pi$ amplitude (at c.m. energy) $E$ is given by $\mathcal{L}_2$ alone to $O(E^2)$.

Explicit symmetry breaking is provided by terms proportional to $\text{Tr}\left(\mathcal{U} + \mathcal{U}^\dagger\right)$, the addition of which means that the vectorial isospin symmetry remains intact, but the axial symmetry is broken, cf. Eq. (43). Then very similarly to the example in Eq. (29) (where $c = \gamma \mu^2 v$)

$$\mathcal{L}_{breaking} = \frac{cv}{4}\text{Tr}\left(\mathcal{U} + \mathcal{U}^\dagger\right) = cv - \frac{c}{2v}\underline{\pi}\cdot\underline{\pi} + \cdots \tag{45}$$

from which we can read off that

$$m_\pi^2 = \frac{c}{v} + \cdots, \tag{46}$$

just as in the first term of Eq. (36). Applying the operator product expansion to QCD, the pion mass can be related to the quark masses by

$$m_\pi^2 = (m_u + m_d)\,B_0 + (m_u + m_d)^2\,C_0 + (m_u - m_d)^2\,D_0 + \cdots. \tag{47}$$

The relative size of the terms in this expansion underlies different orderings [44] of the Feynman rules from the Lagrangians, $\mathcal{L}_n$, of Eq. (44), which we will discuss in the next section.

What we will see is that while the $u$ and $d$ quarks propagate over short distances as though they are very light, over the size of a hadron they move as though their mass is 300 MeV. They behave like ants running through treacle. A short way in they scurry along at almost the speed of light, but as they get more and more syrup on their boots, they slow up becoming increasingly massive, Fig. 18. The vacuum in QCD is more complex than that of QED. It is not just that there is a sea of $q\overline{q}$ pairs, and clouds of gluons, but these interact so strongly that they form condensates, making the vacuum a denser treacle-like medium. What sets the scale for the dominant condensate will be considered later. Low energy $\pi\pi$ scattering allows us to determine this.

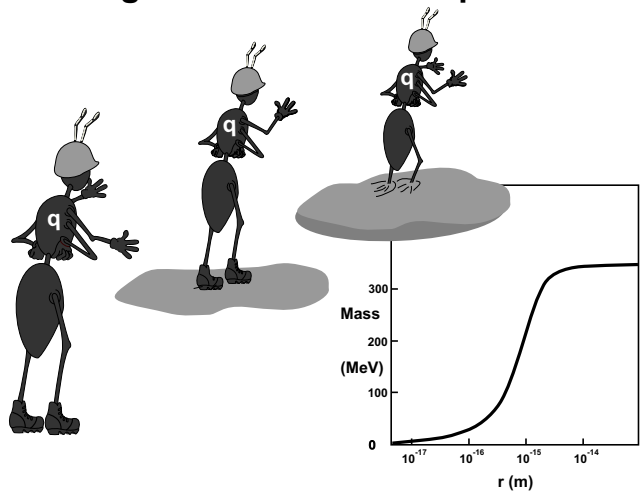

**Figure 18.** A cartoon representation of a quark propagating. Over short distances it is light and moves quickly through the treacle-like medium. Over longer distances its boots become filled with treacle. Consequently, the quark moves much more slowly, corresponding to the dynamical generation of mass. The condensates in the vacuum provide the medium.

## 6. Pion-pion scattering

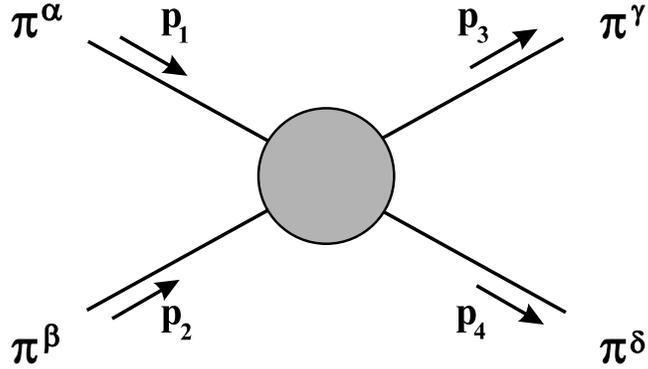

**Figure 19.** Scattering $\pi^\alpha \pi^\beta \to \pi^\gamma \pi^\delta$, where the labels $\alpha - \delta$ denote the isospin state, with 4-momenta $p_1 - p_4$.

Pions being the lightest of all hadrons play a very special role in the QCD. They are the Goldstone bosons of chiral symmetry breaking. Let us see what this predicts for experiment. We begin with a textbook discussion of 2-body scattering of identical spinless particles with isospin 1. First we discuss the kinematics, where the particles have 4-momenta $p_i$ — ($i = 1-4$) labelling the particles as in Fig. 19. Consider scattering in the centre-of-momentum frame, where each particle is on mass-shell so that $p_i^2 = m^2$. In this frame the process is specified by two variables: $E$ the energy of each particle both before and after the scattering, and angle $\theta$ of the scattered particles relative to their initial direction, as illustrated on the left in Fig. 20. The physical region for the scattering process $1\,2 \to 3\,4$ is specified by the allowed values of $E$ and $\theta$ given by

$$E \geq m \quad, \quad 0 \leq \theta \leq \pi \quad, \tag{48}$$

as shown on the right in Fig. 20.

These two variables can be expressed in terms of Lorentz invariants. We first define

$$s = (p_1 + p_2)^2 \quad, \tag{49}$$

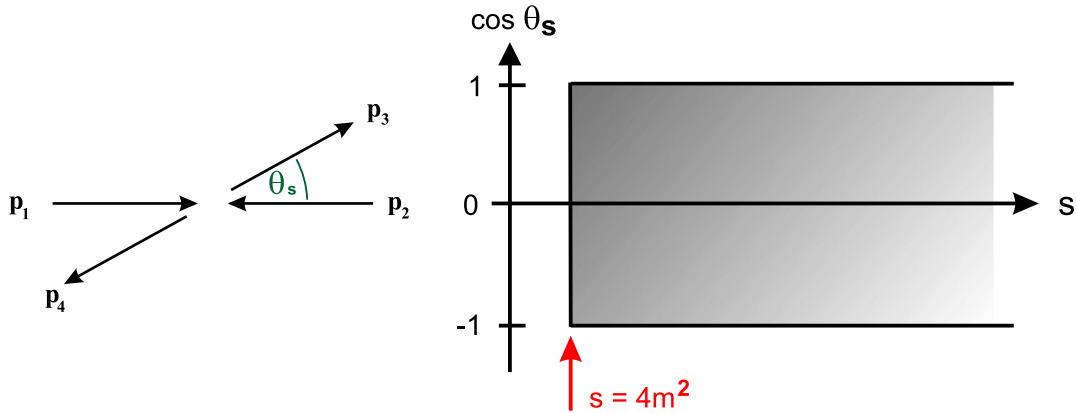

**Figure 20.** Kinematics of two body scattering of particles of equal mass $m$: on the left the scattering in the centre-of-mass frame showing the definition of the scattering angle $\theta$; on the right the region in the $s$-$\cos\theta$ plane allowed for the physical process.

which in the c.m. frame is the square of the total energy, so that $E = \sqrt{s}/2$. The c.m. scattering angle, $\theta$, is then related to 2 invariants $t$, $u$ defined by

$$t = (p_1 - p_3)^2 \quad , \quad u = (p_2 - p_3)^2 \quad , \tag{50}$$

so that

$$t = \frac{1}{2}\left(s - 4m^2\right)(1 - \cos\theta_s) \quad , \quad u = \frac{1}{2}\left(s - 4m^2\right)(1 + \cos\theta_s) \quad . \tag{51}$$

We can readily see in this frame that the 3 Mandelstam variables, $s$, $t$, $u$ are not independent but related by

$$s + t + u = 4m^2 \quad , \tag{52}$$

a relation that holds in every frame. The physical region of the scattering process specified by the allowed values of energy and angle of Eq. (48) is translated into

$$s \geq 4m^2 \, , \qquad t \leq 0 \, , \qquad u \leq 0 \tag{53}$$

displayed in Fig. 21.

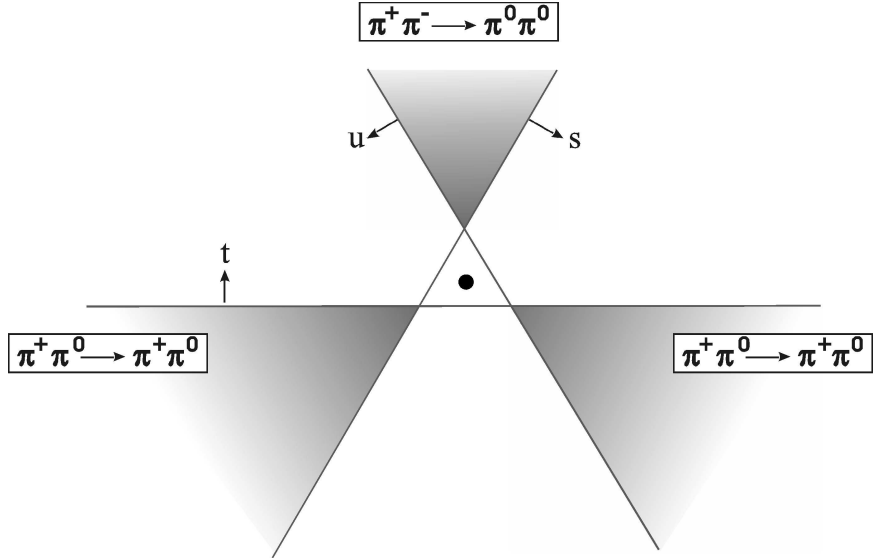

**Figure 21.** The Mandelstam plane defined by Eq. (53) displaying the physically allowed regions (shaded) for $\pi^+\pi^0 \to \pi^+\pi^0$ scattering in the $s$- and $u$-channels and the reaction $\pi^+\pi^- \to \pi^0\pi^0$ in the $t$-channel. The black dot denotes the (unphysical) symmetry point of the Mandelstam triangle, where $s = t = u = 4m_\pi^2/3$.

A key aspect of relativistic dynamics is that the scattering amplitude that describes the process $1\,2 \to 3\,4$ is the boundary value of the very same scattering amplitude that describes the reaction $1\,\bar{3} \to \bar{2}\,4$ and $1\,\bar{4} \to 3\,\bar{2}$, and their time reversed processes. The square of the c.m. energy of each of these channels are $t$ and $u$ respectively. We thus see that the $s$, $t$, $u$ channels are on an equal footing. They each have a physical region that can be displayed on a common diagram, the Mandelstam plane specified by Eq. (53) and its permutations, as shown in Fig. 21. We see that in the centre of the plane is a small triangle of height $4m^2$, which is below the threshold for each of the three channels and so is unphysical. Nevertheless, the behaviour of the $\pi\pi$ scattering amplitude in the Mandelstam triangle is key to chiral dynamics.

Even without chiral considerations, pions have always had a special position in the theoretical study of the strong interaction, because they are the lightest hadrons. From the fundamental axioms of field theory it has been shown that the scattering amplitude for $\pi\pi$ scattering satisfy three essential properties: (i) unitarity from the conservation of probability, (ii) crossing symmetry from relativity and (iii) a domain of analyticity deduced by Martin [45] from causality and the fact the pion is the lightest hadron. For $\pi^0\pi^0 \to \pi^0\pi^0$ scattering, with its manifest three channel crossing symmetry, the symmetry point of the Mandelstam triangle is then clearly a special point. Martin and others have been able to show that the scattering amplitude for $\pi^0\pi^0 \to \pi^0\pi^0$ has a minimum at this point with a value in the range [46, 45]

$$-100 \;\leq\; T(s=t=u=\tfrac{4}{3}m_\pi^2) \;\leq\; 16 \quad, \tag{54}$$

illustrated in Fig. 22. We will see that chiral dynamics, so essential to the nature of the physical pion, fixes the scattering amplitude well within this wide range. (For later it is useful to note that the amplitude for $\pi^0\pi^0 \to \pi^0\pi^0$ scattering, $T(s,t,u)$, pictured in Fig. 22 is related to the $\pi^+\pi^0 \to \pi^+\pi^0$ amplitude, $\mathcal{F}(s,t,u)$, by $T(s,t,u) = \mathcal{F}(s,t,u) + \mathcal{F}(t,u,s) + \mathcal{F}(u,s,t)$.)

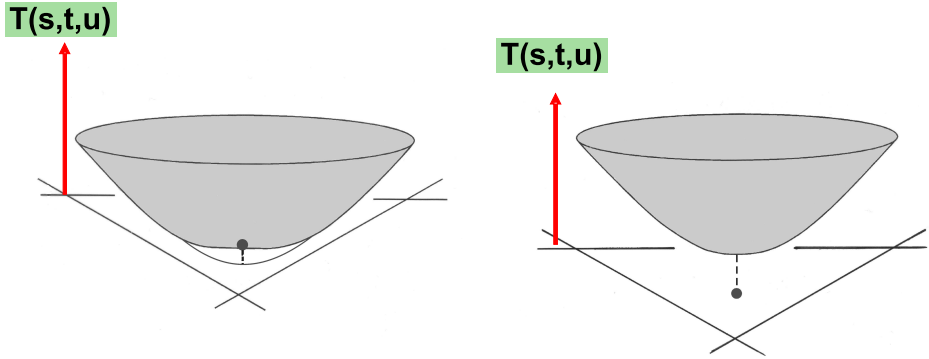

**Figure 22.** 3-dimensional plot of the amplitude, $T(s,t,u)$, for $\pi^0\pi^0 \to \pi^0\pi^0$ scattering over the Mandelstam triangle of Fig. 21. The black dot denotes the symmetry point of the triangle. The two pictures depict the lower and upper limit given in Eq. (54), provable from axiomatic field theory.

If pions are massless, as they almost are, the Mandelstam triangle shrinks to a point and the threshold of the $s$, $t$, $u$-channels all coincide at $s=t=0$. To see the significance of this point, consider the amplitude for pion decay, as shown in Fig. 23. A charged pion decays through an axial vector current $A^\mu$ to a lepton pair. This was long known. Since the '80s we have recognised this axial vector current as the off-shell $W$-propagator. Now the matrix element for this axial current with a pion state is a Lorentz vector. The only vector a spinless particle carries is its momentum $p^\mu$. Then

$$\langle \pi_i \mid A_i^\mu \mid O \rangle \;=\; i\sqrt{2}\, f_\pi\, p^\mu \quad, \tag{55}$$

where the label $i$ denotes the isospin state of the pion and the coefficient $f_\pi$ defines the rate of pion decay, which experiment fixes to be 92 MeV. Comparing with the axial vector current from the general Lagrangian, Eqs. (44,22,26), importantly tells us $v = f_\pi$. Now the pion carrying momentum $p$ has a plane wavefunction of form $\exp(-ip_\mu \cdot x^\mu)$. So if we consider the divergence of this axial vector current, the derivative brings down a factor of $-ip_\mu$. Thus

$$\langle \pi_i \mid \partial_\mu A_i^\mu \mid O \rangle \;=\; \sqrt{2}\, f_\pi\, p^2 \;=\; \sqrt{2}\, f_\pi\, m_\pi^2 \quad, \tag{56}$$

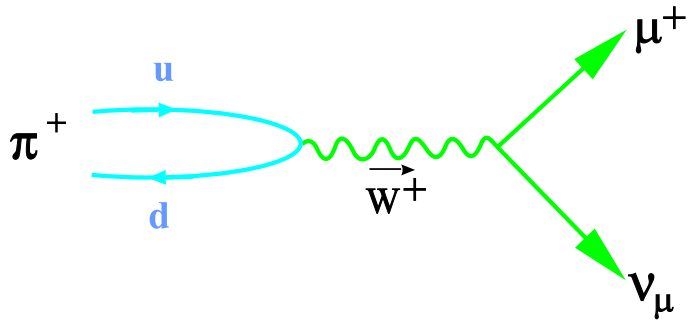

**Figure 23.** Feynman graph representing the decay of a $\pi^+ \to \mu^+ \nu_\mu$.

as the pion is on-shell. Now we see that if the pion is massless, then

$$\partial_\mu A^\mu = 0 \quad . \tag{57}$$

The axial-vector current is conserved. Of course, pions are not massless, but their mass squared is still very small much less than $m_\rho^2$ or $m_N^2$, Fig. 13.

If a current is conserved, this leads to a low energy theorem. The best known example is the conservation of the electromagnetic vectorial current. Consider Compton scattering, the scattering of a photon off an electron. At low energy the photon has long wavelength and only sees the charge of the electron. In this (the Thomson) limit, the scattering amplitude (Fig. 24) is given exactly by just 2 Feynman graphs: electron exchange in the $s$ and $u$ channels, where the coupling of the electron and photon defines the fine structure constant. Though in general the process $\gamma e \to \gamma e$ is given by an infinite number of Feynman graphs, in the Thomson limit all but the lowest order, Born, amplitude vanishes, as a result of the conservation of the vector current. Let us apply this same idea to the scattering of a massless pion off a nucleon, as then the pion will couple through a conserved axial-vector current. In the limit that the pion carries zero momentum, the scattering amplitude is given by the Born amplitude shown in Fig. 24, with nucleon exchange in the $s$ and $u$-channels, where the vertices have strength defined by the $\pi NN$ coupling for massless pions. This fixes the scattering amplitude for $\pi N$ scattering. Of course, pions are not massless and the axial vector current is not exactly conserved. Nevertheless, this is sufficiently close to the physical situation that we expect the corrections to be small $\sim O(m_\pi^2/m_N^2)$.

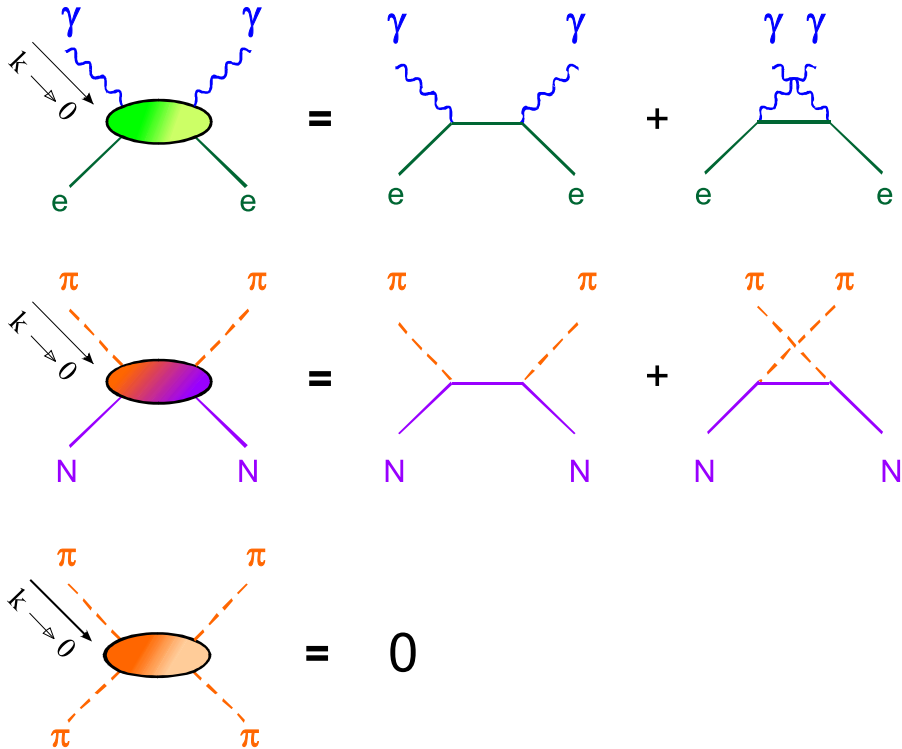

**Figure 24.** Feynman graphs for three processes: (a) Compton scattering, $\gamma e \to \gamma e$, (b) $\pi N \to \pi N$, (c) $\pi\pi \to \pi\pi$ in the limit when the 4-momentum of the photon and massless pion respectively go to zero. Then the low energy theorem of conserved currents requires the relevant scattering amplitudes to be given by the Born amplitudes shown on the right of each equation.

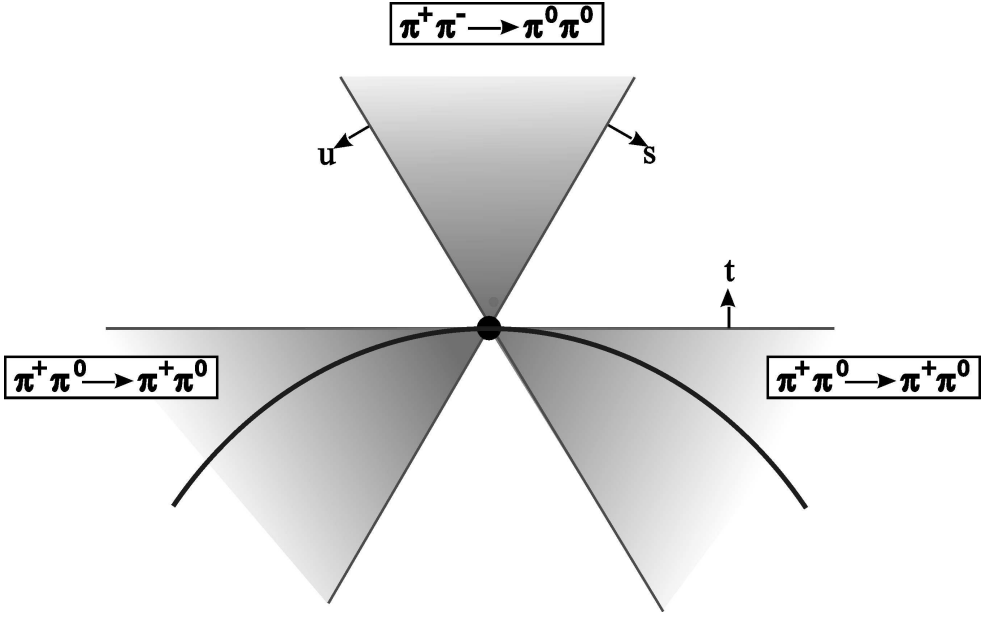

**Figure 25.** Mandelstam plane for $\pi\pi$ scattering of the charges given. In the case of massless pions, chiral dynamics requires the amplitude to vanish at the common threshold $s = t = u = 0$ denoted by the black dot. This zero lies along the line sketched in the $s$ and $u$ physical regions as explained in the text.

Now we consider $\pi\pi$ scattering. There is no Born term for this process, since $G$-parity forbids 3 pions coupling together. Consequently, in the limit of zero pion momentum the $\pi\pi$ scattering amplitude vanishes, Fig. 24, and so

$$\mathcal{F}(s = t = u = 0) = 0 \ . \tag{58}$$

To be concrete let $\mathcal{F}(s,t,u)$ be the amplitude describing $\pi^+\pi^0 \to \pi^+\pi^0$ scattering in the $s$− channel — in fact $\mathcal{F}(s,t,u) = \mathcal{A}(t,u,s)$ where $\mathcal{A}$ is known as the Chew-Mandelstam invariant amplitude. If we look at the Mandelstam plane displaying the physical regions for the scattering of massless pions, this low energy theorem tells us that the strong interaction process is "weak" in the threshold region. The amplitude vanishes at the common threshold $s = t = u = 0$, Fig. 25. Recall that a scattering amplitude is the boundary of an analytic function of the three complex variables $s$, $t$, $u$. The low energy theorem tells us this amplitude is zero at the centre point of the Mandelstam plane. Now an analytic function of several complex variables cannot have an isolated zero, but the zero must lie on a surface. As we shall see this surface has a projection on the plane of real $s$, $t$, $u$, which is represented by the $s-u$ symmetric curve sketched in Fig. 25.

Of course, pions are not quite massless, so we need to get closer to the real physical situation. It was Adler [47], who first appreciated that if in $\pi\pi \to \pi\pi$ scattering, three of the pions had their actual mass, viz. $m_\pi \simeq 140$ MeV, and only one is massless, then, as the 4-momentum of the massless pion goes to zero, the scattering amplitude, $\mathcal{F}$, still vanishes. The Adler condition means that in the world in which $s + t + u = 3m_\pi^2$

$$\mathcal{F}(s = t = u = m_\pi^2) = 0 \quad . \tag{59}$$

The Goldstone nature of the pion requires the amplitude for the scattering of pions to vanish at the unphysical mid-point of the Mandelstam triangle, Figs. 21, 22, 25. A consequence is that $\pi\pi$ scattering with one pion massless is predicted to be weak at low energies, even though it is a strong interaction. This means we can effectively make a Taylor series expansion of the amplitude about the Adler point in powers of the square of momenta. The scale is set by the vacuum expectation value $v$ in Eqs. (25,44), which as we have noted after Eq. (55) is essentially $f_\pi$. Thus the terms in the momentum squared expansion have a scale of $32\pi f_\pi^2$, which like all typical hadronic scales is of order 1 GeV$^2$ as first noted by Weinberg [48]. This expansion was systematised in the classic papers by Gasser and Leutwyler in formulating Chiral Perturbation Theory ($\chi$PT) [49].

Now the world with $s + t + u = 3m_\pi^2$ is not the real one, but it is not far away, because of the smallness of the pion mass, Fig. 13. To encompass the physical world the chiral expansion must include the explicit breaking of chiral symmetry, as in Eq. (45). At lowest order in $E^2$ the amplitude is determined by this and the Lagrangian, $\mathcal{L}_2$ with just two derivatives, Eq. (44). This fixes the amplitude, $\mathcal{F}$, to be [48, 5]

$$\mathcal{F}(s,t) = \frac{(t - m_\pi^2)}{32\pi f_\pi^2} \quad . \tag{60}$$

We see this satisfies the Adler conditions of Eqs. (58,59) and with physical mass has a zero lying along the line $t = m_\pi^2$. It is the higher order corrections needed to ensure unitarity that make the zero contour curve as in Figs. 25, 27. Thus, at order $E^4$, there are contributions from loop graphs with the vertices of $\mathcal{L}_2$ depicted in Fig. 26, plus the tree level term of $\mathcal{L}_4$, Eq. (44). The loop graphs give an imaginary part to the amplitude required by unitarity to this order.

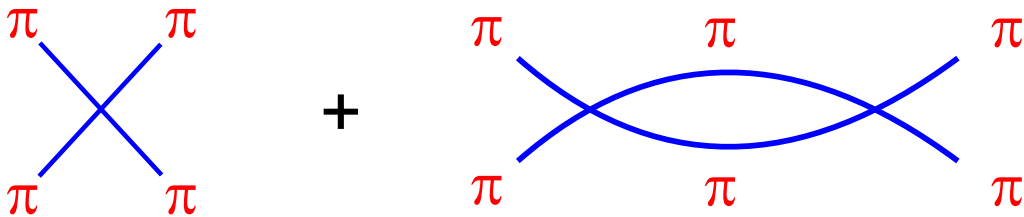

**Figure 26.** Examples of tree level and one loop contributions to $\pi\pi \to \pi\pi$ scattering. Such contributions have Feynman rules given by $\mathcal{L}_2$ of Eq. (44), for example, and its first iteration.

The manner in which chiral symmetry is explicitly broken affects the ordering of the chiral expansion. As already remarked, on quite general grounds we know that the pion mass can be related to the current quark masses using the Operator Product Expansion [50] to give Eq. (47), which we repeat here:

$$m_\pi^2 = (m_u + m_d)\, B_0 + (m_u + m_d)^2\, C_0 + (m_u - m_d)^2\, D_0 + \cdots . \tag{61}$$

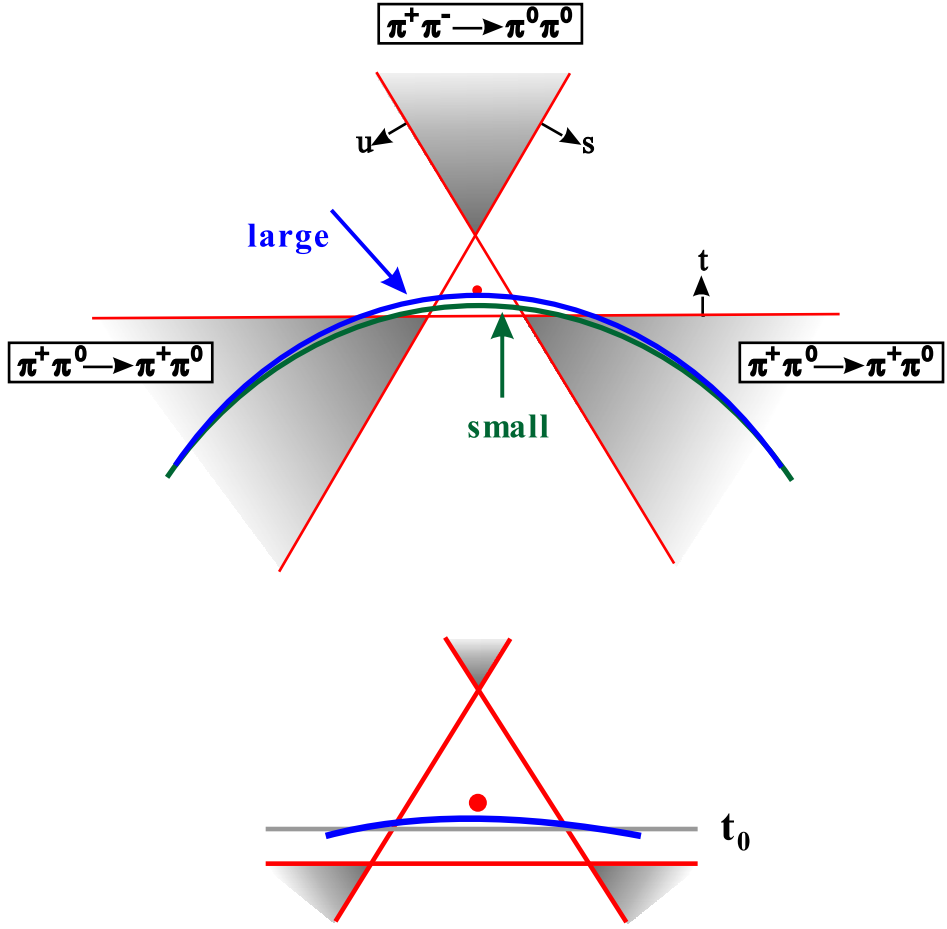

**Figure 27.** Mandelstam plane for $\pi^+\pi^0 \to \pi^+\pi^0$ scattering the $s$ and $u$-channels. The lines on the upper graph denote the on-shell appearance of the Adler zero depending on whether the explicit breaking of chiral symmetry is generated by a *large* or *small* $\bar{q}q$ condensate. Below is an enlargement of the Mandelstam triangle indicating that there are values of $t = t_0$ within the triangle where the amplitude will be zero.

Soft pion techniques relate $B_0$ to the value of the $\bar{q}q$ condensate by

$$B_0 = -\frac{1}{f_\pi^2} \langle \bar{q}q \rangle \quad, \tag{62}$$

where because of isospin symmetry

$$\langle \bar{q}q \rangle = \langle O \mid \bar{u}u \mid O \rangle = \langle O \mid \bar{d}d \mid O \rangle. \tag{63}$$

In Standard $\chi$PT, the first term in Eqs. (47,61) dominates the expansion and $m_\pi^2 \propto m_q$. The alternative known as Generalized $\chi$PT allows a more conservative view of which terms control this expansion. Indeed, in the extreme $B_0$ is small and $m_\pi^2 \propto m_q^2$. The relative size of the terms in this expansion underlies different orderings of the Feynman rules from the Lagrangians, $\mathcal{L}_{2n}$, of Eq. (44).

The amplitude at the mid-point of the Mandelstam triangle is then predicted to be [50]

$$\mathcal{F}\left(s = t = u = \frac{4}{3}m_\pi^2\right) \simeq \frac{\alpha\, m_\pi^2}{96\pi f_\pi^2} \tag{64}$$

where the key parameter $\alpha$ depends on the mechanism of explicit chiral symmetry breaking. In the standard version with a sizeable $q\bar{q}$ condensate, when the term $B_0$ dominates the expansion of Eqs. (47,61), $\alpha \simeq 1$, while a small $q\bar{q}$ condensate (when quark-gluon condensates control Eqs. (47,61)) increases $\alpha$ to 4 [50, 51]. Can experiment tell the difference between these possibilities?

In either case, the amplitude in the middle of the Mandelstam triangle is small, between 0.01 and at most 0.03. If pions were not Goldstone bosons and did not couple through the axial-vector current, the amplitude could be much bigger as allowed by Eq. (54). In contrast the Goldstone nature of pions means the amplitude is very small. Indeed, $\mathcal{F}$ has a zero, in fact a line of zeros, which cut through the Mandelstam triangle. The exact position of this line depends on the value of the $\bar{q}q$ condensate, in the way shown in Fig. 27.

The fact that there is a zero passing through the Mandelstam triangle provides us with a remarkable connection between low energy chiral dynamics and the asymptotic behaviour of the same scattering amplitude. Let us see how. One can prove that the amplitude $\mathcal{F}(s,t)$ is an analytic function in the cut-$s$ plane and satisfies a fixed-$t$ dispersion relation for $4m_\pi^2 > t$. The amplitude at fixed $t$ in this range will then fulfill Cauchy's theorem with

$$\mathcal{F}(s,t) = \frac{1}{2\pi i} \oint_C ds' \frac{\mathcal{F}(s',t)}{s' - s} \tag{65}$$

round any closed contour $C$ in the complex $s$-plane that avoids the cuts. Let us first assume the amplitude decreases at high energies so

$$|\mathcal{F}(s,t)| < \text{constant}, \qquad \text{as } |s| \to \infty \ . \tag{66}$$

Then when we stretch the contour out along the cuts, the contribution from the "almost" closed circle at infinity will vanish and so we have

$$\mathcal{F}(s,t) = \frac{1}{\pi} \int_{4m_\pi^2}^{\infty} ds' \left( \frac{1}{s'-s} + \frac{1}{s'-u} \right) \operatorname{Im} \mathcal{F}(s',t) \ , \tag{67}$$

since the amplitude $\mathcal{F}$ for $\pi^+\pi^- \to \pi^0\pi^0$ in the $t$-channel is $s - u$ symmetric (see Figs. 21, 25 27). Now for $0 \leq t < 4m_\pi^2$ the imaginary part of $\pi^+\pi^0 \to \pi^+\pi^0$ scattering amplitude must be positive, since by the optical theorem it is related to the total cross-section for this process. Consider the amplitude inside the Mandelstam triangle, which has been enlarged in the lower part of Fig. 27, there $0 < s, t, u < 4m_\pi^2$, and the factors of $(s' - s)$ and $(s' - u)$ in Eq. (67) must both be positive since in the integration has $s' \geq 4m_\pi^2$. Then the unsubtracted dispersion relation of Eq. (67) tells us that

$$\mathcal{F}(s,t) > 0 \qquad \text{inside the Mandelstam triangle} \ . \tag{68}$$

The amplitude cannot therefore have the Adler zero shown in Fig. 27 if it decreases at high energies. It immediately follows that the on-shell appearance of the Adler zero requires the amplitude $\mathcal{F}(s,t)$ for $t < 4m_\pi^2$ grows at high energies. Its growth is nevertheless restricted by the Froissart bound. This bound implies [45] the partial wave expansion of the $s$-channel amplitude must converge for $0 \leq t < 4m_\pi^2$ and so

$$\text{constant} < \mathcal{F}(s,t) < |s|^2 \ , \qquad \text{as } |s| \to \infty \ . \tag{69}$$

Because of the $s - u$ symmetry of the amplitude $\mathcal{F}(s,t)$, this bound is sufficient to prove that the difference of two dispersion relations like that of Eq. (67) for $\mathcal{F}(s,t)$ and $\mathcal{F}(s_0,t)$ converges to

$$\mathcal{F}(s,t) = \mathcal{F}(s_0,t) + \frac{(s-s_0)}{\pi} \int_{4m_\pi^2}^{\infty} ds' \left( \frac{1}{s'-s} + \frac{1}{s'-u} \right) \frac{\operatorname{Im} \mathcal{F}(s',t)}{s'-s_0} \ , \tag{70}$$

for fixed values of $t < 4m_\pi^2$. We can now check that this is consistent with an Adler zero passing through the Mandelstam triangle. If we choose the value of $t = t_0$ as in the lower part of Fig. 27, then the positivity of the integral in Eq. (70) for $s$, $t$, $u$, $s_0$, $t_0$ all with values inside the Mandelstam triangle, means that the amplitude along the line $t = t_0$ in Fig. 27 has a minimum on the symmetry axis at $s = u$ and grows monotonically to right and left away from there. Thus it can be negative at $s = u$ and must then eventually pass through zero and become positive. We see that the on-shell appearance of the Adler zero is only consistent with the positivity of total cross-sections if the amplitude grows asymptotically, as in Eq. (69). Experiments on hadron interactions, for which $pp$ and $\bar{p}p$ collisions have been observed at the highest accessible energies, are of course totally in keeping with this requirement.

Let us now return to our discussion of low energy $\pi\pi$ scattering. One can immediately check whether there is indeed a zero that crosses the line $u = 0$ (or equally $s = 0$) in the unphysical region (between the two physical regions) of Fig. 27, because if there is then the amplitude must have different signs on either side of the Mandelstam triangle. Information on $\pi\pi$ scattering can be gained by studying high energy dipion production in $\pi N$ collisions. At small momentum transfers this process is dominated by one pion exchange and the $\pi\pi$ scattering amplitude can be determined. From the data of Ref. [52] we show in Fig. 28 the real part of the corresponding amplitudes in the two physical regions as we travel down the line $u = 0$ in Fig. 27. We see that indeed the amplitude must change sign somewhere in the unphysical region, exactly where is specified by analytic continuation, but we have no doubt there is a zero. This is the zero required by the Goldstone nature of the pion.

We can track the dynamical significance of this zero more closely by noting from Fig. 27 that the line of zeros must enter the $s$ and $u-$ channel physical regions. As already remarked unitarity forces this. Above $\pi\pi$ threshold the zero becomes complex, so for a given $s$-channel energy the zero is at a complex value of $t$ (or equally a complex value of $\cos\theta$, Eq. (51)). Such a complex zero produces a dip in the physical cross-section. If at a fixed, real value of the c.m. energy, $\sqrt{s}$, the zero is at $\cos\theta = z_0$, then the dip occurs at $\cos\theta \simeq \text{Re}(z_0)$ and the height of the dip is proportional to $\text{Im}(z_0)$. Why is easy to see. If $z_0$ is real, then there is a real zero in the amplitude and its modulus squared will dip down to the axis (i.e. to zero). Consequently, the depth of the dip reflects how far the zero is from the real axis [53].

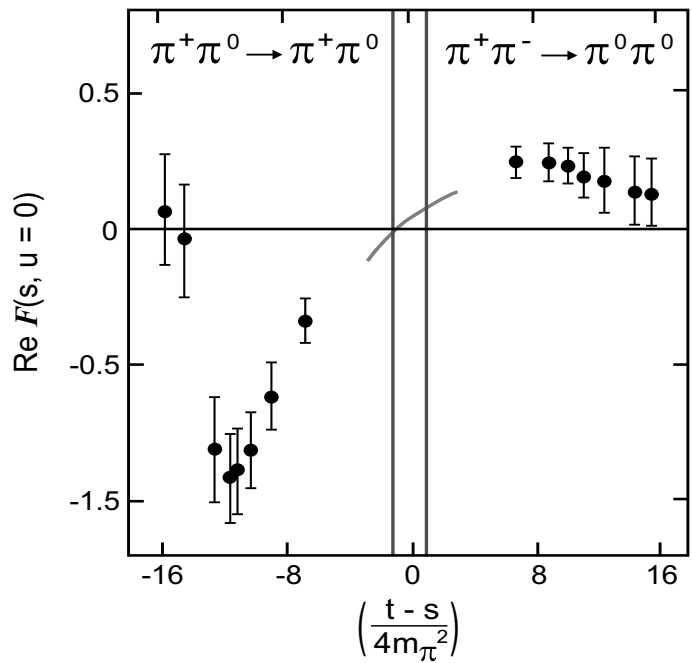

**Figure 28.** Real part of the scattering amplitude $\mathcal{F}$, from data of Ref. [52], along the line $u = 0$ on the Mandelstam plane of Fig. 27. The vertical lines delineate the unphysical domain of Fig. 27, in which the amplitude is expected to have a zero, as indicated by the curve that has to analytically interpolate the data.

At low energies the scattering is dominated by resonances and the cross-section has a series of peaks. Moreover, the angular distribution has dips. The number and position of these tell us about the quantum numbers of the contributing resonances. If the external particles are spinless, as pions are, then the number of dips equals the spin of any resonance produced. Thus for a spin-0 resonance, the angular distribution is flat. For spin-1 there is one dip, spin-2 two dips, etc. The actual distribution and depth of dips depends on the non-resonating waves. Now let us turn to the amplitude for the process shown in Figs. 21, 27. In the $s$ (and $u$)-channels the reaction is $\pi^+\pi^0 \to \pi^+\pi^0$, which has isospin 1 and 2. The $I = 2$ $S-$wave is known to be small [54] (there are no $\overline{q}q$ resonances with $I = 2$ and this channel is wholly repulsive). The main feature of the $I = 1$ amplitude at low energies is the $\rho-$resonance at $\sqrt{s} = m_\rho \simeq 770$ MeV. In the neighbourhood of this resonance the cross-section and angular distribution look like:

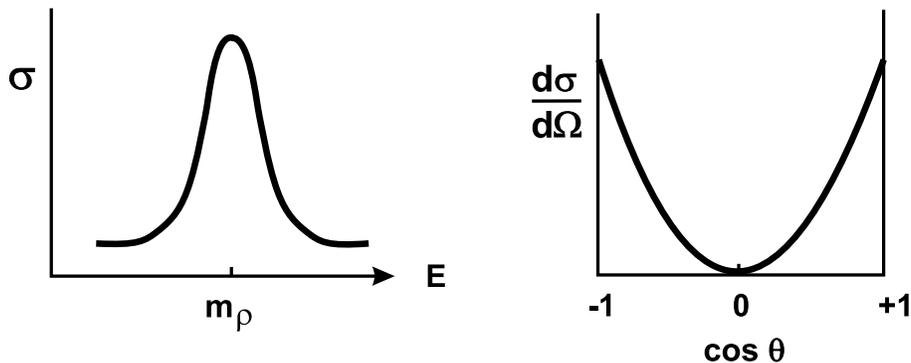

**Figure 29.** $I = 1$ $\pi\pi$ cross-section shows a peak at the $\rho$ mass, while the angular distribution at this energy has a dip at $\theta = \pi/2$ because of the $\cos^2\theta$ shape expected for a spin-1 dominance.

since the amplitude is controlled by the $\rho$ and given by

$$\mathcal{F}(s, \cos\theta) \simeq \frac{3}{2} \frac{m_\rho \, \Gamma_\rho \, \cos\theta}{m_\rho^2 - s - im_\rho\Gamma_\rho} \quad , \tag{71}$$

where $\Gamma_\rho$ is the width of the $\rho$ and $m_\rho$ its mass. We see that in the region of $\sqrt{s} \simeq m_\rho$, the angular distribution has a zero for $\theta \simeq \pi/2$, Figs. 29, 30, as the $\rho$ has spin-1. Such zeros are essential in generating the spin of the resonance. Since they correspond to the zeros of the Legendre polynomials, $P_J(z)$, they are referred to as Legendre zeros. Zeros have to move continuously across the Mandelstam plane [53]. They do not suddenly appear or disappear. As we go higher and higher in energy in the $s-$channel a new zero enters the physical region and when it lines up with the other zeros already present so that $P_J(\cos\theta) = 0$, the spin $J$ wave resonates.

We can track the on-shell appearance of the Adler zero using experimental data. Its path is sketched in Fig. 30. We see the Adler zero becomes the Legendre zero of the $\rho$. It is this zero that makes the $\rho$ have spin one. The Goldstone nature of the pion provides this zero at low energies, through the Adler condition. It is then immediately clear why the early models of $\pi\pi$ scattering developed by Chew, Mandelstam and Noyes [55] could not generate a $\rho-$resonance, while those based on chiral dynamics automatically do. The scale of the resonance, the ratio of its width to the cube of its mass $\Gamma_\rho/m_\rho^3$, is also fixed by chiral dynamics through the KSFR relation [56], or the Olsson sum rule [57], which is essentially a dispersion relation like those of Eqs. (67,70) for the $I = 1$ amplitude in the $t$-channel, which we do not discuss here but is related to $1/32\pi f_\pi^2$. That there is a line of zeros generating a spin-1 resonance connected to the Adler condition reflects chiral dynamics and the fact that the pion is the Goldstone boson of chiral symmetry breaking [58]. Where the surface of zeros cuts through the Mandelstam triangle provides us with a unique insight into the structure of the QCD vacuum and the *treacle*-like medium through which the quarks, from which we are made, propagate, Fig. 18.

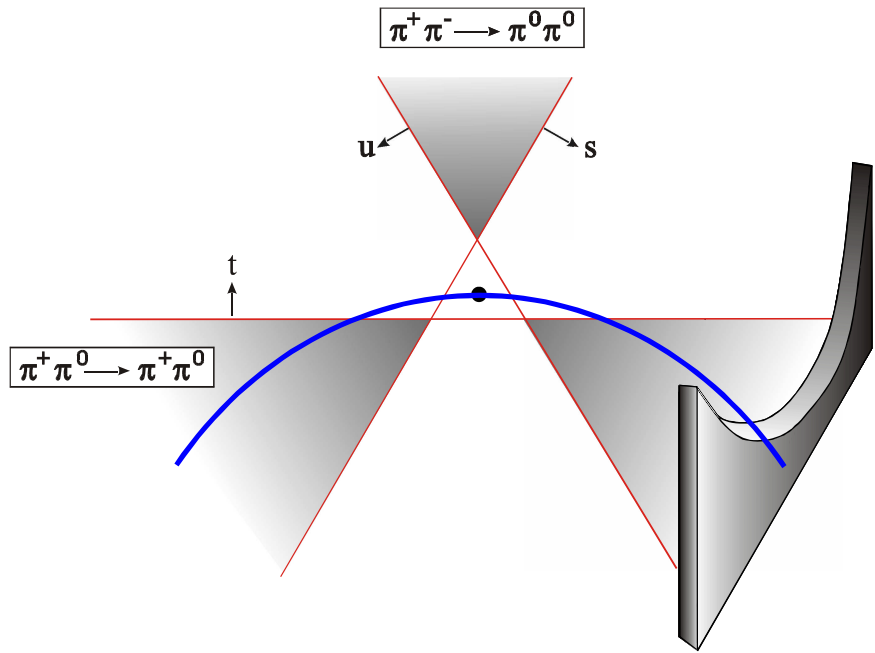

**Figure 30.** Mandelstam plane for the amplitude $\pi^+\pi^0 \to \pi^+\pi^0$, which has a line of zeros passing close to the symmetry point of the Mandelstam triangle, which produces the dip in the angular distribution in the $\rho$ region, as shown on the right in Fig. 29.

Though the structure of the vacuum determines the size of the amplitude in the unphysical region at the centre of the Mandelstam triangle, the chiral expansion about this point involves parameters that have to be fixed by experiment, for instance in the $\rho$−region. So though the predictions for the amplitude differ by up to a factor of 4 in the unphysical region depending on the mechanism of chiral symmetry breaking, they differ by only 30% at threshold, when $\sqrt{s} = 2m_\pi \simeq 280$ MeV and hardly at all at $\sqrt{s} \simeq 700$ MeV. We thus require precision data close to threshold to be able to distinguish these possibilities. Fortunately, two recent experiments have the potential to discriminate. One is the DIRAC experiment at CERN [59]. This aims to measure the lifetime of pionium. In pionium, a $\pi^+\pi^-$ system binds by exchanging photons. How long this survives is determined by how readily the charged pions annihilate into a lighter $\pi^0\pi^0$ pair by the strong interaction, Fig. 31. This annihilation is determined by the square of the amplitude $\mathcal{F}$ at $t = 4m_\pi^2$, $u = s = 0$, Figs. 27, 30. This is a most difficult experiment. Running over several years, the DIRAC experiment has detected pionic atoms [59]. While (as just mentioned) the two zero contours of Fig. 27 make the amplitudes differ by 30% at threshold, the DIRAC experiment hopes to measure the lifetime of pionium to an accuracy of 10% and so determine the amplitude at threshold to 5% and thereby distinguish the extremes given in Fig. 27. In the next year we should learn if this expectation is realised.

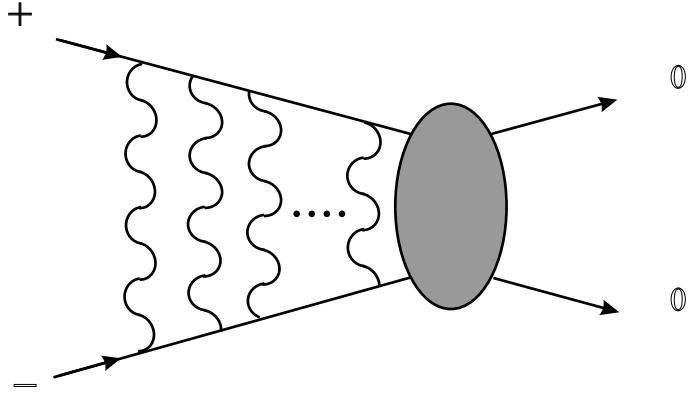

**Figure 31.** Pionic atom is formed by the electromagnetic interaction of two oppositely charged pions. This eventually decays to the lighter $\pi^0\pi^0$ state.

As already discussed most information on $\pi\pi$ scattering in past decades came from large statistics experiments on high energy, small momentum transfer di-pion production like $\pi N \to \pi\pi N$. There the low $\pi\pi$ masses necessary to distinguish the mechanisms of chiral symmetry breaking are not readily accessible. Fortunately, the conservation of probability, viz. unitarity, requires that the lightest hadrons scatter universally irrespective of how they are produced. Thus a precision way to learn about low energy $\pi\pi$ scattering is provided by the so-called $K_{e4}$ decay, e.g. $K^+ \to \pi^+\pi^- e^+ \nu_e$ shown in Fig. 32.

The decay proceeds by the $\overline{s}$ quark in a $K^+$ turning into a $\overline{u}$ and an off-shell $W^+$. This leaves a $u\overline{u}$ which converts into a $\pi^+\pi^-$ pair, while the $W^+$ decays to $e^+\nu_e$. The process depends on 5 kinematic variables. The choice of these proposed by Cabibbo and Maksymowicz [60] is indicated in the lower part of Fig. 32. Let us define them. In its rest frame, the kaon decays into two pions of mass $M_{\pi\pi}$ and a lepton pair of mass $M_{e\nu}$, moving back-to-back in what defines the $x-$direction. We then go to the di-pion rest frame and each pion is produced at an angle $\theta_\pi$ to the $x-$axis. Similarly, in the di-lepton rest frame, the electron moves off at an angle $\theta_e$ to the same $x-$axis. The two pions define a plane in the di-pion rest frame and the two leptons define another plane in the di-lepton rest frame. These two planes are at an angle $\phi$ to each other, Fig. 32. This specifies the 5 variables: $M_{\pi\pi}$, $M_{e\nu}$, $\theta_\pi$, $\theta_e$ and $\phi$.

Once produced the two pions scatter in different angular momentum states. By studying the decay distribution as a function of all 5 variables, one can measure the interference between the $\pi\pi$ system in an $S$ and a $P-$wave. This determines the phase difference $\delta_S - \delta_P$. Now the weak decay proceeds by the $\Delta I = 1/2$ rule, which means that while the $P-$wave has $I=1$, the $S-$wave has only $I=0$. By Watson's final state interaction theorem [61], the phase differences in $K_{e4}$ decay are those of $\pi\pi$ elastic scattering. If the $\pi\pi$ system has isospin $I$ and angular momentum $J$, then its phase-shift is labelled by $\delta_J^I$. Watson's theorem requires

$$\delta_S - \delta_P = \delta_0^0 - \delta_1^1 \quad . \tag{72}$$

The previous significant $K_{e4}$ decay experiments are by the Pennsylvania collaboration [62] with 7000 events and the Geneva-Saclay group [63] with 30,000 events, while the most recent BNL-E865 experiment [64] has 400,000 events presently analysed. While the Geneva-Saclay experiment could not discriminate between $q\overline{q}$ condensate values from 0 to $-(350\text{MeV})^3$, the increased precision of the BNL-E865 experiment shown in Fig. 33 does far better.

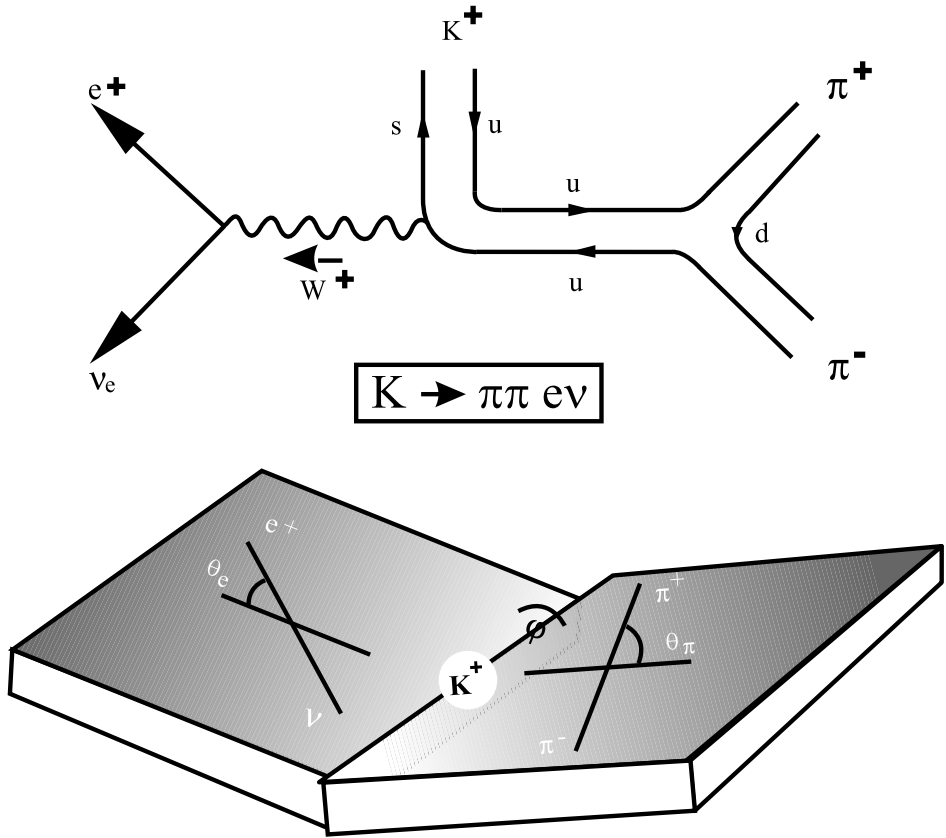

**Figure 32.** $K^+ \to e^+ \nu_e \pi^+ \pi^-$ decay : (a) showing the weak decay of the strange quark, (b) the kinematics discussed in the text.

The extrapolation of data to threshold, and even below, requires an analytic continuation which for $\pi\pi$ scattering is constrained by threefold crossing symmetry. The Roy equations [65] embody these general properties for each partial wave amplitude. Colangelo et al. [66] have expanded on the earlier phenomenological analyses [67] of these equations to incorporate recent experimental information. The inclusion of the new $K_{e4}$ decay results permit a reasonably precise determination of the parameter $\alpha$ in Eq. (64). The Adler zero lies along the upper curve in Fig. 27. This yields a condensate of size [68]

$$\langle \overline{q}q \rangle \simeq -(270 \text{ MeV})^3 \,, \tag{73}$$

which implies that at least 94% of the expansion in quark mass in Eqs. (47,61) is given by the very first term, just as Gell-Mann, Oakes and Renner proposed long ago [69].

$$m_\pi^2 f_\pi^2 \;=\; -(m_u + m_d)\langle \overline{q}q \rangle \;+\; \cdots \quad . \tag{74}$$

The $\overline{q}q$ condensate drives the dynamical breakdown of chiral symmetry and provides the syrupy medium through which quarks propagate, Fig. 18.

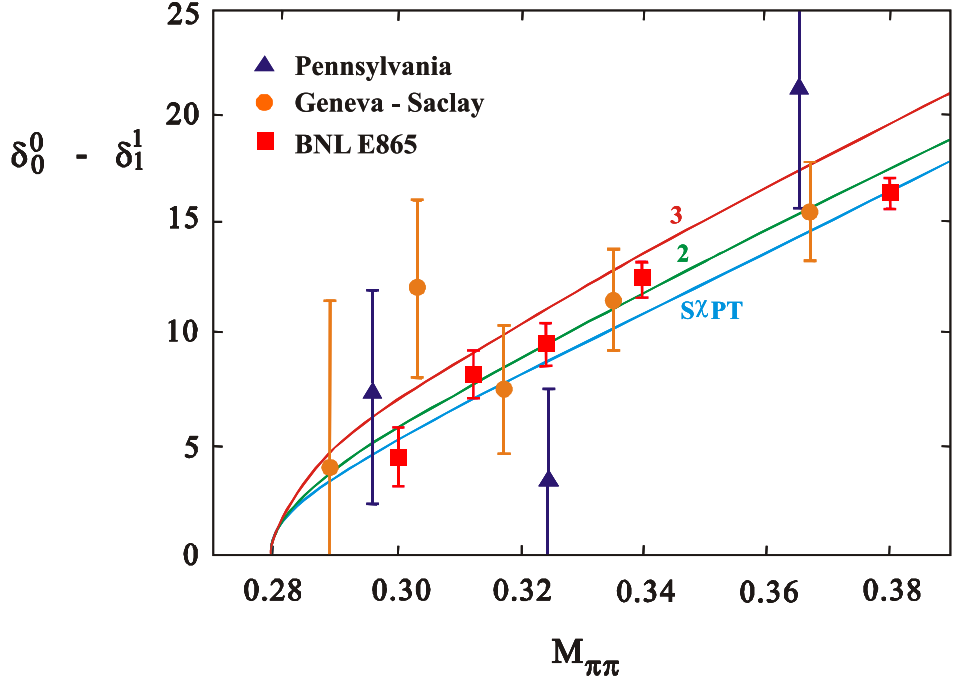

**Figure 33.** The phase difference $\delta_0^0 - \delta_1^1$ of Eq. (72) as a function of dipion mass. The triangles, circles and squares are from the Pennsylvania [62] and Geneva-Saclay [63] experiments and the BNL-E865 $K_{e4}-$ decay results [64], respectively. The curves are the predictions of two loop chiral perturbation theory: the standard result [66] for which $\alpha \simeq 1.2$ and for $\alpha = 2$ and 3 of the generalised theory [50, 51].

## 7. Dynamical mass generation

We have seen that the *up* and *down* quarks have almost zero current masses, yet in a hadron they have constituent masses of 300-350 MeV. As we have discussed experiment tells us that this implies a non-trivial structure for the vacuum. But can we really see whether this accords with theoretical calculations of this structure in QCD? We begin by asking when can a particle with zero bare mass have a mass dynamically generated by its interactions? We want to study this problem in QCD, but we start with QED as the simpler prototype gauge theory [70]. Consider a fermion with bare mass $m_0$. Let us compute its mass function in perturbation theory from the Feynman graphs of Fig. 34, then

$$\mathcal{M}(p) = m_0 \left( 1 + c_1 \alpha \ln \left( \frac{p^2}{\mu^2} \right) + c_2 \alpha^2 \ln^2 \left( \frac{p^2}{\mu^2} \right) + \cdots \right) \quad , \tag{75}$$

where $\alpha$ is the coupling defined at some momentum scale $\mu$. We see that at every finite order the mass function at all momenta is proportional to the bare mass. If the bare mass is zero then so is the *dressed* mass. This tells us that if a mass is to be generated, we must study the problem beyond perturbation theory, either to all orders in the coupling $\alpha$ or genuinely non-perturbatively. Non-perturbative Green's functions can be studied on the lattice. Calculation at a discrete set of space-time points does represent the exact continuum theory when the lattice spacing, $a$, goes to zero and the lattice has a huge number of points, $N_L$. But in practice $a$ is not very small and $N_L$ not very large. To minimise the dependence on $a$, different versions of discretized QCD are used with a preference for what are called *improved actions*. Nevertheless, the lattice has finite size and so the momentum $p$ range studied is limited to $1/a > p > 1/(N_L a)$. The large momentum region can be made accessible by using the renormalization group in the case of QCD. However, the infrared region is problematical. Massless particles clearly cannot fit on a finite size lattice. This is indeed a serious issue for computations of physics of $u$ and $d$ quarks [71] and at present calculations never have this less than $1/3 - 1/2$ the strange quark mass. We necessarily have to solve the problem in the continuum using the field equations of the theory — the Schwinger-Dyson equations.

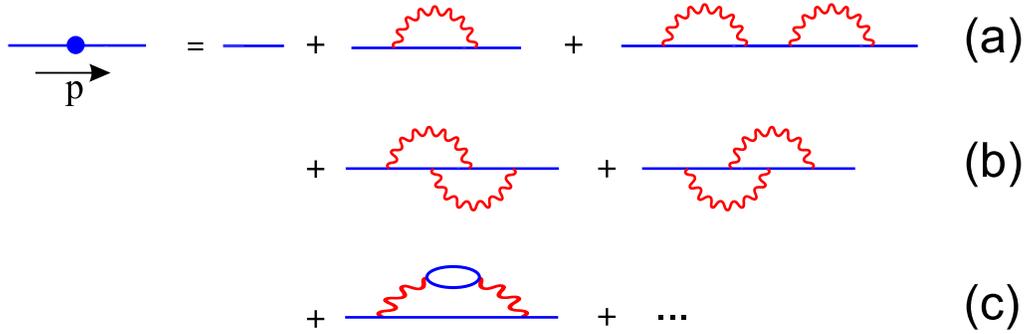

**Figure 34.** Feynman graphs that need to be summed to determine the behaviour of a fermion propagator. Those in (a) renormalise the fermion propagator, (b) renormalise the vertices and (c) correct the boson propagator, and all their iterations.

To understand these equations, particularly for the fermion propagator, let us consider the derivation as a means of summing all orders in perturbation theory, the first few graphs of which are shown in Fig. 34. This expansion of course involve an infinite number of Feynman graphs. These we need to sum. To order this summation, it is helpful to note that the graphs are of three types: (a) those that renormalise the fermion propagator, (b) those that renormalise the fermion-boson vertices, and (c) those that correct the boson propagator, and of course their many iterations. It is then natural to define a quantity usually called the full "self energy" $\Sigma(p)$ shown in Fig. 35, which has no external propagators. Technically, this is called 'one particle irreducible'.

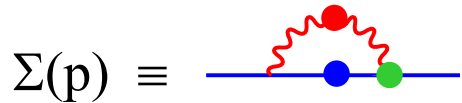

**Figure 35.** Self-energy corrections for the fermion propagator. The dots indicate a complete summation of all contributing insertions. Only one vertex is renormalised to avoid double counting. $\Sigma(p)$ is defined with no external propagators.

The solid dots mean that every possible insertion in the fermion and boson propagators and the fermion-boson vertex are included. This loop diagram like all others involves 4-dimensional integration over all loop momenta as these are unconstrained by the external momenta. With this definition, it can be shown that the fermion propagator now involves a calculable infinity of graphs displayed in Fig. 36. These give

$$S_F(p) = S_F^0(p) + S_F^0(p)\,\Sigma(p)\,S_F^0(p) + S_F^0(p)\,\Sigma(p)\,S_F^0(p)\,\Sigma(p)\,S_F^0(p) + \cdots \quad . \tag{76}$$

We see this is just a geometric series that we can sum to give:

$$S_F(p) = \frac{S_F^0(p)}{1 - \Sigma(p)\,S_F^0(p)} \quad . \tag{77}$$

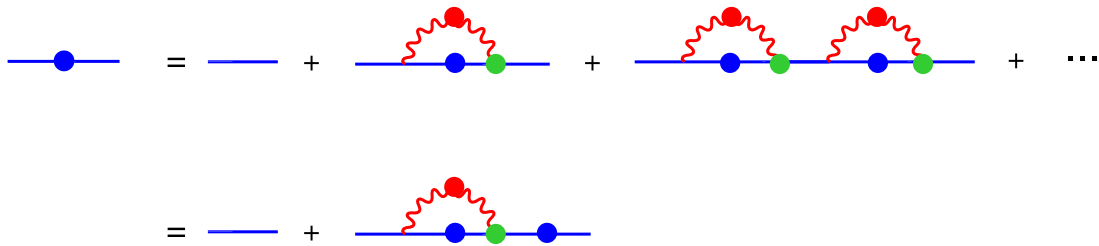

**Figure 36.** Using the definition of self-energy displayed in Fig. 35, the full fermion propagator is given by the summation given in the top line of this figure. The lower figure depicts the result of this summation. The solid dots indicate the Green's functions are fully dressed.

This can equally well be written as

$$S_F(p) \;=\; S_F^0(p) \;+\; \frac{S_F^0(p)\,\Sigma(p)\,S_F^0(p)}{1 - \Sigma(p)\,S_F^0(p)} \quad. \tag{78}$$

Substituting Eq. (77) into Eq. (78), we see that Eq. (78) can be expressed as:

$$S_F(p) \;=\; S_F^0(p) \;+\; S_F^0(p)\,\Sigma(p)\,S_F(p) \quad, \tag{79}$$

displayed diagrammatically in the lower part of Fig. 36. If we multiply this equation by $S_F^0(p)^{-1}$ from the left and $S_F(p)^{-1}$ on the right and rearrange, we obtain the field equation for the inverse fermion propagator

$$S_F(p)^{-1} \;=\; S_F^0(p)^{-1} \;-\; \Sigma(p) \quad, \tag{80}$$

shown in Fig. 37. This is the fermion Schwinger-Dyson equation, sometimes called the Dyson-Schwinger equation [72]!

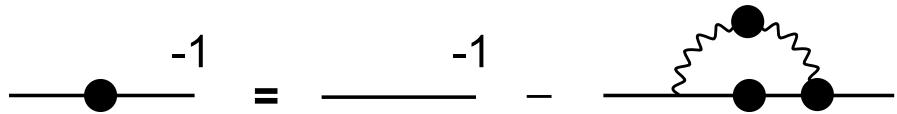

**Figure 37.** Schwinger-Dyson equation for the inverse fermion propagator. The solid dots mean the Green's functions are fully dressed. Compared with Fig. 34 the graphs here are one-particle irreducible and have no external propagators.

As one can readily appreciate there is a similar equation for the inverse photon propagator, involving the full fermion loop, seen in Fig. 38. Now one can imagine solving these two coupled equations. The first determines the full fermion propagator in terms of the bare propagator and a loop of the full fermion and boson propagators and the full fermion-boson vertex, while the second determines the full photon propagator in terms of the bare propagator and a loop of the same full fermion and the full fermion-boson vertex. These can be solved together provided we know the full fermion-boson interaction. This 3-point vertex in turn satisfies its own Schwinger-Dyson equation that relates it to the two 2-point functions, the 3-point function and now the 4-point function, see Fig. 38. The 4-point function, for which there is no bare vertex in QED satisfies its own Schwinger-Dyson equation too, relating it to 2, 3, 4 and 5-point functions. Thus we see that the Schwinger-Dyson equations are a nested set of integral equations. They are the complete field equations of the theory and if we could solve them, we would have the exact solution of the theory. However, to learn about even the 2-point functions seems to require knowledge of all the higher point functions. Can it really be necessary to know the complete 110-point function to learn about the behaviour of the full fermion propagator? Fortunately, no it is not. The fermion Schwinger-Dyson equation involves a very particular projection of the higher point functions and we can use this to advantage as we will see later.

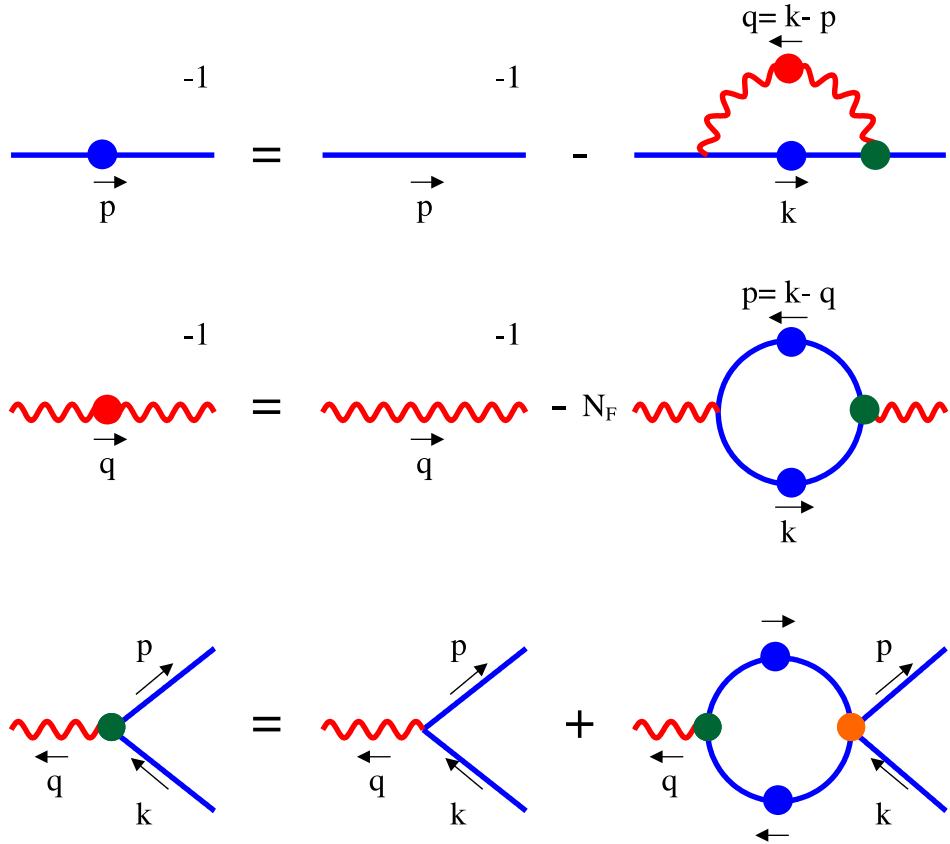

**Figure 38.** Schwinger-Dyson equations for the inverse of the fermion propagator, the inverse of the boson propagator, and the full fermion-boson vertex, with the momenta defined for later. The solid dots mean the Green's functions are fully dressed.

Though we have motivated the Schwinger-Dyson equations by appealing to the idea of summing all orders in perturbation theory, it is important to recognise that they are genuinely non-perturbative equations. They can be derived as the field equations of the theory using a functional integral approach. Indeed, expanding each Green's function in the nested set of Schwinger-Dyson equations in powers of the coupling is what perturbation theory is. To study dynamical mass generation we know we have to go beyond the perturbative approach. For the fermion propagator $S_F(p)$ the Schwinger-Dyson equation depicted in Figs. 37, 38 is given by the integral equation:

$$S_F(p)^{-1} = \not{p} - m_0 - \frac{\alpha}{4\pi} \int d^4k \, \gamma_\mu \, S_F(k) \, \Gamma_\nu(k,p) \, \Delta^{\mu\nu}(q) \quad , \tag{81}$$

where we have substituted for the inverse bare propagator

$$S_F^0(p)^{-1} = \not{p} - m_0 \quad . \tag{82}$$

Factors of $\pm i$ have been dropped and can be found in any textbook, like Bjorken & Drell [73] or Itzykson & Zuber [74]. The photon propagator in a covariant gauge, specified by the parameter $\xi$, is given by

$$\Delta^{\mu\nu}(q) = \frac{\mathcal{G}(q)}{q^2}\left(g^{\mu\nu} - \frac{q^\mu q^\nu}{q^2}\right) + \xi\,\frac{q^\mu q^\nu}{q^4} \tag{83}$$

with its independent transverse and longitudinal tensors. To solve the fermion equation in full would require us to know both the full photon wavefunction renormalization function $\mathcal{G}(q)$ and the complete structure of the fermion-boson interaction vertex, $\Gamma^\mu(k,p)$. These in turn satisfy their own coupled Schwinger-Dyson equations. This would be to solve QED exactly. As we shall see the fermion and boson equations illustrated in Fig. 38 are each really two equations, while the vertex equation contains 12 independent equations, *etc.*, each representing the independent tensor structures possible. Thus the network of coupled Schwinger-Dyson equations is even more complicated than Fig. 38 would at first indicate. To understand the prerequisites for dynamical fermion mass generation this is far too (unnecessarily) difficult, so let us make a series of simplifications, which though little justified, provide us with a framework for studying the more general problem.

We first treat the photon propagator as bare and so in Eq. (83) simply set

$$\mathcal{G}(q) \equiv 1 \quad . \tag{84}$$

The coupling $\alpha$ is then a constant and does not run. To signify this we denote the coupling by $\alpha_0$. Next we replace the vertex structure by its bare form

$$\Gamma^\nu(k,p) \quad \to \quad \gamma^\nu \quad . \tag{85}$$

Then only the fermion propagator is treated non-perturbatively. This approximation is shown graphically in Fig. 39 for the inverse fermion propagator. This sums the graphs in the lower part of Fig. 39 for the propagator itself and is consequently known as the *rainbow* approximation.

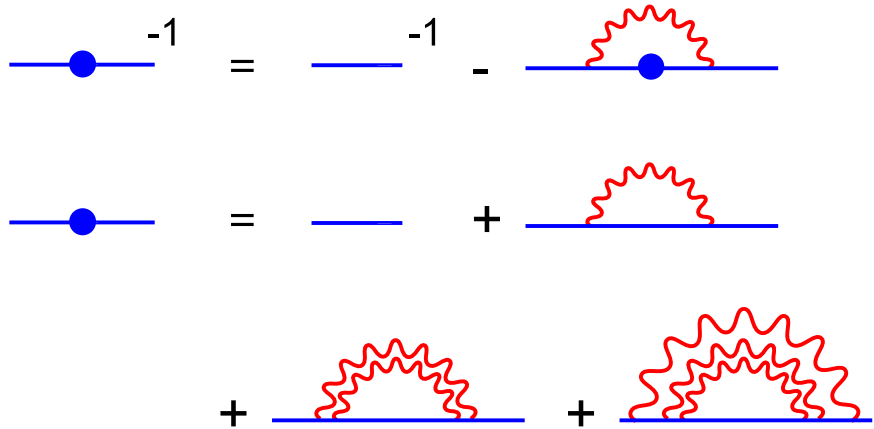

**Figure 39.** Approximation to the Schwinger-Dyson equation for the inverse fermion propagator where the boson and fermion-boson vertices are bare. This sums the *rainbow* graphs displayed and all their iterations. The solid dots mean the fermion propagator is fully dressed.

Now in general the Dirac structure of the fermion propagator means it depends on two independent functions, the wavefunction renormalization $\mathcal{F}(p)$ and the mass function $\mathcal{M}(p)$,

so that
$$S_F(p) = \frac{\mathcal{F}(p)}{\not{p} - \mathcal{M}(p)} \quad . \tag{86}$$

The bare propagator is the special case where
$$\mathcal{F}(p) = 1 \quad , \qquad \mathcal{M}(p) = m_0 \quad . \tag{87}$$

Substituting these forms, Eqs. (82,83,85), into the Schwinger-Dyson equation, Eq. (81) we readily deduce two equations on tracing with the unit matrix and with $\not{p}$. Wick rotating to Euclidean space we can perform the angular integrals and so deduce the two coupled fermion equations:

$$\frac{\mathcal{M}(p)}{\mathcal{F}(p)} = m_0 + \frac{\alpha_0}{4\pi}(3+\xi) \int_0^{\kappa^2} dk^2 \; \frac{\mathcal{F}(k)\mathcal{M}(k)}{k^2 + \mathcal{M}(k)^2} \left[\theta_+ \frac{k^2}{p^2} + \theta_-\right] \tag{88}$$

$$\frac{1}{\mathcal{F}(p)} = 1 + \frac{\alpha_0 \xi}{4\pi} \int_0^{\kappa^2} dk^2 \; \frac{\mathcal{F}(k)}{k^2 + \mathcal{M}(k)^2} \left[\theta_+ \frac{k^4}{p^4} + \theta_-\right] \tag{89}$$

where
$$\theta_+ = \begin{cases} 1 & (p^2 - k^2) > 0 \\ 0 & (p^2 - k^2) < 0 \end{cases} , \qquad \theta_- = \begin{cases} 1 & (k^2 - p^2) > 0 \\ 0 & (k^2 - p^2) < 0 \end{cases} . \tag{90}$$

Only the radial integral remains to be done. To render this finite we have introduced an ultraviolet cutoff, $\kappa$. Though this regulates the integrals, they are not translationally invariant, which we will see is important for later, but for now this provides a simple pedagogical example of how to ensure the equations define finite quantities. What we see is that if the bare mass $m_0 = 0$, there is always the solution $\mathcal{M}(p) = 0$. In fact this is always true if the vertex involves an odd number of gamma-matrices. But what we want to know is, if $m_0 = 0$ when can there be a solution with a non-zero mass function [70, 75]?

We can see immediately that we can simplify these two coupled equations further by working in the Landau gauge, because then
$$\xi = 0 \qquad \Longrightarrow \qquad \mathcal{F}(p) = 1 \tag{91}$$

and our infinite set of nested Schwinger-Dyson equations is reduced to just one for the fermion mass function, $\mathcal{M}(p)$. Eq. (88) then becomes

$$\mathcal{M}(p) = m_0 + \frac{3\alpha_0}{4\pi}\left[\frac{1}{p^2}\int_0^{p^2} dk^2 \; \frac{k^2 \mathcal{M}(k)}{k^2 + \mathcal{M}(k)^2} + \int_{p^2}^{\kappa^2} dk^2 \; \frac{\mathcal{M}(k)}{k^2 + \mathcal{M}(k)^2}\right] \quad , \tag{92}$$

Since the approximations we have made mean that the coupling does not run, the only scale for any mass that is generated when $m_0 \to 0$ is provided by the ultraviolet cutoff $\kappa$, as we will see. Now to solve this integral equation it is simpler to convert it to a differential equation. One can always do this, but it does not always lead to a small number of derivatives. Here with the simplifications and assumptions we have built in, the equivalent differential equation is just second order. We begin by differentiating Eq. (92) with respect to $p^2$ to give:

$$\frac{d}{dp^2}\mathcal{M}(p) = -\frac{3\alpha_0}{4\pi}\frac{1}{p^4}\int_0^{p^2} dk^2 \; \frac{k^2 \mathcal{M}(k)}{k^2 + \mathcal{M}(k)^2} \quad . \tag{93}$$

We then multiply by $p^4$ and differentiate again to obtain:

$$\frac{d}{dp^2}\left(p^4 \frac{d\mathcal{M}}{dp^2}\right) = -\frac{3\alpha_0}{4\pi}\frac{p^2 \mathcal{M}(p)}{p^2 + \mathcal{M}(p)^2} \quad . \tag{94}$$

While $\mathcal{M}(p) = 0$ is always possible, is there a non-zero solution? To understand the behaviour of such a solution to this equation, let us first consider what happens at large momenta when $p^2 \gg \mathcal{M}(p)^2$, then

$$\frac{d}{dp^2}\left(p^4 \frac{d\mathcal{M}}{dp^2}\right) \simeq -\frac{3\alpha_0}{4\pi}\mathcal{M}(p) \ . \tag{95}$$

Now it is easy to see that this homogeneous equation is satisfied by power behaviour for the mass function:

$$\mathcal{M}(p) \sim (p^2)^s \ . \tag{96}$$

Substituting this form into Eq. (95) yields:

$$s(s+1) = -\frac{3\alpha_0}{4\pi} \ . \tag{97}$$

Adding $1/4$ to each side one has:

$$\left(s+\frac{1}{2}\right)^2 = \frac{1}{4}\left(1 - \frac{3\alpha_0}{\pi}\right) \ . \tag{98}$$

Thus,

$$s = \frac{1}{2}\left[-1 \pm \sqrt{1 - \frac{3\alpha_0}{\pi}}\right] \ . \tag{99}$$

We see that the behaviour of the solutions depends crucially on whether the coupling $\alpha_0$ is weaker or stronger than the critical coupling $\alpha_c = \pi/3$. If the coupling is less than $\alpha_c$ the mass function is power behaved, while if $\alpha > \alpha_c$ the mass function oscillates at large momenta [70, 75].

How do we know which is the right form of the solution? When one converts an integral equation to a differential equation, one, of course, loses information. This is encoded in the boundary conditions, which we now deduce. We return to Eq. (92) and substitute for

$$\frac{3\alpha_0}{4\pi}\frac{\mathcal{M}(k)}{k^2 + \mathcal{M}(k)^2} \to -k^4 \frac{d}{dk^2}\mathcal{M}(k) \tag{100}$$

as specified by Eq. (94) to give

$$\mathcal{M}(p) = m_0 - \frac{1}{p^2}\int_0^{p^2} dk^2 \frac{d}{dk^2}\left(k^4 \frac{d\mathcal{M}}{dk^2}\right) - \int_{p^2}^{\kappa^2}\frac{dk^2}{k^2}\frac{d}{dk^2}\left(k^4 \frac{d\mathcal{M}}{dk^2}\right) \ . \tag{101}$$

On integrating we find:

$$\mathcal{M}(p) = m_0 + \mathcal{M}(p) + \frac{1}{p^2}\left[k^4\frac{d\mathcal{M}}{dk^2}\right]_{k^2=0} - \left[\mathcal{M}(k) + k^2\frac{d\mathcal{M}}{dk^2}\right]_{k^2=\kappa^2} \ . \tag{102}$$

Thus the boundary conditions, when $m_0 = 0$, are:

$$\left[k^4\frac{d\mathcal{M}}{dk^2}\right]_{k^2=0} = 0 \tag{103}$$

$$\left[\mathcal{M}(k) + k^2\frac{d\mathcal{M}}{dk^2}\right]_{k^2=\kappa^2} = 0 \ . \tag{104}$$

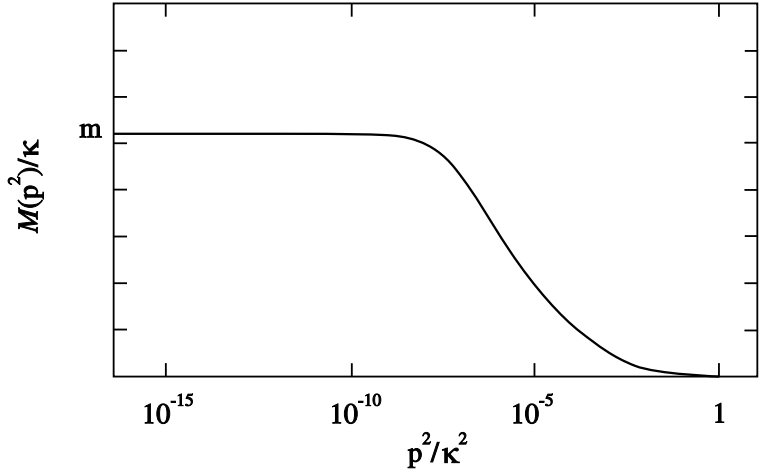

**Figure 40.** The mass function, $\mathcal{M}(p^2)$, for some coupling $\alpha_0$ above its critical value. In this calculation with a constant coupling the ultraviolet cut-off, $\kappa$, sets the scale.

The $k^2 \to 0$ condition is in fact easily satisfied since the solution to Eq. (94) gives $\mathcal{M}(k) \to$ constant, in this limit. However, the second condition cannot be satisfied by simple power behaviour, but demands the oscillatory form

$$\mathcal{M}(p) = C \frac{\kappa^2}{p} \left[ \sin\left(\frac{1}{2}\sqrt{\frac{\alpha_0}{\alpha_c} - 1} \ln\frac{p^2}{\kappa^2}\right) - \sqrt{\frac{\alpha_0}{\alpha_c} - 1} \cos\left(\frac{1}{2}\sqrt{\frac{\alpha_0}{\alpha_c} - 1} \ln\frac{p^2}{\kappa^2}\right) \right] \quad (105)$$

with $C$ a dimensionless constant. The presence of the square root means this form is only

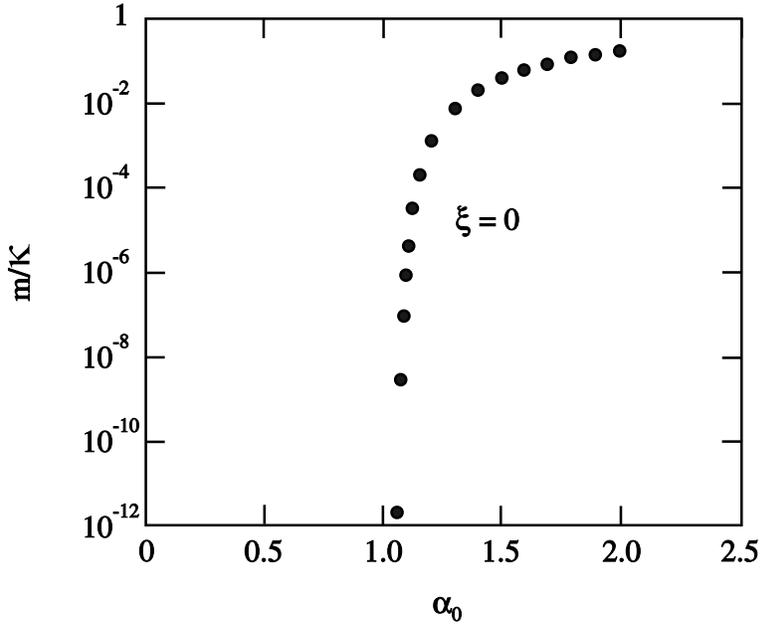

**Figure 41.** The Euclidean mass $m$ as a function of the coupling $\alpha_0$ in the *rainbow* approximation in the Landau gauge. $\kappa$ is the ultraviolet cut-off.

possible if

$$\alpha_0 \geq \alpha_c \equiv \frac{\pi}{3} \quad . \tag{106}$$

Thus a dynamical mass can be generated only if the interaction is strong with coupling greater than unity. The factor of 3 and 1/3 in Eqs. (92,106), respectively, can be traced to back to Eqs. (81,88) where it appears as the (*number of dimensions - 1*). In physical terms the basic problem can be thought of in terms of the behaviour of a massless electron in the field of a series of nuclei of different atomic numbers. Imagine a world of *quenched* QED. What we have learnt is that in the field of heavy nuclei of charge $\geq 140$ the electron would gain a mass and no longer travel at the speed of light.

We can now solve Eq. (92), or equivalently Eqs. (94,103,104) numerically. The dependence of the generated mass function on momentum is illustrated in Fig. 40. To quantify the mass that is generated we can use either

$$M_0 \equiv \mathcal{M}(0) \qquad \text{or} \quad M_E \equiv \mathcal{M}(p^2 = M_E^2) \quad . \tag{107}$$

The latter can be regarded as the "Euclidean" mass. $M_0$ and $M_E$ have almost identical behaviour with the coupling, as seen from Fig. 40, so we call them simply $m$. In Fig. 41 we plot $m/\kappa$ as a function of $\alpha_0$. We see that for $\alpha_0 \leq \alpha_c$, there is only one solution $\mathcal{M}(p) = 0$. But as $\alpha_0$ increases above $\pi/3$, a second solution with non-zero mass bifurcates away. Though these results have been obtained in the rainbow approximation in the Landau gauge we expect qualitatively similar behaviour in all gauges, so we expect dynamical mass generation to occur if the coupling is of $\mathcal{O}(1)$. However, what does actually happen in other covariant gauges? We return to Eq. (88) with $\xi \neq 0$. Then we have to solve the coupled equations for $\mathcal{M}(p)$ and $\mathcal{F}(p)$. Again we plot, Fig. 42, the results for $m$ in units of $\kappa$ as the coupling varies in several covariant gauges. While the behaviour is seen to be qualitatively similar, the critical value of the coupling depends on the gauge. But this cannot be correct for a physical quantity — the behaviour of an electron in the field of a heavy nucleus cannot possibly depend on the gauge we choose for our computation. Clearly our *rainbow* approximation violates gauge invariance. Let us understand why.

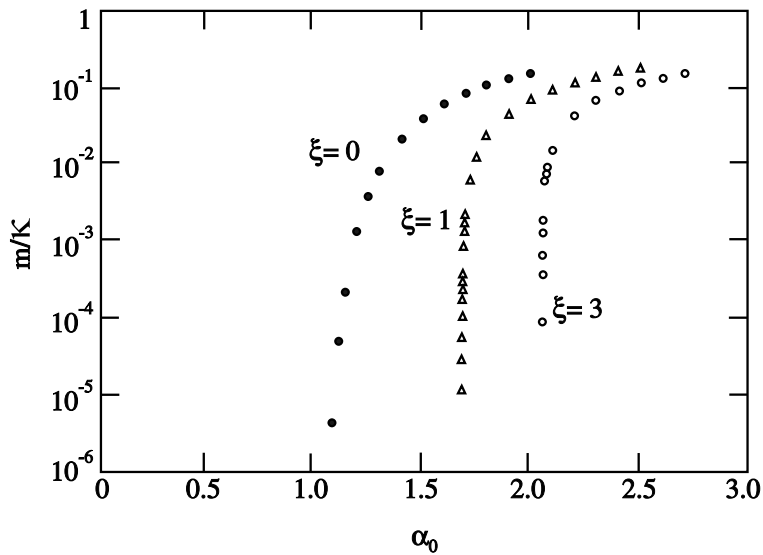

**Figure 42.** The Euclidean mass $m$ as a function of the coupling $\alpha_0$ in the *rainbow* approximation in different covariant gauges, denoted by $\xi$. $\kappa$ is the ultraviolet cut-off.

## 8. Ward-Green-Takahashi identity

The Ward-Green-Takahashi identity [76, 77] is one of the most important consequences of gauge invariance in QED. It states

$$S_F^{-1}(k) - S_F^{-1}(p) = q_\mu \Gamma^\mu(k,p) \quad . \tag{108}$$

Its origin starts with its form when $q^\mu = 0$. In 1952 J.C. Ward [76] noted that there is a very simple relation between the bare fermion propagator and the fermion-boson vertex, which are

$$S_F^{0\,-1}(p) = \slashed{p} - m_0 \,, \qquad \Gamma^0_\mu = \gamma_\mu \,, \tag{109}$$

connected by:

$$\frac{\partial}{\partial p_\mu} S_F^{0\,-1}(p) = \gamma^\mu = \Gamma^\mu_0 \quad . \tag{110}$$

This generalises to all orders in perturbation theory and indeed genuinely non-perturbatively to

$$\frac{\partial}{\partial p_\mu} S_F^{-1}(p) = \Gamma^\mu(p,p) \quad , \tag{111}$$

relating the full fermion propagator to the fermion-boson vertex, when the photon has zero momentum. It was Herb Green [77], who first recognised that this could be generalised further to the case of the photon carrying any momentum $q$. This is commonly known as the Ward-Takahashi identity. At the lowest order one readily sees that

$$S_F^{0\,-1}(k) - S_F^{0\,-1}(p) = \slashed{k} - \slashed{p} = \slashed{q} = q_\mu \Gamma^\mu_0 \quad .$$

This is the first step in recognising that the full propagator and vertex are related by

$$S_F^{-1}(k) - S_F^{-1}(p) = q_\mu \Gamma^\mu(k,p) \quad .$$

The Ward-Green-Takahashi identity has important consequences for both the fermion and photon propagators. Two of these we now discuss. We first consider the photon propagator $\Delta^{\mu\nu}(q)$ and its inverse $\Pi^{\mu\nu}(q)$ which are related by

$$\Pi_{\lambda\mu}(q) \Delta^{\mu\nu}(q) = g_\lambda^\nu \quad . \tag{112}$$

We can decompose both into transverse and longitudinal components:

$$\Delta^{\mu\nu}(q) = \Delta_T^{\mu\nu}(q) + \Delta_L^{\mu\nu}(q) \,, \qquad \Pi^{\mu\nu}(q) = \Pi_T^{\mu\nu}(q) + \Pi_L^{\mu\nu}(q) \,, \tag{113}$$

where by definition the transverse components are orthogonal to the photon momentum $q^\mu$. Thus we can write quite generally in a covariant gauge specified by the parameter $\xi$:

$$\Delta^{\mu\nu}(q) = \frac{\mathcal{G}(q)}{q^2} \left( g^{\mu\nu} - \frac{q^\mu q^\nu}{q^2} \right) + \xi \frac{q^\mu q^\nu}{q^4} \quad . \tag{114}$$

as in Eq. (83), so its inverse is given by

$$\Pi^{\mu\nu}(q) = \frac{1}{\mathcal{G}(q)} \left( q^2 g^{\mu\nu} - q^\mu q^\nu \right) + \frac{1}{\xi} q^\mu q^\nu \quad . \tag{115}$$

The function $\mathcal{G}(q^2) = 1$ at lowest order in perturbation theory and so represents the renormalisation of the photon propagator at higher orders. This function multiplies the transverse part, while the longitudinal propagator is unrenormalised at all orders.

To see how this is fulfilled by the full photon propagator, we consider its Schwinger-Dyson equation, Fig. 38,

$$\Pi^{\mu\nu}(q) = \Pi_0^{\mu\nu}(q) - \frac{e^2}{(2\pi)^n} \int d^n k \, \mathrm{Tr} \left\{ \gamma^\mu \, S_F(k) \, \Gamma^\nu(k,p) \, S_F(p) \right\} \quad . \tag{116}$$

To show that the self-energy corrections are wholly transverse, let us contract the inverse propagator of Eq. (116) with $q^\nu$

$$\Pi^{\mu\nu}(q) \, q_\nu = \Pi_0^{\mu\nu}(q) \, q_\nu - \frac{e^2}{(2\pi)^n} \int d^n k \, \mathrm{Tr} \left\{ \gamma^\mu \, S_F(k) \, \Gamma^\nu(k,p) \, S_F(p) \right\} q_\nu \quad , \tag{117}$$

which using Eq. (115) and the Ward-Green-Takahashi identity, Eq. (108), under the integral gives

$$\frac{q^\mu q^2}{\xi} = \frac{q^\mu q^2}{\xi} - \frac{e^2}{(2\pi)^n} \int d^n k \, \mathrm{Tr} \left\{ \gamma^\mu \, S_F(k) \left[ S_F^{-1}(k) - S_F^{-1}(p) \right] S_F(p) \right\} \quad . \tag{118}$$

This means that

$$\int d^n k \, \mathrm{Tr} \left\{ \gamma^\mu \left[ S_F(p) - S_F(k) \right] \right\} \tag{119}$$

must equal zero. If the integrals are regularised in a translationally invariant way then we can rewrite this as

$$\int d^n p \, \mathrm{Tr} \left\{ \gamma^\mu \, S_F(p) \right\} - \int d^n k \, \mathrm{Tr} \left\{ \gamma^\mu \, S_F(k) \right\} \tag{120}$$

where $p = k - q$. Since in 4-dimensions the integrals diverge we do have to ensure we can regularise these appropriately, but if this is done then the integrals being of infinite range exactly cancel and the correction to the photon propagator is transverse as required.

The Ward-Green-Takahashi identity has an equally important effect on the behaviour of the fermion propagator. We start with its Schwinger-Dyson equation, where the loop graph is shown in Fig. 43

$$S_F^{-1}(p) = S_F^{0\,-1}(p) - \frac{e^2}{(2\pi)^n} \int d^n k \, \gamma^\mu \, S_F(k) \, \Gamma^\nu(k,p) \, \Delta_{\mu\nu}(q = k-p) \quad . \tag{121}$$

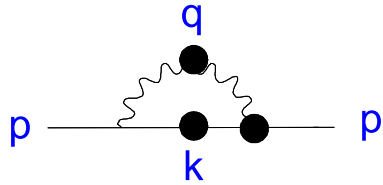

**Figure 43.** Momenta defined by the self-energy corrections in Eq. (121) are shown.

Its gauge dependence comes from the longitudinal part of photon propagator, Eq. (114). This component gives the following contribution to the loop integral

$$\int d^n k \, \gamma^\mu \, S_F(k) \, \Gamma^\nu(k,p) \, \xi \, \frac{q^\mu q^\nu}{q^4} \quad . \tag{122}$$

Substituting the Ward-Green-Takahashi identity of Eq. (108) gives

$$\xi \int d^n k \frac{\not{q}}{q^4} S_F(k) \left(S_F^{-1}(k) - S_F^{-1}(p)\right) = \xi \int d^n k \frac{\not{q}}{q^4} \left(1 - S_F(k) S_F^{-1}(p)\right) \quad (123)$$

Again by translation invariance

$$\int d^n k \frac{\not{q}}{q^4} = 0 \quad , \quad (124)$$

so the first term in the integral of Eq. (123) is zero. For a physical fermion there is a pole in the propagator at a momentum that defines its mass, $M$, namely at $p^2 = M^2$, $S_F^{-1}(p) = 0$. Thus the gauge dependent contribution to the full fermion propagator vanishes at $p^2 = M^2$ [78]. The Ward-Green-Takahashi identity ensures that the pole position (which is a physical observable) is gauge independent!

Whilst in sect. 7 we regularised the loop integrals by introducing an ultraviolet cut-off, $\kappa$, we have seen in our discussions here in imposing the consequences of gauge invariance, in particular Eqs. (120,124), requires an infinite range for the integration to ensure translational invariance. Dimensional regularisation in $n = 4 - \epsilon$ dimensions is a highly effective way to implement this in perturbative calculations, where the integrands are completely specified by the Feynman rules. However, in non-perturbative calculations, like those here, unknown renormalization functions, like $\mathcal{F}$ and $\mathcal{G}$ of Eqs. (86,83), appear in the integrals and make dimensional regularisation much more difficult to use. Nevertheless, numerical studies are encouraging [79].

Since the Ward-Green-Takahashi identity is so crucial to ensuring that solutions to the Schwinger-Dyson equations are physical, we want to build it into any ansatz for the full fermion-photon vertex, $\Gamma^\mu(k,p)$. Let us make the simplest guess as to how to solve the Ward-Green-Takahashi identity, Eq. (108), viz.

$$\Gamma^\nu(k,p) = \frac{q^\nu}{q^2} \left(S_F^{-1}(k) - S_F^{-1}(p)\right) \quad . \quad (125)$$

Now substitute this into the integrand for the fermion self-energy in the Landau gauge ($\xi = 0$) when the photon propagator is wholly transverse, then in Eq. (121)

$$\gamma^\mu S_F(k) \Gamma^\nu(k,p) \Delta_{\mu\nu}(q = k - p) = \gamma^\mu S_F(k) \Gamma^\nu(k,p) \frac{1}{q^2} \left(g_{\mu\nu} - \frac{q_\mu q_\nu}{q^2}\right)$$
$$= 0 \quad (126)$$

and we see the interaction makes no contribution to the fermion propagator. It remains unrenormalised. Can this be right?

To see that it is not correct, let us consider the limit $k \to p$ of Eq. (125):

$$\Gamma^\nu(p+q,p) = \frac{q^\mu q^\nu}{q^2} \frac{\partial S_F^{-1}(p)}{\partial p^\mu} \quad , \quad (127)$$

but this is not equal to the required Ward identity of Eq. (111):

$$\neq \frac{\partial S_F^{-1}(p)}{\partial p^\mu} \quad . \quad (128)$$

The reason for this is that the hypothetical solution, Eq. (125), has a kinematic singularity at $q^2 = 0$, while the fact that the Ward identity is the $q \to 0$ limit of the Green-Takahashi identity means that there can be no such singularity.

This provides us with an important clue as to how to solve the Ward-Green-Takahashi identity, by starting from the Ward identity. From its Dirac structure we know quite generally that the fermion propagator involves two functions which specify its wavefunction renormalization and its mass function, Eq. (86). This can be written equivalently as:

$$S_F^{-1}(p) = A(p^2)\,\not{p} + B(p^2)\,\mathbf{1} \quad , \tag{129}$$

where $\mathbf{1}$ is a $4{\times}4$ unit matrix and $A(p^2),\,B(p^2)$ are related to the wavefunction renormalization and mass function we introduced earlier by $A(p^2) = 1/\mathcal{F}(p)$ and $B(p^2) = -\,\mathcal{M}(p)/\mathcal{F}(p)$ in Minkowski space, but it is easier to use $A, B$ for the present purpose. Substituting this general expression into the Ward identity, Eq. (111), gives

$$\Gamma^\mu(p,p) = \frac{\partial}{\partial p_\mu} S_F^{-1}(p) = A(p^2)\gamma^\mu + A'(p^2)\,2\,\not{p}p^\mu + B'(p^2)\,2\,p^\mu \ . \tag{130}$$

This is the vertex when $k = p$. We then write a form for $\Gamma^\mu(k,p)$ so that this limit is automatic, namely:

$$\begin{aligned}\Gamma^\mu(k,p) = &\ \frac{1}{2}\left(A(k^2) + A(p^2)\right)\gamma^\mu + \frac{1}{2}\frac{(A(k^2) - A(p^2))}{k^2 - p^2}(\not{k}+\not{p})(k^\mu + p^\mu) \\ &\ + \frac{(B(k^2) - B(p^2))}{k^2 - p^2}(k^\mu + p^\mu) + \Gamma_T^\mu(k,p) \quad .\end{aligned} \tag{131}$$

Then the Green-Takahashi identity **and** Ward identity are satisfied provided:

$$\Gamma_T^\mu(p,p) = 0 \quad , \qquad q_\mu \Gamma_T^\mu(k,p) = 0 \quad . \tag{132}$$

This form is free of kinematic singularities, unlike the incorrect representation of Eq. (125). Having deduced the vertex that fulfills both the Ward and Green-Takahashi identities, it is useful to translate this in terms of the wavefunction renormalization and mass function in Minkowski space. Eq. (131) then becomes

$$\begin{aligned}\Gamma^\mu(k,p) = &\ \frac{1}{2}\left(\frac{1}{\mathcal{F}(k)} + \frac{1}{\mathcal{F}(p)}\right)\gamma^\mu + \frac{1}{2}\left(\frac{1}{\mathcal{F}(k)} - \frac{1}{\mathcal{F}(p)}\right)\frac{(\not{k}+\not{p})(k^\mu + p^\mu)}{k^2 - p^2} \\ &\ - \left(\frac{\mathcal{M}(k)}{\mathcal{F}(k)} - \frac{\mathcal{M}(p)}{\mathcal{F}(p)}\right)\frac{k^\mu + p^\mu}{k^2 - p^2} + \Gamma_T^\mu(k,p)\end{aligned} \tag{133}$$

as first deduced by Ball and Chiu [80]. At present we know nothing about the transverse component of the vertex $\Gamma_T^\mu$ beyond the two conditions of Eq. (132).

As noted by Bernstein [81] the vertex $\Gamma^\mu(k,p)$ representing the coupling of two spin 1/2 particles to a vector boson can be decomposed in terms of **12** basis vectors. 4 of these are longitudinal:

$$\gamma^\mu \ ; \quad (k^\mu + p^\mu) \ ; \quad (k^\mu + p^\mu)(\not{k}+\not{p}) \quad \text{and} \quad \sigma^{\mu\nu}(k_\nu + p_\nu) \ . \tag{134}$$

The coefficients of each of these are fixed by the Ward-Green-Takahashi identity, the fourth being zero — see Eq. (108). The eight transverse $T_i^\mu(k,p)$ form a basis for writing

$$\Gamma_T^\mu(k,p) = \sum_i \tau_i(k^2, p^2, q^2)\,T_i^\mu(k,p) \ , \tag{135}$$

where for instance

$$T_2^\mu(k,p) = (p^\mu k \cdot q - k^\mu p \cdot q)(\not{k} + \not{p}) ,$$

$$T_3^\mu(k,p) = q^2 \gamma^\mu - q^\mu \not{q} ,$$

$$T_6^\mu(k,p) = \gamma^\mu \left(k^2 - p^2\right) - (k+p)^\mu (\not{k} - \not{p}) ,$$

$$T_8^\mu(k,p) = -\gamma^\mu p^\nu k^\rho \sigma_{\nu\rho} + p^\mu \not{k} - k^\mu \not{p} , \tag{136}$$

each satisfies

$$q_\mu T_i^\mu(k,p) = 0 , \qquad T_i^\mu(p,p) = 0 . \tag{137}$$

The coefficients $\tau_i(k^2, p^2, q^2)$ are determined by the Schwinger-Dyson equation for the vertex in Fig. 38, which involves the full 4-fermion interaction which is too complicated to solve exactly.

## 9. Multiplicative Renormalizability

Of course, the introduction of an ultraviolet cutoff $\kappa$ to regulate the loop momenta is unphysical. One can think of this as defining where new interactions of which we know nothing act at very short distances to cut off the theory we are studying. Physical quantities cannot depend on our choice of cutoff, in much the same way as Newton's law of gravity cannot depend on Planck's constant. To take account of this we renormalize each Green's function by defining the physical value of some quantities at some momentum scale $\mu_R$. In a gauge theory only a small number of physical observables are required to specify every Green's function and hence provide predictability.

The act of renormalization is performed by multiplying each Green's function by a small number of $Z$-factors that themselves depend on both $\kappa$ and $\mu_R$ and ensure the renormalized Green's functions (denoted by the subscript $R$) are independent of the cutoff $\kappa$. Thus for the fermion and boson propagators and the fermion-boson vertex we have three factors $Z_i$ defined by [74]

$$S_R(p, \mu_R) = Z_2^{-1}(\kappa, \mu_R) S(p, \kappa) \tag{138}$$

$$\Delta_R^{\mu\nu}(q, \mu_R) = Z_3^{-1}(\kappa, \mu_R) \Delta^{\mu\nu}(q, \kappa) \tag{139}$$

$$\Gamma_R^\mu(k, p, \mu_R) = Z_1(\kappa, \mu_R) \Gamma^\mu(k, p, \kappa) \tag{140}$$

These in turn define the renormalized coupling $e_R$ in terms of the bare coupling $e$:

$$e_R = \frac{Z_2}{Z_1} \sqrt{Z_3} \, e \tag{141}$$

We have to ensure that the Ward-Green-Takahashi identity, which is such a key consequence of gauge invariance, holds both for the bare and renormalized Green's functions. Let us recall the bare relationship:

$$q_\mu \Gamma^\mu(k, p, \kappa) = S^{-1}(k, \kappa) - S^{-1}(p, \kappa) , \tag{142}$$

which using Eqs. (138-140) implies the following relation for the renormalized vertex and propagator:

$$q_\mu Z_1^{-1} \Gamma_R^\mu(k, p, \mu_R) = Z_2^{-1}(\kappa, \mu_R) \left(S_R^{-1}(k, \mu_R) - S_R^{-1}(p, \mu_R)\right) , \tag{143}$$

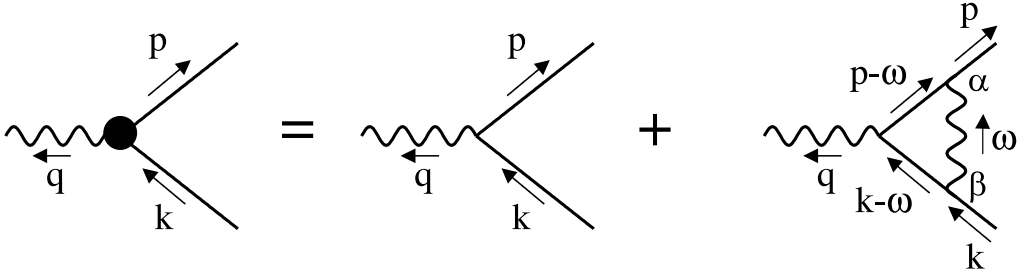

**Figure 44.** One loop perturbative corrections to the fermion-boson vertex, $\Gamma^\mu(k,p)$.

which implies the Ward-Green-Takahashi identity for the renormalized vertex and fermion propagator:

$$q_\mu \Gamma^\mu_R(k,p,\mu_R) = S_R^{-1}(k,\mu_R) - S_R^{-1}(p,\mu_R) \tag{144}$$

provided

$$Z_1(\kappa,\mu_R) = Z_2(\kappa,\mu_R) \quad . \tag{145}$$

To see what the consequences of multiplicative renormalizability might be for the structure of the full fermion-boson vertex, $\Gamma^\mu_R$, let us see what multiplicative renormalizability means for the fermion functions $\mathcal{F}$ and $\mathcal{M}$. We rewrite the condition:

$$S_R(p,\mu_R) = Z_2^{-1}(\kappa,\mu_R) S(p,\kappa) \quad, \tag{146}$$

using the general form of Eq. (86) then

$$\frac{\mathcal{F}_R(p,\mu_R)}{\not{p} - \mathcal{M}_R(p,\mu_R)} = Z_2^{-1}(\kappa,\mu_R) \frac{\mathcal{F}(p,\kappa)}{\not{p} - \mathcal{M}(p,\kappa)} \quad, \tag{147}$$

which implies

$$\mathcal{F}_R(p,\mu_R) = Z_2^{-1}(\kappa,\mu_R) \mathcal{F}(p,\kappa) \quad, \qquad \mathcal{M}_R(p,\mu_R) = \mathcal{M}(p,\kappa) \quad, \tag{148}$$

while

$$\Gamma^\mu_R(k,p,\mu_R) = Z_1(\kappa,\mu_R) \Gamma^\mu(k,p) \quad . \tag{149}$$

Thus, we expect from multiplicative renormalizability that all components of the vertex

$$\Gamma \sim \frac{1}{\mathcal{F}} \tag{150}$$

just like the part determined by the Ward-Green-Takahashi identity in the Ball-Chiu construction of Eq. (133). This is the clue that the transverse component of the full fermion-boson vertex being multiplicatively renormalized by the same factor $Z_1$ as the longitudinal component, must similarly be proportional to $1/\mathcal{F}$.

The coefficients $\tau_i(k^2,p^2,q^2)$ of Eq. (135) are fixed, of course, in perturbation theory when the 4-fermion interaction is given by one boson exchange at $O(\alpha)$ shown in Fig. 44. Even at this order the $\tau_i$ are far from trivial [82]. However, multiplicative renormalizability of the fermion propagator is related to the behaviour of the loop diagram of Fig. 37 at large loop momentum, $k$. This is dependent on the form of the vertex when $k^2 \gg p^2$. In this limit

$$\Gamma^\mu_T(k,p) = -\frac{\alpha\xi}{8\pi} \ln\frac{k^2}{p^2} \left[\gamma^\mu - \frac{k^\mu \not{k}}{k^2}\right] \quad . \tag{151}$$

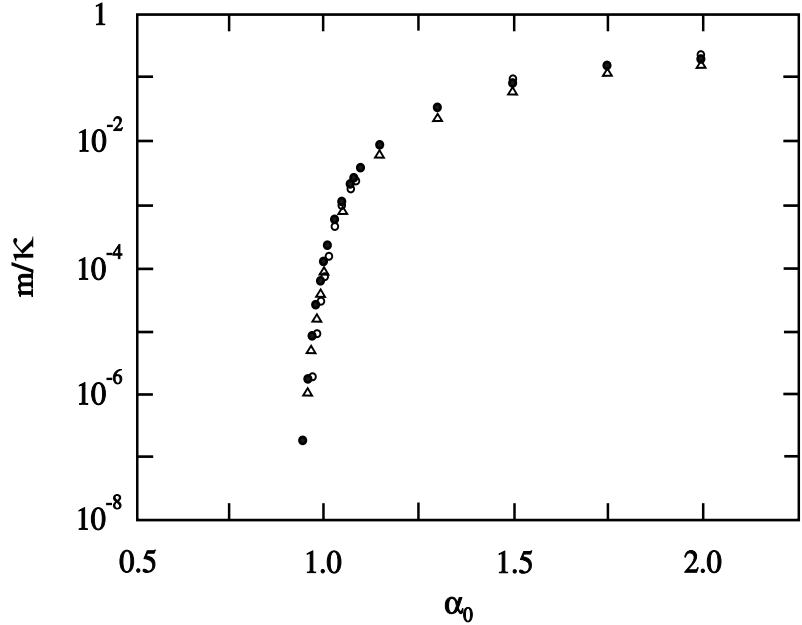

**Figure 45.** The Euclidean mass $m$ [84] as a function of the coupling $\alpha_0$ with the CP vertex in the same covariant gauges as Fig. 42. $\kappa$ is the ultraviolet cut-off.

This can be written non-perturbatively as noted in Ref. [83] as

$$\Gamma_T^\mu(k^2 \gg p^2) = \frac{1}{2}\left(\frac{1}{\mathcal{F}(k)} - \frac{1}{\mathcal{F}(p)}\right)\frac{(k^2+p^2)}{(k^2-p^2)^2}T_6^\mu \quad, \tag{152}$$

where $T_6^\mu$ is defined in Eq. (136). Including this transverse vertex in Eqs. (133,135) and substituting into Eqs. (81,121), we can again calculate the coupled fermion system for $\mathcal{F}(p^2)$ and $\mathcal{M}(p^2)$. The mass that is now generated is shown in Fig. 45 and remarkably we see the result is almost gauge independent. We have achieved an almost physical result with this simple approximation to the full fermion-boson interaction.

While the CP vertex of Eq. (152) does lead to a multiplicatively renormalizable result for the fermion propagator and near gauge independence for any critical behaviour [84], it does this only in the quenched approximation, when the coupling does not run and there are no corrections to the free boson propagator. To determine the vertex structure that ensures both the fermion and boson propagators are multiplicatively renormalizable is a trickier problem. Renormalizability depends on the behaviour of the loops at large loop momenta, with the external momentum fixed. The multiplicative renormalizability of the fermion and boson propagators probe different momentum regions of the vertex, Fig. 38. With the vertex defined as in Eqs. (151,152) where $k$, $p$ are the fermion momenta 'in' and 'out', and $q = k - p$ is the photon momentum, the fermion propagator probes the vertex when $k^2 \gg p^2$, while for the boson propagator it is $k^2 \gg q^2$, i.e. when $k^2 \simeq p^2$, that is critical. A solution to this problem has been worked out by Ayşe Kızılersü and myself [85], but that is beyond the scope of these lectures.

A Green's function in one gauge is related to its form in any other gauge. Landau and Khalatnikov [86] worked out this connection. Though formally very simple in coordinate space, the Landau-Khalatnikov transformations do require knowledge of the complete non-perturbative propagators and vertices in some gauge before we can determine what they are in another gauge. Only relatively recently has significant progress been made in solving these transformations, particularly by Bashir and Raya [87]. Again the discussion of these interesting constraints is a topic beyond these lectures.

## 10. Irrelevant Operators

We have seen that in a renormalizable QED-like gauge theory with Lagrangian

$$\mathcal{L} = \overline{\psi}\,(i\gamma_\mu\,D^\mu - m_0)\,\psi - \frac{1}{4}\,\mathcal{F}_{\mu\nu}\,\mathcal{F}^{\mu\nu} \quad, \tag{153}$$

that if the bare mass $m_0 = 0$, the fermion field $\psi$ can have a non-zero mass if the dimensionless coupling implicit in the covariant derivative $D^\mu$ is larger than some critical value $\alpha_c$ of order unity. But QED is often regarded as incomplete, for reasons that are discussed in Ref. [70, 88]. To see why let us notice that something surprising happens as $\alpha \to \alpha_c$.

To be renormalizable we do not include four-fermion interactions in the QED Lagrangian. Such interactions would be described by a term given by:

$$\mathcal{L}_{4f} = \frac{G}{\kappa^2}\,(\overline{\psi}\psi)^2 \quad, \tag{154}$$

where we introduce a mass scale $\kappa$ so that the four-fermion coupling $G$ is dimensionless. $\kappa$ can be thought of as the ultraviolet cutoff of Eqs. (88,89). When $\kappa \to \infty$ such interactions disappear and are consequently said to be *irrelevant operators*. However, let's see what happens when $\alpha \to \alpha_c$.

First recall from Eq. (96) the large momentum behaviour of the mass function we deduced:

$$\mathcal{M}(p) \simeq a_1\,\kappa\,\left(\frac{p^2}{\kappa^2}\right)^{s_1} + a_2\,\kappa\,\left(\frac{p^2}{\kappa^2}\right)^{s_2} \quad, \tag{155}$$

where $s_1, s_2$ are the two roots of Eq. (98). When the interaction is switched off by taking $\alpha \to 0$, gives the canonical behaviour

$$\mathcal{M}(p) \simeq a_1\,\kappa + a_2\,\frac{\kappa^3}{p^2} \quad. \tag{156}$$

The first term corresponds to the bare mass component, which here is zero, and so the general behaviour is given by

$$\mathcal{M}(p) \sim a_2\,\frac{\kappa^3}{p^2}\,\left(\frac{\kappa}{p}\right)^{\sqrt{1-\alpha/\alpha_c}-1} \quad. \tag{157}$$

Now let us note that from the operator product expansion (cf. Eq. (167) later)

$$\mathcal{M}(p) \sim \frac{\langle\overline{\psi}\psi\rangle}{p^2} \quad. \tag{158}$$

Then Eq. (157) implies that

$$\langle\overline{\psi}\psi\rangle^2 \sim \kappa^{4+2\sqrt{1-\alpha/\alpha_c}}\,p^{2-2\sqrt{1-\alpha/\alpha_c}} \quad. \tag{159}$$

From this we see that

$$\langle\overline{\psi}\psi\rangle^2 \sim \kappa^6 \qquad \text{when} \quad \alpha \to 0 \quad, \tag{160}$$

which in Eq. (154) makes the 4-fermion operator irrelevant for a renormalizable gauge theory. However, the anomalous dimensions change its behaviour to

$$\langle\overline{\psi}\psi\rangle^2 \sim \kappa^4\,p^2 \qquad \text{when} \quad \alpha \to \alpha_c \quad. \tag{161}$$

The $\kappa^4$ behaviour of the Lagrangian density makes it a relevant operator. Consequently, electrodynamics without 4-fermion interactions is incomplete in studying the phase structure of the theory. Nevertheless, we have learnt that if the fermion-boson interaction is strong enough a mass can be generated dynamically for the matter particles. This is key to how chiral symmetry breaking works in QCD.

## 11. QCD and Confinement

It is an experimental fact that free quarks are not seen. When an electron and positron annihilate and create hadrons, they do this by first creating a $q\bar{q}$ pair, which subsequently radiate gluons which in turn create new $q\bar{q}$ pairs. Though at high energies jets of hadrons are observed that follow the direction of the initially produced quark and antiquark, remembering their spin and additive quantum numbers, free quarks are not observed. The quarks never leave the femto-universe. The fact that the strong coupling becomes of $\mathcal{O}(1)$ over such distance scales, as seen in Fig. 17, is believed to stop coloured states propagating beyond a few fermis. How do we calculate this in QCD?

A simple picture is provided by studying heavy quark systems, which being non-relativistic allow us to consider the bound states in a potential model. At short distances this is very similar to positronium, since there the potential is Coulomb-like being generated by one gluon exchange. However, at large distances the differences between QED and QCD arise. While in QED the potential goes to a constant at large distances, for a system of very heavy quarks the potential rises indefinitely because of confinement. Phenomenology, as indicated first by Richardson [89], is well approximated by

$$V(r) = -c\frac{\alpha(r)}{r} + \sigma r \quad , \qquad (162)$$

shown in Fig. 46, where $c$ is dimensionless constant and $\sigma$, the "string tension". This picture underlies the Wilson area law for confinement of infinitely heavy quarks.

How is this related to QCD? For such static sources, the potential is the Fourier transform

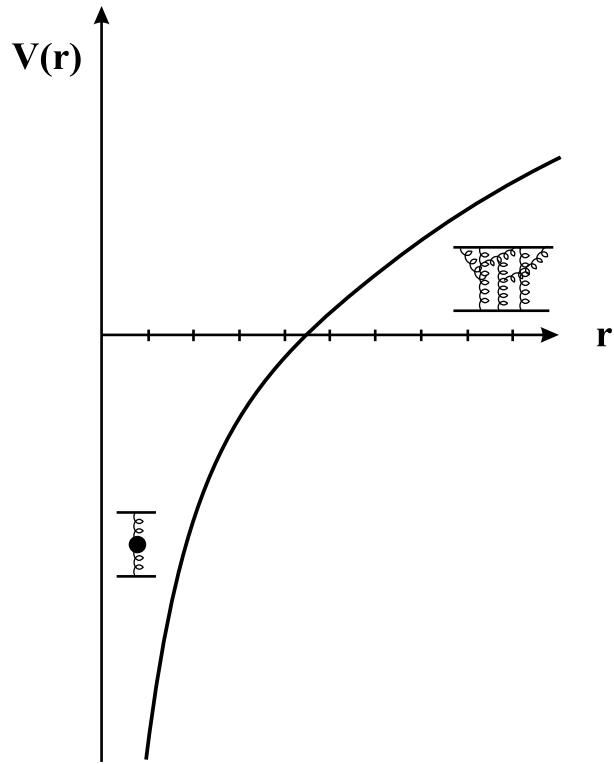

**Figure 46.** Sketch of the potential between heavy quarks deduced from the spectrum of charmonium and bottomonium, Eq. (162). At short distances this is dominated by one gluon exchange, whereas at larger distances this corresponds to a network of gluon exchanges.

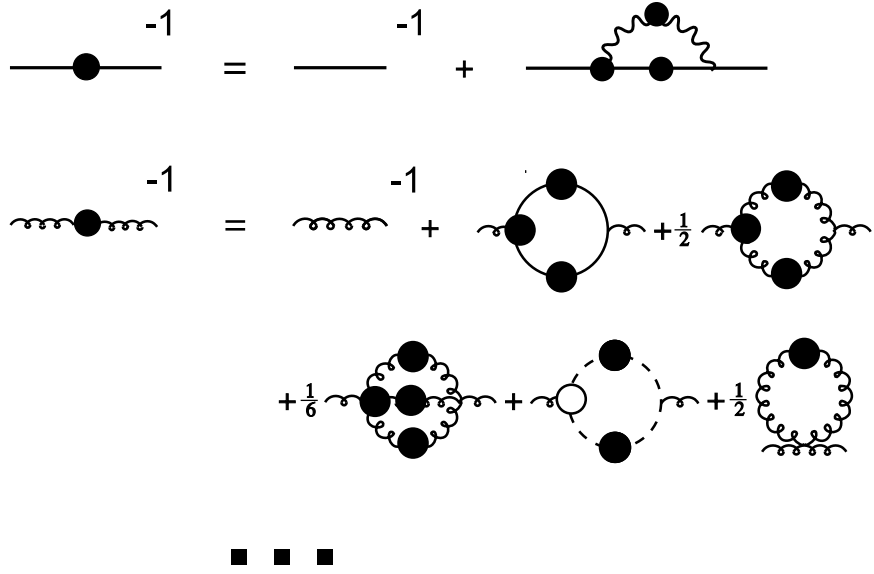

**Figure 47.** Schwinger-Dyson equation for the quark and gluon propagators. The solid dots mean the Green's functions are fully dressed. The dashed loop indicates the ghost component.

of the time-time component of the gluon propagator. This tells us that if the potential behaves like
$$V(r) \simeq r^a \quad , \quad \text{then} \quad \Delta_{00}(q) \simeq q^{-3-a} \quad , \tag{163}$$
for $qr \sim 1$. At short distances when $r \ll 1$ fm, the potential is Coulombic and $a = -1$, then $\Delta_{00}(q) \sim q^{-2}$, like the bare propagator when $q \gg 1\,\text{GeV}$, as expected from asymptotic freedom. However, at longer distances when $r > 1$ fm, the Richardson potential is linearly rising with $a = +1$, which means that $\Delta_{00}(q) \sim q^{-4}$, when $q < 250$ MeV. Such behaviour is not possible at any finite order in perturbation theory. As indicated earlier, the natural vehicle for investigating the strong coupling behaviour is through the Schwinger-Dyson equations applied now to QCD.

The Schwinger-Dyson equations for the quark and gluon propagators provide a coupled system rather like that for the electron and photon we considered in sect. 7. However, the equation for the boson propagator shown in Fig. 47 is just that bit more complicated because of the self-interaction of the gluon fields. We expect colour confinement to occur even if the number of quark flavours goes to zero. Then the source of confinement like that of asymptotic freedom is generated by the gluon self-interactions. The first approximation is to neglect loops with 'dressed' quartic gluon interactions, i.e. we keep only the tadpole graph with the bare quartic coupling, which of course is fixed by the QCD Lagrangian. In axial gauges the gluon has only the two transverse degrees of freedom. These are defined by the inverse of the gluon propagator, $\Pi^{\mu\nu}(q, n)$, not only being orthogonal to its momentum $q^\mu$, but $\Delta^{\mu\nu}$ being orthogonal to the 'axial' vector, $n^\mu$. Consequently we have two renormalization functions $\mathcal{G}_1$ and $\mathcal{G}_2$ with

$$\Pi^{\mu\nu}(q) = \frac{q^2}{(\mathcal{G}_1(q,n) + \mathcal{G}_2(q,n))(\mathcal{G}_1(q,n) + \mathcal{G}_2(q,n)\gamma)} \times$$
$$\left[\mathcal{G}_1(q,n)\left(g^{\mu\nu} - \frac{q^\mu q^\nu}{q^2}\right) + \gamma\mathcal{G}_2(q,n)\left(g^{\mu\nu} - \frac{n^\mu q^\nu + n^\nu q^\mu}{n \cdot q} + \frac{n^\mu n^\nu q^2}{(n \cdot q)^2}\right)\right] \tag{164}$$

where $\gamma = (n \cdot q)^2/(n^2 q^2)$. At zeroth order in perturbation theory, $\mathcal{G}_1 = 1$ and $\mathcal{G}_2 = 0$.

Just as $n$-point Green's functions in QED are related by a series of Ward identities, in QCD they are connected by their non-Abelian generalization, the Slavnov-Taylor identities [90]. The key "dressed" triple gluon interaction, $\Gamma^{\lambda\mu\nu}(p,q,r)$, where $p$, $q$, $r$ are the incoming 4-momenta with $p+q+r=0$, shown in Fig. 48, satisfies the simple Slavnov-Taylor identity:

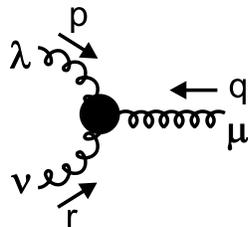

$$p_\lambda \Gamma^{\lambda\mu\nu}(p,q,r) \;=\; \Pi^{\mu\nu}(q,n) \;-\; \Pi^{\mu\nu}(r,n) \quad . \tag{165}$$

**Figure 48.** The momenta and Lorentz indices for the 3-gluon vertex, which, in axial gauges defined by the vector $n^\mu$, satisfies the Slavnov-Taylor identity stated in Eq. (165).

While axial gauges seem much simpler, one never gets anything for nothing. The axial gauge gluon propagator has the two dressing functions of Eq. (164). Baker, Ball and Zachariasen [91] solved the corresponding Schwinger-Dyson equation, assuming that $\mathcal{G}_2 \equiv 0$ non-perturbatively. They then found that the propagator behaved like $1/q^4$ at low momenta (i.e. $\mathcal{G}_1 \sim 1/q^2$). However, West [92] subsequently showed that setting $\mathcal{G}_2 = 0$ was an inconsistent approximation in a rather dramatic way. He proved that the gluon propagator in axial gauges, which have only positive norm states, could not be more singular than $1/q^2$, and so studies in axial gauges came to a halt.

Turning to covariant gauges, the gluon dressing involves just one function, as for the photon in Eq. (83), but ghosts are essential in removing the unphysical scalar components of the 4-dimensional gluon. These ghosts have their own dressing function, but the first studies either neglected this completely, or approximated it by its bare form [93, 94, 95]. The ghosts substantially complicate the Slavnov-Taylor identity for the 3-gluon vertex. Eq. (165) becomes

$$p_\lambda \Gamma^{\lambda\mu\nu}(p,q,r) \;=\; H(p^2)\left[G_{\mu\sigma}(r,-p)\,\Pi^T_{\sigma\nu}(q) \;-\; G_{\nu\sigma}(q,-p)\,\Pi^T_{\sigma\mu}(r)\right] \quad , \tag{166}$$

where $\Pi^T_{\alpha\beta}$ refers to the first term of Eq. (115), $G_{\mu\nu}$ is a ghost-gluon scattering kernel and $H(p^2)$ is the ghost dressing function. Assuming the ghost is bare simplifies this drastically. Keeping the ghosts to regulate the ultraviolet behaviour of the gluon then matches the perturbative result. Nevertheless, this approximation presumes that ghosts play no significant role in the infrared confinement dynamics. The first investigations were performed by Pagels [96], by Mandelstam [93] and by Bar-Gadda [94]. These all indicated that the gluon was highly singular in the infrared region with a behaviour very like the $1/q^4$ form we anticipated before, that is linked to a linearly rising heavy quark potential of Eq. (162) and Fig. 46.

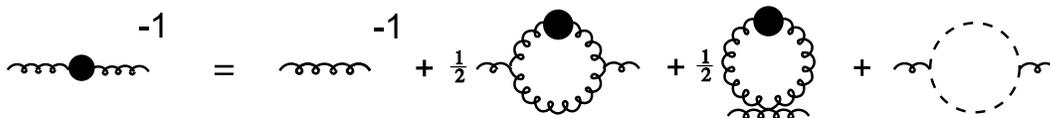

**Figure 49.** The "Mandelstam" approximation to the full Schwinger-Dyson equation for the gluon propagator used by Brown et al. [95] with a bare ghost propagator.

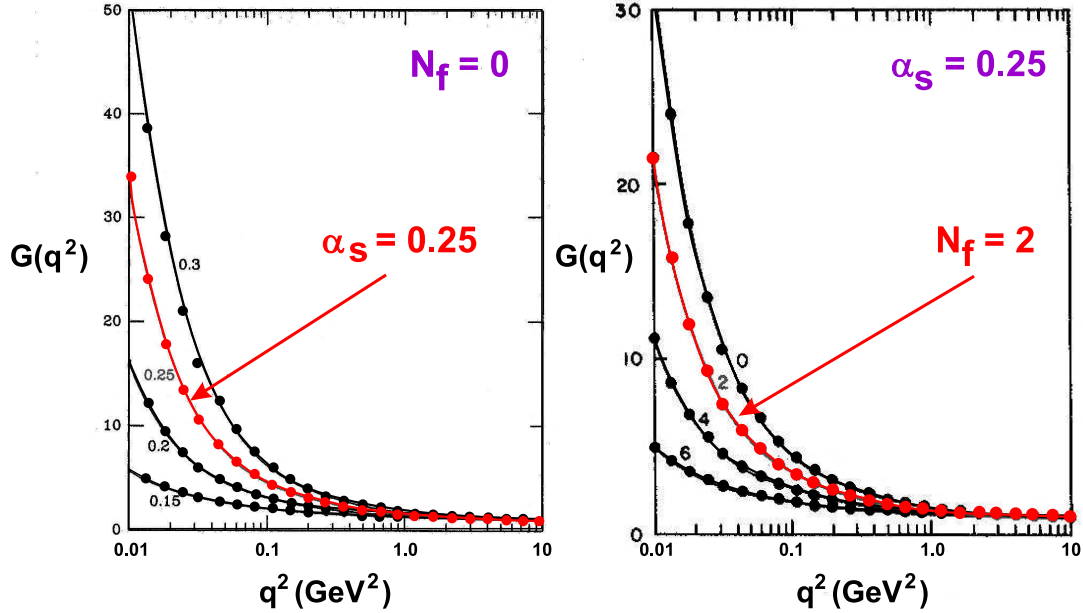

**Figure 50.** Renormalization function, $G(q^2)$, for the gluon propagator as a function of $q^2$ determined by solving the Schwinger-Dyson equation in the approximation shown in Fig. 49 [95]. The input to the right hand side of Fig. 49 are shown as curves, while the dots are the output on the left of the equation. The results in the graph on the left above are for $N_f = 0$ for various values of the strong coupling $\alpha_s$ at 10 GeV$^2$. On the right are the results of solving the coupled equations of Fig. 51 self-consistently with $\alpha_s = 0.25$ for different numbers of massless fermions. The solutions for $\alpha_s = 0.25$ with $N_f = 0$ and 2 are highlighted.

The largest numerical study was performed in the second half of the 1980's by Nick Brown and myself [95] in the Landau gauge. Setting the ghost function to unity, and neglecting quartic gluon couplings, the gluon Schwinger-Dyson equation shown in Fig. 49 was solved. The gluon dressing function $G(q^2)$ is defined as in Eq. (83). The self-consistent solutions for $G$ at Euclidean momenta (i.e. in the spacelike region of Minkowski space) are displayed on the left in Fig. 50. The different curves are labelled by the value of the strong coupling $\alpha_s(\mu^2)$ at $\mu^2 = 10$ GeV$^2$. The Fourier transform of the time-time component of this gluon propagator is then remarkably close [95] to the phenomenological non-relativistic potential for heavy quark systems of Fig. 46. For heavy quarks, only gluon interactions matter for confinement. But what about the role of light quarks?

**Figure 51.** Schwinger-Dyson equations for the gluon propagators in the "Mandelstam" approximation with $N_f$ flavours of fermion used in the analysis of Ref. [95]. The solutions for the gluon for different numbers of massless fermion flavours are shown on the right in Fig. 50.

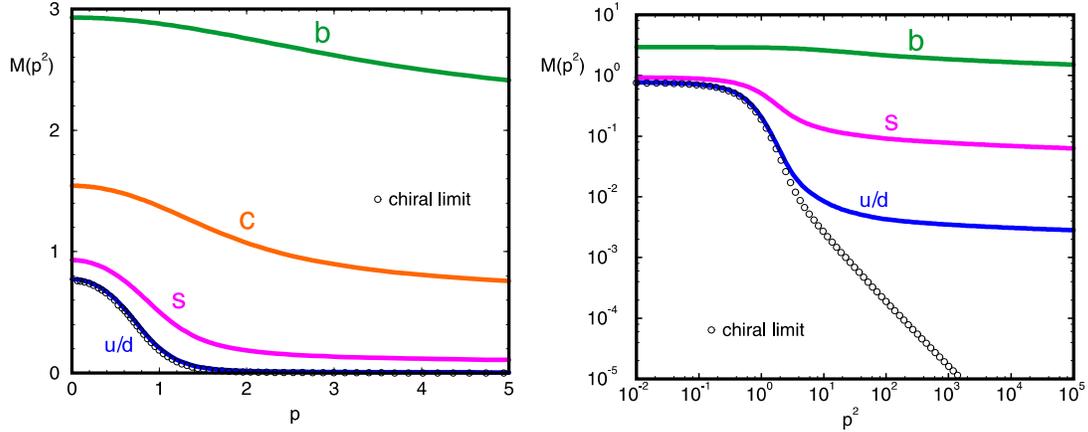

**Figure 52.** Mass function $\mathcal{M}(p^2)$ as a function of momentum, $p$, on linear and logarithmic scales for different current masses corresponding to $b$, $s$ and $u, d$ quarks. The chiral limit of massless current quarks is shown by the circles, from the results of Ref. [97].

To study this we coupled the gluon to the quark equation and considered just massless quarks, as in Fig. 51. On the right in Fig. 50 is shown how the gluon dressing function, $G(q^2)$, now changes as the number of massless quarks is added. Quarks are seen to have a deconfining effect — the gluon propagator is less singular at infrared momenta the more massless fermions there are. Though Brown and myself [95] only studied the chirally symmetric case, we found that the enhanced infrared behaviour of the gluon removed the pole in the quark propagator, so that the fermion renormalization function $\mathcal{F}(p^2) \to 0$ as $p^2 \to 0$. Exactly as one would expect for a confined field.

In our earlier lecture on dynamical mass generation, we anticipated that the enhancement of the coupling at low momentum is sufficient to produce an effective quark mass. Indeed, Maris and Roberts [97] approximated the product $[\,\alpha\,\Gamma^\nu\,\Delta^{\mu\nu}\,]$ of Eq. (81) to show how 300 MeV of mass is generated for each quark by the enhanced gluon interactions. In Fig. 52, we display their results for the quark mass at Euclidean momenta for different quark flavours. The flavours are specified by the intrinsic (current) quark mass at large spacelike momenta. The $u$ and $d$ quarks have current masses of just a few MeV. On the same plot are shown the results for zero quark mass, which differ little on the linear scale. However, on a logarithmic scale we see a significant difference. The operator product expansion tells us that the fermion mass function has its momentum dependence given by:

$$\mathcal{M}(p^2) \;=\; m_0 \left[\ln(p^2/\Lambda_{QCD}^2)\right]^{d_1} \;+\; C\,\frac{\langle \overline{q}q \rangle}{p^2}\left[\ln(p^2/\Lambda_{QCD}^2)\right]^{d_2} \quad, \tag{167}$$

where the anomalous dimensions, $d_1$, $d_2$, and the constant $C$ are calculable. Langfeld *et al.* [98] proved that in the chiral limit (i.e. when $m_0 \to 0$), the $\overline{q}q$ condensate is gauge independent. They show that the numerical value of the condensate in this limit from the results of Fig. 52 is

$$\langle \overline{q}q \rangle_0 \equiv \langle\, O \,\mid\, \overline{q}q \,\mid\, O\, \rangle_0 \simeq -\,(240\text{ MeV})^3 \quad, \tag{168}$$

very close to the value of Eq. (73) phenomenology requires.

While the strong coupling generated by the infrared enhanced gluon gives a potential that confines quarks and removes the pole in the fermion propagators, how are gluons themselves confined [99]? Their pole is not suppressed, but strongly reinforced at zero momentum. The gluon confines, but is it confined? These alternatives are illustrated in Fig. 53?

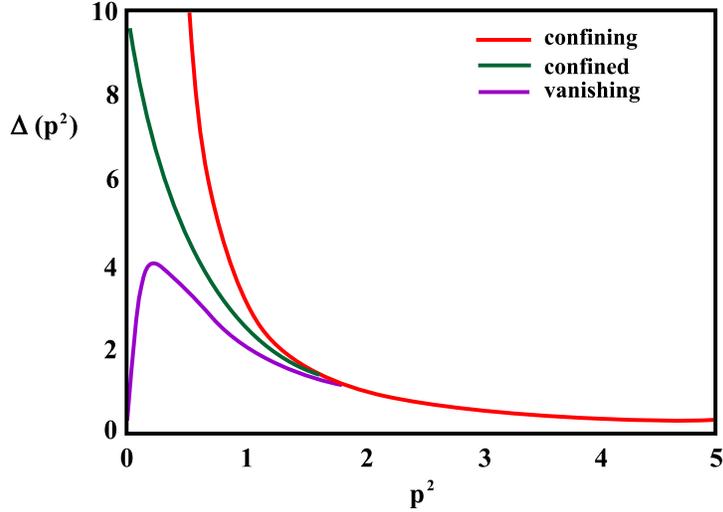

**Figure 53.** Three different behaviours of the Landau gauge gluon propagator, $\Delta(p^2)$ defined as the 00 component of Eq. (83), deduced in studies of the Schwinger-Dyson equations of Fig. 47 with different approximations, as discussed in Ref. [99]. Each has the same large momentum behaviour, but as $p^2 \to 0$, they have different values of the exponent $a$ in Eq. (163) : the curve labelled *confining* has $a \simeq 1$ as found in Refs. [93, 94, 95], *vanishing* has $a < -3$ as in Ref. [100], while *confined* is the limiting behaviour with $a$ just less than $-1$.

All these studies neglected any infrared role for ghosts. The Tübingen group of Alkofer *et al.* [100, 101] have addressed this issue. The ghost function itself is not constrained by Slavnov-Taylor identities but plays a role in constraining the triple gluon and quark-gluon interaction, as in Eq. (166). Watson [102] performed the most complete study to date of what is known about the ghost interactions. The simplest fact is that the ghost dressing function is one in the Landau gauge. This is the gauge the Tübingen group use to solve the coupled ghost, gluon Schwinger-Dyson equations with results shown in Fig. 54. Alkofer and co-workers [100] find that contrary to previous investigations the gluon does not become enhanced at low momenta, but in fact its dressing function goes to zero. In contrast, it is the ghost dressing function that becomes enhanced at infrared momenta. Comparison of these calculations with Monte Carlo lattice simulations shows excellent agreement [101, 100]. The combined behaviour of the gluon and ghost functions means the effective coupling increases at low momenta, with a constant infrared limit, as illustrated in Fig. 54. Though the exact value of this limit is known to be dependent on the approximations made, the fact that it is larger than one is sufficient to ensure that dynamical chiral symmetry breaking occurs.

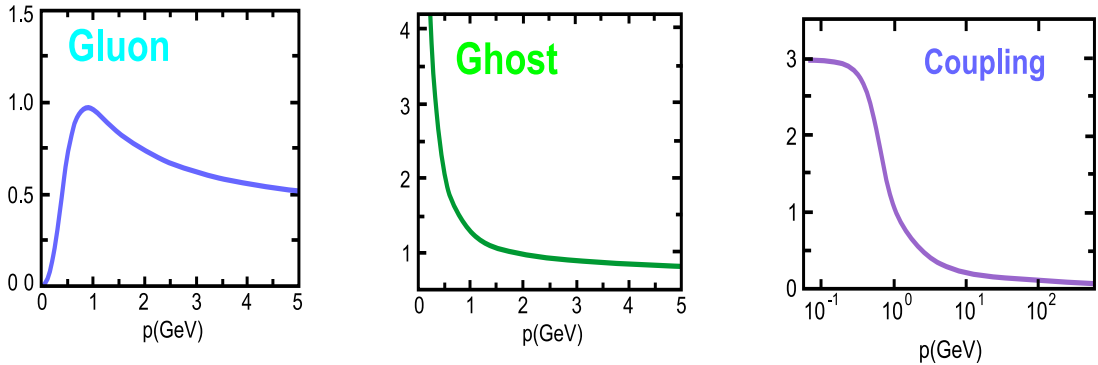

**Figure 54.** Momentum dependence of the Landau gauge gluon and ghost dressing functions from Schwinger-Dyson studies [100]. These give the *effective* quark-gluon coupling shown.

This scheme appears to solve the mystery of whether gluons are confined or confining [99]? If a particle is confined then one does not expect it to have a mass-shell and so the numerator of its propagator should cancel any potential pole in the denominator. This the Tübingen scheme [100] achieves for the transverse gluons. Nevertheless, we know the effective coupling must be enhanced in the infrared if confinement is to be generated. Here this is produced by the unphysical degrees of freedom of the gluon (and ghost) propagators. Thus this scheme for the ghost/gluon sector potentially resolves the dilemma and both confines and is confined. However, how this mechanism can actually generate the linearly confining potential for heavy quark systems of Eq. (162) and Fig. 46 is quite unclear. This is presently under study.

In summary, the Schwinger-Dyson approach outlined here has made considerable progress in providing an understanding of the strong coupling regime responsible for confinement and dynamical chiral symmetry breaking, but we still have many unresolved issues for future work.

## 12. The Higgs of the strong interaction

We have seen that experiment confirms that the near masslessness of the pion is because it is the Goldstone boson of chiral symmetry breaking. This breaking is generated by the non-zero vacuum expectation value for a scalar field, but what scalar field? In the Nambu model this is a single field, the $\sigma$, but nature as we shall see indicates there are a series of scalars. Which of these is the Higgs of the strong interaction? Are all of them essential to chiral symmetry breaking?

We expect there should be a conventional scalar $\overline{q}q$ nonet, as in Fig. 55. To have $J^{PC} = 0^{++}$ quantum numbers the spin of the quark system must be $S = 1$ with relative orbital angular momentum $L = 1$, as can be seen from Eq. (1). Which of the scalar states that experiment reveals fit into this multiplet? Do they conform to the ideal mixing pattern so plainly evident for vectors and tensor mesons, as reviewed in sect. 1?

Before we can establish this, let us ask how we know when we have a state in the spectrum of hadrons. The simplest picture is that we see a peak in a cross-section, rather like that in Fig. 29, shown again in Fig. 56. The position of the resonance peak tells us the mass, $M$, of the state and its width, $\Gamma$, is inversely proportional to the lifetime of the state. Such a simple

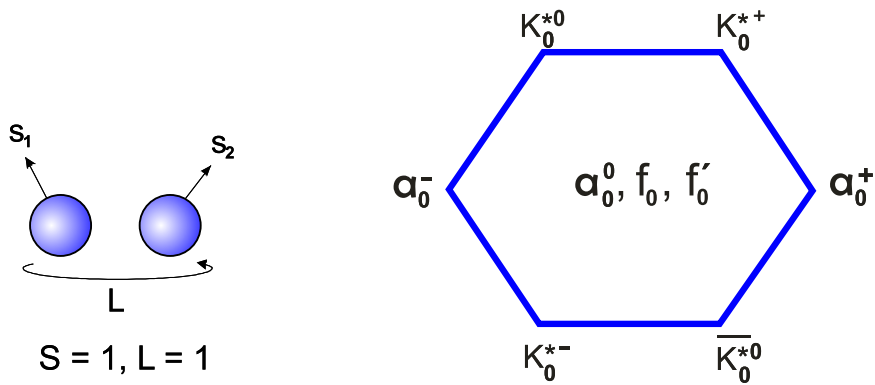

**Figure 55.** Expected nine $\overline{q}q$ states with $J^{PC} = 0^{++}$.

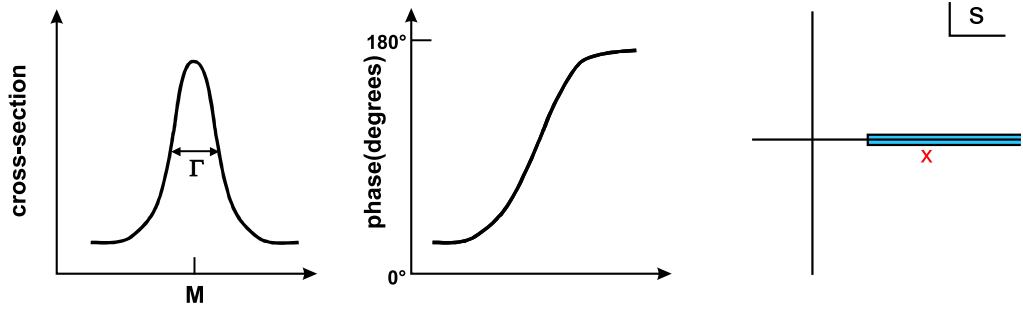

**Figure 56.** Cross-section and phase variation expected of an amplitude with an isolated resonance pole at $s = M^2 - iM\Gamma$ in the complex $s$-plane.

structure is represented by a Breit-Wigner form, where the underlying amplitude is given by

$$T(s) \simeq \frac{g_1\, g_2}{M^2 - s - iM\Gamma} \quad . \tag{169}$$

In some circumstances the phase of the amplitude can also be measured as shown in Fig. 56 and this varies by up to $180^o$ as the energy, $\sqrt{s}$, traverses the resonance energy. In turn the underlying amplitude is seen to have a pole in the complex energy plane on the nearby unphysical sheet at a position whose real part gives the mass and its imaginary part is related to its width. This simple situation applies when the resonance is isolated — isolated from other resonances with the same quantum numbers and isolated from strongly coupled thresholds. However, in the scalar sector, particularly that with the quantum numbers of the vacuum, the resonances overlap one with another and with strongly coupled thresholds. Then resonances are not simply identified by peaks, indeed they can even produce dips. What determines that we have a state in the spectrum of hadrons is a pole in the corresponding $S$-matrix element [8].

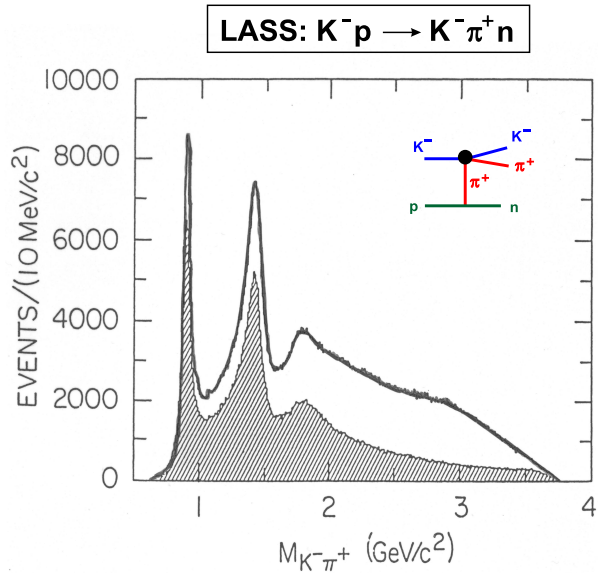

**Figure 57.** Mass distribution for $K^-\pi^+$ system produced in $K^-p \to K^-\pi^+n$ interactions from the LASS experiment [103] at 11 GeV/$c$ incident kaon momentum and small momentum transfers $|t| < 1$ GeV$^2$. There the process is dominated by the one pion exchange as shown in the inset Feynman diagram. The unshaded histogram shows all the events, while the shading depicts events with $n\pi^+$ mass $> 1.7$ GeV. The two peaks are the $K^*(892)$ and the $K_2^*(1430)$.

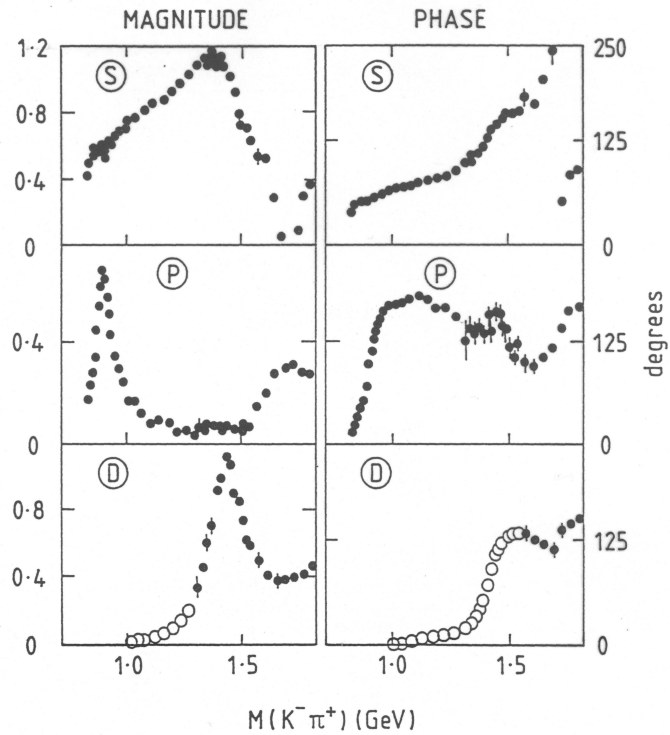

**Figure 58.** Partial wave analysis of the $K^-\pi^+ \to K^-\pi^+$ amplitudes deduced from the LASS results of Fig. 57, showing the magnitude and phase for the $S$, $P$ and $D$-waves [103].

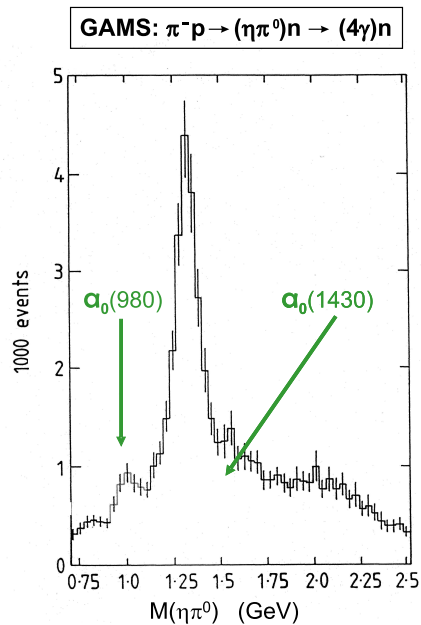

**Figure 59.** Mass distribution for $\pi^0\eta$ from the GAMS experiment [11] on $\pi^-p \to (\pi^0\eta)n$, where both $\pi^0$ and $\eta$ are detected in the $\gamma\gamma$ decay mode. Partial wave analysis reveals that the $J=0$ wave has two possible resonances $a_0(980)$ and $a_0(1430)$ in this mass region.

With this preamble, let us first look at the nine lightest scalar states that experiment reveals. We begin with the four strange states. These are seen in $K\pi$ scattering. The highest statistics experiment on high energy $K\pi$ production comes from the LASS Collaboration [103]. At small momentum transfers this process is dominated by one pion exchange and this allows the $\pi K \to \pi K$ amplitude to be determined. The LASS mass distribution is shown in Fig. 57. We see clear peaks at 890 MeV and 1430 MeV. To establish that these correspond to the spin-1 $K^*$ and spin-2 $K_2^*$ requires a partial wave separation, and then we can deduce the $S$-wave underneath these peaks, which is what we want to uncover. In Fig. 58 are shown the magnitudes and phases of the $S$, $P$ and $D$-waves. We see that the narrow peaks in both the $P$ and $D$-waves coincide with rapid phase variations we identify with resonances. Under these is a broad $S$-wave structure with a slowly increasing phase change matched to a state of mass 1430 MeV and width 294 MeV, the $K_0^*$. The LASS data start at 825 MeV and so can say little about speculations that there is another possible broad scalar state, the $\kappa$, down near $K\pi$ threshold. Only analytic continuation can address this issue. Whatever the conclusion there is definitely a $K_0^*(1430)$.

We now turn to the $I = 1$ scalars. These appear in the $\eta\pi$ channel. The neutral mode has been made accessible by the development of precision photon detectors (as discussed in sect. 1) that allow the $\pi^0$, $\eta$ and $\eta'$ to be observed in their $\gamma\gamma$ decay modes. In Fig. 59 is seen the $\pi^0\eta$ mass distribution from di-meson production in $\pi^-p$ scattering at small momentum transfer. The cross-section is dominated by a peak that is largely the $a_2(1320)$ of the $\overline{q}q$ tensor multiplet. Partial wave analysis [11] reveals that there is a strong scalar signal identified as the $a_0(980)$ with another possible scalar under the tensor peak, the $a_0(1430)$.

We next consider the $I = 0$ sector. There the states couple to the $\pi\pi$ channel. In the absence of pion targets $\pi\pi \to \pi\pi$ scattering has been inferred from a number of the classic experiments on high energy dipion production in $\pi N$ scattering, which at small momentum transfers is again controlled by one pion exchange. Supplementing this information with more recent results on reactions like $\overline{p}p$ annihilation to multi-meson final states, we can deduce the $I = J = 0$ $\pi\pi \to \pi\pi$ cross-section as shown in Fig. 60 [104]. We see a very broad structure with a very narrow dip close to 1 GeV. That this narrow structure is inextricably linked to the $\overline{K}K$ channel, is not just the proximity of the dip to this threshold, but as shown in Ref. [105] the $\pi\pi \to \overline{K}K$ cross-section rises dramatically as soon as the channel opens. It rises to 50% of the maximum unitarity allows for an inelastic cross-section. This means there must be a dynamical feature near the threshold: in fact a state that is the $f_0(980)$. More recent analyses [106] which include results from $\pi\pi$ final states in $\bar{p}p$ annihilation at rest from the Crystal Barrel experiment at LEAR [19] show the $f_0(980)$ is just one of a series of scalar states as illustrated in Fig. 60: $f_0(400-1200)$, $f_0(980)$, $f_0(1370)$, $f_0(1500)$, $f_0(1710)$. Which one of these is the Higgs of the strong interaction? or is it all of them? Which forms the expected $S = L = 1$ $q\bar{q}$ multiplet? Which is the expected ground state glueball? These are questions we cannot yet definitively answer. Nevertheless, we can pose further questions that may help to test the various hypotheses.

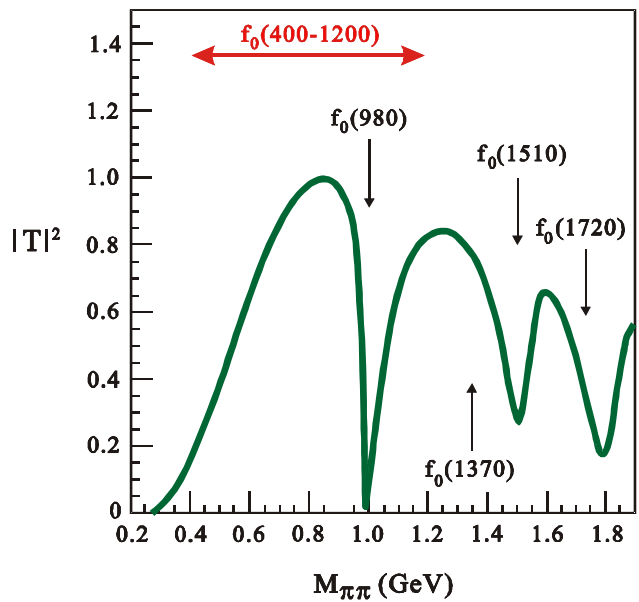

**Figure 60.** Sketch of our current knowledge of the modulus squared of the $I = J = 0$ $\pi\pi \to \pi\pi$ scattering amplitude from Zou [104]. This shows a series of dips and bumps from which a number of states the $f_0(400 - 1200)$, $f_0(980)$, $f_0(1370)$, $f_0(1500)$ and $f_0(1710)$ [13] have been deduced.

In sect. 1, we presumed a close relationship between hadrons and underlying quark model states. Here we question when this identification can be made. As we shall see, the complexity of the scalar sector throws this question into sharp relief. For orientation we first consider the relationship between the underlying *bare* quark model states and the hadrons experiments reveal for the the vector meson multiplet shown in sect. 1 in Fig. 1. The *bare* states do not decay and so have propagators that are real being simply $1/(m_0^2 - s)$, where $s$ is the square of the 4-momentum it carries, and $m_0$ its *bare* mass. This is of course not the propagator of a real hadron, that is short-lived and decays. A real hadron has a more complicated Fock space. The $\rho^+$ is not just $u\bar{d}$, but also $\pi^+\pi^0$. It is through these $\pi\pi$ components that it decays. The $\phi$ is not just $s\bar{s}$, but also $K\overline{K}$. Let us attempt to calculate these effects. The hadron propagator is related to that of the underlying bares state by hadronic loop corrections, as illustrated in Fig. 61. We will just consider those contributions from two pseudoscalars that dominate their decays. While the pole of the bare propagator is on the real axis, the pole of the hadron is at a complex position, see Fig. 62. The $s$-plane now has a cut along the real axis. A cut that is generated by each hadronic channel with the same quantum numbers. In the case of the $\rho$ the lowest threshold is of course $\pi\pi$. The pole is now on the nearby unphysical sheet as shown in Fig. 62. How far the pole moves is determined by the strength of the coupling to the open channels. For the vector mesons, like the $\rho$ and $\phi$, these are $P$-wave couplings to two pseudoscalars and so the discontinuity across the cut (seen in the right hand part of Fig. 62) is suppressed towards threshold, and the poles do not move far. The observed hadrons and the underlying quark states are very similar. Not surprising then that the underlying quark model picture could be readily inferred despite the inevitable presence of hadronic dressing.

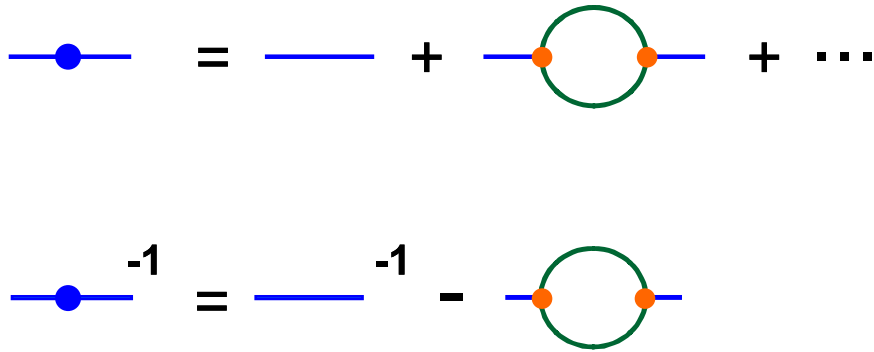

**Figure 61.** Feynman graphs for a hadron propagator with corrections from its decay channels, and its Dyson summation for the inverse propagator.

In contrast, for scalar mesons, the couplings are much stronger — that is a matter of phenomenology and there is no angular momentum suppression at threshold. The Adler condition provides some "suppression" and we can take that into account. Nevertheless, a scalar can freely couple to two pseudoscalars and the effects are much greater. This is illustrated by a series of calculations by van Beveren *et al.* [107] and by Tornqvist [108]. Using the Dyson summation of Fig. 61, we compute the inverse of the *dressed* propagator:

$$\mathcal{M}^2(s) - s = m_0^2 - s - \Pi(s) \quad . \tag{170}$$

When this is zero defines when there is a state in the spectrum of hadrons. In the neighbourhood of this zero, the inverse propagator can be represented by

$$s_R - s \equiv M_R^2 - iM_R\Gamma_R - s \quad , \tag{171}$$

the Breit-Wigner approximation.

Let us consider what happens if the underlying bare scalar multiplet is ideally mixed like the vectors and tensors. We begin with the $I = 1/2$ strange sector. The results [108, 109] are sketched in Fig. 63. We plot the real and imaginary part of the square of the mass, the $\mathcal{M}^2(s)$ of Eq. (170), as a function of $\sqrt{s}$. A hadron emerges when this function is equal to $s$ as in Eq. (171). When there is no coupling to the decay channels, the mass is a constant. We then switch on the decay channels. Importantly we include only real decays and not the infinity of virtual loops. Here the decay channels are $K\pi$ and $K\eta'$, with $K\eta$ very weak as $SU(3)_F$ requires. Each of these thresholds is clearly seen in the imaginary part, and these in turn induce a cusp in the real part. The strength of the coupling to the pseudoscalar channels and the value of the bare mass are fixed by the requirement that the pole should reproduce the $K_0^*(1430)$. Knowing that the strange quark has a mass of $\sim 120$ MeV more than that of the $u$ and $d$ quarks determines all the parameters.

For the $I = 1$ sector, the bare state is then near 1420 MeV. Turning on the coupling to $\pi\eta$, $K\overline{K}$ and $\pi\eta'$, we find as shown in Fig. 64 the hadron has a mass shifted down to 980 MeV with a very large $K\overline{K}$ component. This is in remarkable agreement with the pattern of the $a_0(980)$ [108, 109]. The $I = 0$ sector is more complicated, since there are bare states with not just $u\overline{u}$ and $d\overline{d}$, but also $s\overline{s}$ [108], and even pure glue [109]. The common decay channels mix these states and the hadrons result after diagonalising the mass matrix. One of these states emerges again around 980 MeV with a strong $K\overline{K}$ coupling [108, 109], recognisably the $f_0(980)$. So though we started from a system that is ideally mixed with the mass and coupling relations we expect from such an arrangement, they seed hadrons with quite different compositions. In particular, readily identifiable $a_0(980)$ and $f_0(980)$ emerge with $\sim 40\%$ $K\overline{K}$ in their Fock space. No wonder these states have many of the properties of $K\overline{K}$ molecules [110]. When the creation of quark loops that can lead to real (rather than virtual) intermediate states is unimportant, as with the vectors and tensors, the quark model and the hadrons we observe are closely related. In the language of lattice calculations "unquenching" is unimportant, Fig. 65. In contrast, when the creation of quark loops is important, the intermediate hadronic states that are created play a crucial role in the composition of the resulting resonance, Fig. 65. While it is argued that internal quark loops are higher order in an expansion in terms of $1/N_c$, where $N_c$ is the number of colours as in Eq. (6), in the scalar sector there is no such suppression. Indeed the fact that here an $\overline{s}s$ system couples very strongly to $\overline{u}u$ and $\overline{d}d$ is determined by the flavour structure of the QCD vacuum, a topic under current study [111].

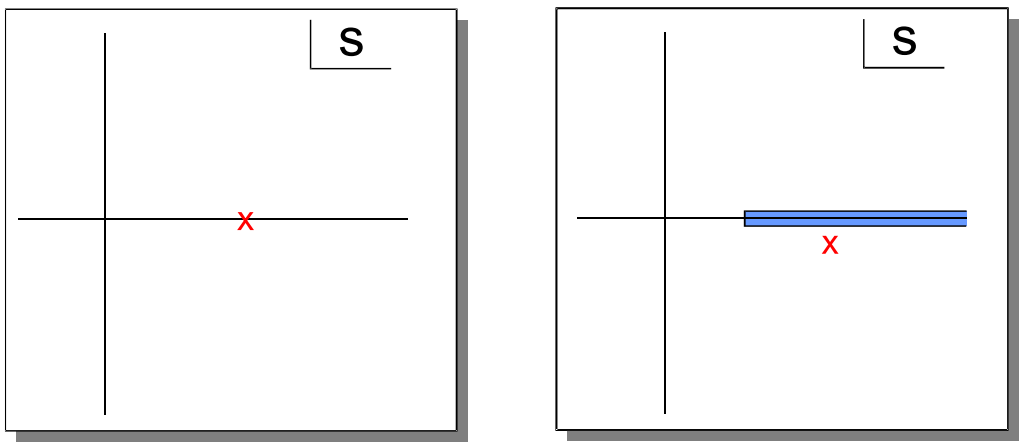

**Figure 62.** The crosses denote the poles in the complex plane for a bare "non-interacting" hadron on the left and for a hadron dressed by its decay channels on the right. In this latter case the propagator has cuts corresponding to each decay channel.

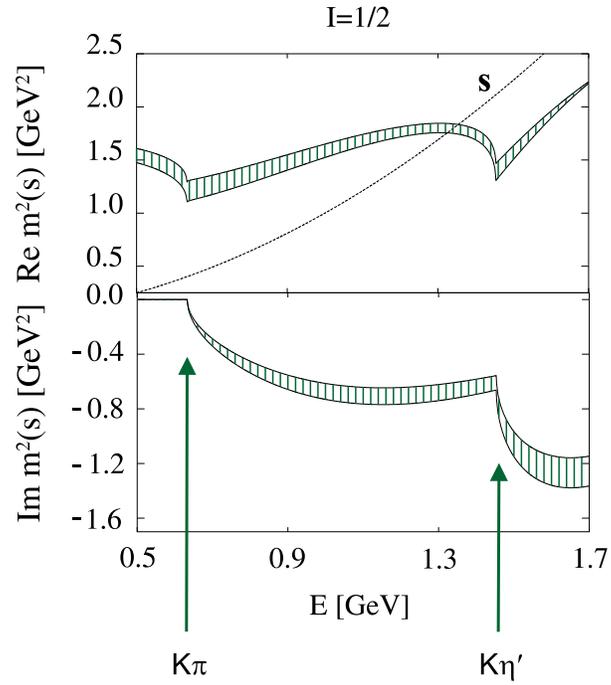

**Figure 63.** The real and imaginary parts of $\mathcal{M}^2(s)$ as functions of $E = \sqrt{s}$ for the $I = 1/2$ propagator. Here for the "bare" strange quark mass $m_0 = 1540$ MeV [108, 109].

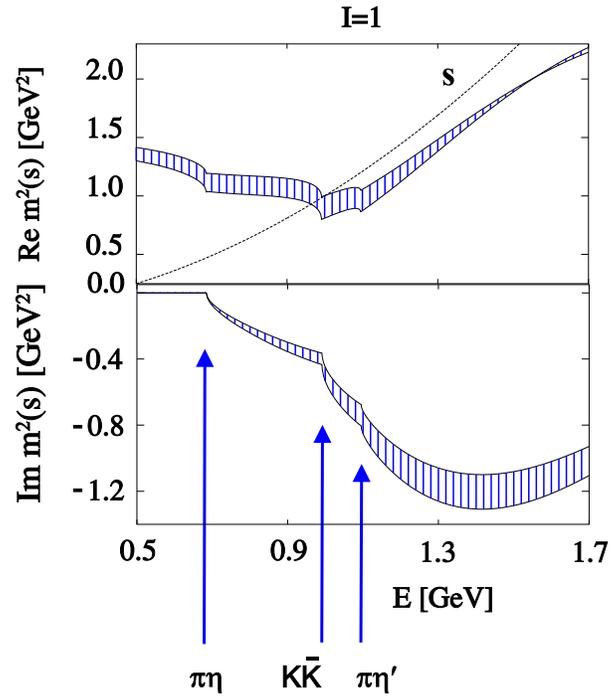

**Figure 64.** The real and imaginary parts of $\mathcal{M}^2(s)$ as functions of $E = \sqrt{s}$ for the $I = 1$ propagator. Here for the "bare" non-strange quark mass $m_0 = 1420$ MeV [108, 109].

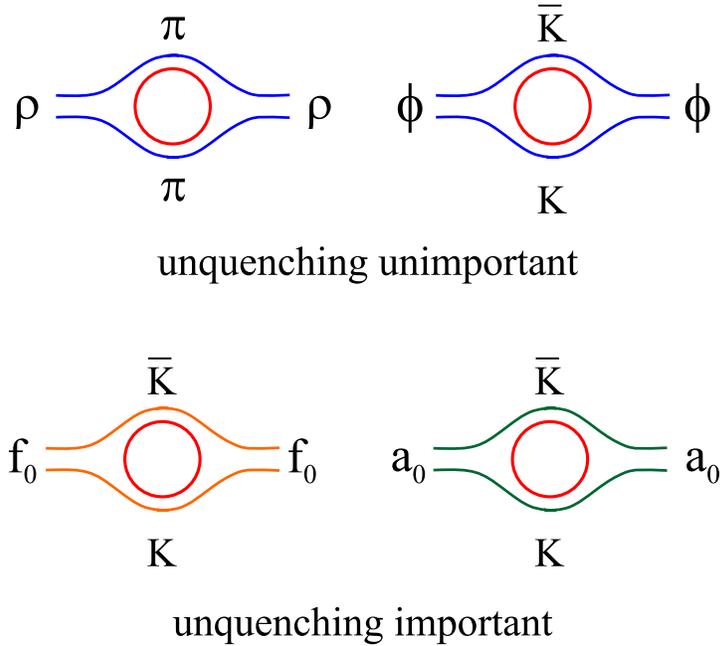

**Figure 65.** For vector mesons, like the $\rho$ and $\phi$, the fact that they decay to $\pi\pi$ and $\overline{K}K$ has relatively little effect on their Fock space. In contrast, for the $f_0(980)$ and $a_0(980)$ their coupling to $\overline{K}K$ shifts their masses by hundreds of MeV and gives them a large $\overline{K}K$ component.

That the $f_0(980)$ has a strong affinity for hidden strangeness in $K\overline{K}$ and $s\overline{s}$ is clearly seen by comparing in Fig. 66 its appearance in reactions with little or no strangeness, like $\pi\pi$ production in elastic scattering or in $pp \to pp\pi\pi$, where the $f_0(980)$ creates a dip or a shoulder, whilst in the decay $J/\psi \to \phi(\pi\pi)$, the $f_0(980)$ appears as a peak. In the latter process one can think of the $J/\psi$ as a source of energy, while the $\phi$ ensures it is an $s\overline{s}$ ( or $K\overline{K}$ ) system that creates the final $\pi\pi$ state [105]. A similar peak occurs in $D_s \to 3\pi$ decay as also discussed in Ref. [105]. This pattern confirms the structure expected from the strong effects of decays found in the calculations reported in Figs. 63, 64.

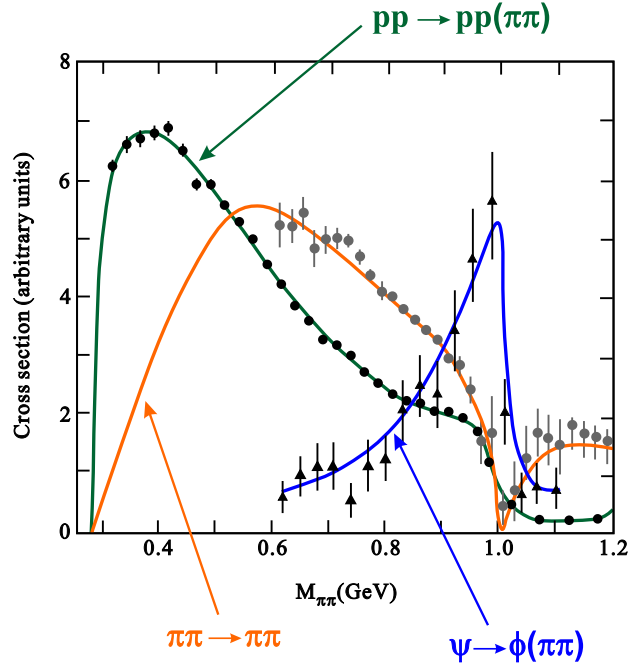

**Figure 66.** $I = J = 0$ cross-section and decay distibutions for $\pi\pi \to \pi\pi$ (red), central production of dimeson final state in $pp \to pp(\pi\pi)$ (green) and for $J/\psi \to \phi(\pi\pi)$ (blue). This comparison illustrates how in the last process with its effective $\bar{s}s$ or $\overline{K}K$ initial state the $f_0(980)$ appears as a peak, whilst in the former two with their effective $\bar{n}n$ (where $n = u, d$) or $\pi\pi$ initial state the $f_0(980)$ appears as a dip or shoulder.

Having earlier determined the behaviour of quarks and gluons and their interactions in the strong coupling regime of QCD allows us to address the problem of bound states. The formalism with which to discuss these [112] are the Bethe-Salpeter equations illustrated in Fig. 67. The basic ingredients are the QCD Green's functions we discussed in sect. 11 together with a modelling of the quark-quark scattering kernel. The simplest model for this is the ladder approximation.

**Figure 67.** Bethe-Salpeter (bound state) equation for $\bar{q}q$ systems with P(seudoscalar) and V(ector) quantum numbers. The key to solving these is the approximation made for the quark-quark scattering kernel shaded in the picture.

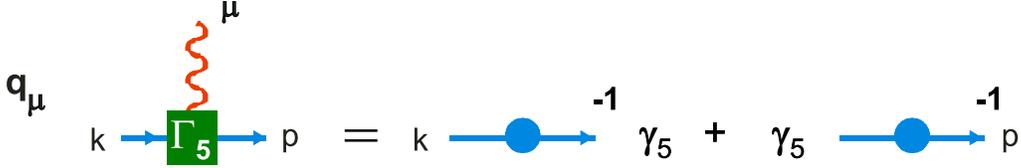

**Figure 68.** Diagramatic representation of the axial Ward identity.

What makes the pseudoscalar pion distinct from a scalar is the fact that its interaction satisfies the axial Ward identity depicted in Fig. 68. As shown by Roberts and collaborators [113], this is sufficient to ensure that the pion is massless in the limit of zero quark mass. It is the Goldstone boson of chiral symmetry breaking. While the pion depends on the current quark, the scalar knows of the dressed quark. However, as found in Ref. [114], calculation of the scalar spectrum is critically dependent on the exact form of the quark-quark scattering kernel and so we cannot yet solve QCD to provide clear insights into the role of the plethora of $f_0$ states in Fig. 60.

Fortunately the vector states are easier to understand. Using the ladder approximation for the scattering kernel and the form of the Tübingen gluon function of Fig. 54 in the Bethe-Salpeter equation we can compute the pion and rho masses. Calculations in lattice QCD can do the same, but only for large quark mass, like 150 MeV or greater. At large mass both the pion and rho are linear in the quark's current mass, $m_q$. At small $m_q$, while the rho remains linear, the chiral nature of the pion means it depends on $\sqrt{m_q}$, as discussed extensively in sects. 5, 6. There has been a major effort [115, 71] to extrapolate lattice simulations at larger quark mass to physical values by incorporating analytic forms from chiral perturbation theory. These inevitably have to confront the issue of where the transition from linear dependence on $m_q$ to that on $\sqrt{m_q}$ occurs. For some this is a parameter in the modelling. Using the solutions of the Bethe-Salpeter equations we can however calculate the pion and rho masses for any value of $m_q$. Though we cannot yet claim to be solving QCD, but merely a reasonable approximation to it, we do not need to ask where the transition to chiral dynamics occurs. It is built into the system of equations.

Let us start from the lattice QCD results of CP-PACS [116], shown in Fig. 69, for the relation between the vector and pseudoscalar mesons masses. Using the Tübingen modelling of the strong coupling limit of QCD, we can then calculate the very same relationship for all quark masses in the continuum approach. Following the work of Maris and Tandy [117], the gluon propagator dressing can be modelled as

$$\Delta^{\mu\nu}(p) \;=\; C\,\frac{p^2}{\omega^2}\,\exp\left(-\frac{p^2}{\omega^2}\right)\Delta_0^{\mu\nu}(p) \tag{172}$$

where $\Delta_0$ is the bare transverse gluon propagator. The parameter $C$ is related to the strength of the interaction, while $\omega$ corresponds to the momentum at which the gluon dressing peaks as seen in Fig. 54 [118]. From the work of Ref. [117] this is known to be around 0.5 GeV. Using the simple rainbow-ladder approximation to solve the Bethe-Salpeter equation for the $\rho$ and $\pi$-mesons one finds the behaviour shown as the line in Fig. 69. Optimal agreement [119] between Bethe-Salpeter (BS) and CP-PACS results occurs for $\omega = 0.425$ GeV, very close to the expected value. The continuation from a pseudoscalar mass of 500 MeV down below 140 MeV to zero mass is specified. For this continuation one need not estimate where the transition to the dominance of chiral dynamics occurs. This is built in. The non-perturbative approach in the continuum knows — or at least that is the principle.

While the lattice results in Fig. 69 show little sign of anything but a linear relation between $m_V$ (the vector mass) and $m_{PS}$ (the pseudoscalar mass) for different values of the lattice coupling parameter, $\beta$, the Bethe-Salpeter solutions beautifully interpolate these and then extrapolate them to physical masses. While the scattering kernel approximation used does not include all details of chiral perturbation theory, it clearly does incorporate the essence. This highlights how the continuum approach [112, 119, 120] can reproduce hadron masses for any quark mass, and is the natural vehicle for incorporating chiral dynamics, which is so much a feature of the hadron world. While we cannot yet answer which of the many scalar states, all called $f_0$, are approximated by the sigma field in Nambu's model of chiral symmetry breaking, we are getting very close to discovering the secrets of this key Higgs sector of the strong interaction, and its intimate relation to the structure of the QCD vacuum.

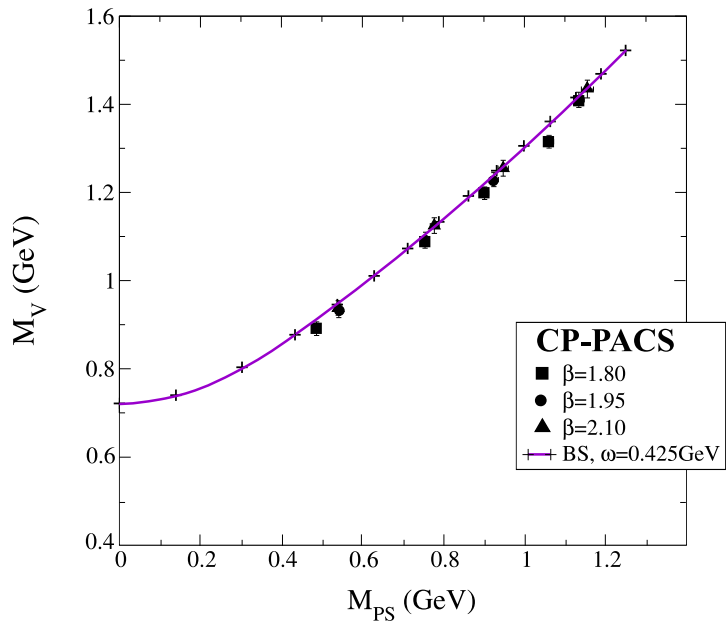

**Figure 69.** Vector meson mass as a function of pseudoscalar meson mass. The data points are from the lattice calculations of CP-PACS [116] with parameters shown in the inset. The curve that connects the crosses is the result of the Bethe-Salpeter calculation using the Tübingen modelling of the full quark and gluon functions by Watson *et al.* [119]. The parameter $\omega$ is explained in the text.

These lectures have highlighted how combining theoretical, phenomenological and experimental studies has allowed considerable progress in understanding the workings of strong coupling QCD to be achieved. We can expect further progress to come, hopefully by some of the young people at this School.

### Acknowledgments

It is a very great pleasure to thank the Mexican Physical Society, Division of Particles and Fields, and in particular Myriam Mondragon, Adnan Bashir and Jens Erler for inviting me to give these lectures and for their warm hospitality. I want to thank them and the students for their enthusiastic participation in this School, which took place at the wonderful Anthropological Museum in Xalapa. I am grateful to Richard Williams for his reading of this manuscript. I acknowledge the support of the EU-RTN Programme, Contract No. HPRN-CT-2002-00311, "EURIDICE" for part of the research work discussed here.

[1] Halzen F and Martin A D 1984 *Quarks and leptons: An Introductory Course in Modern Particle Physics* (J. Wiley).
[2] Cheng T-P and Li L-F 1984 *Gauge theory of elementary particle physics* (Oxford University Press).
[3] Leader E and Predazzi E 1982 *An Introduction to Gauge Theories and the 'New Physics'* (Cambridge University Press).
[4] Close F E 1979 *An Introduction to Quarks and Partons* (Academic Press).
[5] Donoghue J F, Golowich E and Holstein, B R 1992 *Dynamics of the Standard Model* (Cambridge University Press).
[6] Hayne C and Isgur N 1982 *Phys. Rev.* **D25** 1944.
[7] Jaffe R L 1977 *Phys. Rev.* **D15** 267; Jaffe, R L 1978 *Phys. Rev.* **D17** 1444.
[8] Pennington M R 1990 *In search of hadrons beyond the quark model*, Proceedings of the Dalitz Conference, Oxford, "Plots, Quarks & Strange Particles", (ed. Aitchison I J R *et al.*; pub. World Scientific) pp. 66-107.
[9] Close F E and Tornqvist N A 2002 *J. Phys.* **G28** R249.
[10] Amsler C and Tornqvist N A 2004 *Phys. Rept.* **389** 61.
[11] Alde D *et al* (GAMS) 1988 *Phys. Lett.* **205B** 397.
[12] Thompson D R *et al.* (BNL-E852) 1997 *Phys. Rev. Lett.* **79** 1630; Adams G S. *et al.* 1998 *Phys. Rev. Lett.* **81** 5760.
[13] Eidelman S *et al.* (PDG) *Phys. Lett.* **B592** 1.
[14] Popov A V (BNL-E852) 2002 *Proceedings of Hadron 2001 A.I.P. Conf. Proc.* **619** 135; Dzierba A R *et al.* 2003 *Phys. Rev.* **D67** 094015.
[15] Amelin D V *et al.* (VES) 2002 *Proceedings of Hadron 2001 A.I.P. Conf. Proc.* **619** 143.
[16] Dzierba A R 2003 *Int. J. Mod. Phys.* **A18** 397.
[17] Bali G S *et al.* (UKQCD) 1993 *Phys. Lett.* **B389** 378.
[18] Pennington M R 1999 *Glueballs: the naked truth*, Proc. Workshop on Photon Interactions and Photon Structure, Lund, Sweden, Sept. 1998 (ed. Jarlskog G and Sjostrand T ; pub. Lund) pp. 313-328.
[19] Amsler C 1998 *Rev. Mod. Phys.* **70** 1293.
[20] Augustin J *et al.* (DM2) 1987 *Z. Phys.* **C36** 369; Augustin J *et al.* (DM2) 1988 *Phys. Rev. Lett.* **60** 2238.
[21] Baltrusaitis R M *et al.* (Mark III) 1986 *Phys. Rev. Lett.* **56** 107.
[22] Çakir M B and Farrar G R 1994 *Phys. Rev.* **D50** 3268.
[23] Close F E, Farrar G R and Li Z 1997 *Phys. Rev.* **D55** 5749.
[24] Bai J Z *et al.* (BES) 1996 *Phys. Rev. Lett.* **76** 3502.
[25] Nakano T *et al.* 2003 *Phys. Rev. Lett.* **91** 012002.
[26] Stepanyan S *et al.* (CLAS) 2003 *Phys. Rev. Lett.* **91** 252001-1; Kubarovsky V *et al.* (CLAS) 2004 *Phys. Rev. Lett.* **92** 032001-1.
[27] Diakonov D, Petrov V and Polyakov M 1997 *Z. Phys.* **359** 305; Diakonov D, Petrov V and Polyakov M 2004 *Preprint* hep-ph/0404212.
[28] Jaffe R L and Wilczek F 2003 *Phys. Rev. Lett.* **91** 232003.
[29] Karliner M and Lipkin H J 2003 *Preprint* hep-ph/0307243; and *Phys. Lett.* B**575** 249; Karliner M and Lipkin H J 2004 *Phys Lett.* B**586** 303.
[30] Bodek A *et al.* 1964 *Phys. Lett.* **51B** 417.
[31] Dzierba A R, Meyer C A and Szczepaniak A P 2004 *Preprint* hep-ex/0412077.
[32] Maltman K 2004 *Preprint* hep-ph/0408144.
[33] Close F E and Dudek J J 2004 *Phys. Lett* **B586** 75 (*Preprint* hep-ph/0401192).
[34] Aubert B *et al* (BaBar) 2004 *Preprint* hep-ex/0408064.
[35] Aktas A *et al* (H1) 2004 *Phys. Lett.* **B588** 17.
[36] Chekanov S *et al.* (ZEUS) 2004 *Phys. Lett.* **B591** 7.
[37] Antipov Yu *et al.* (SPHINX) 2004 *Eur. Phys. J.* **A21** 455 and *Preprint* hep-ex/0407026.
[38] Choi S-K *et al.* (Belle) 2003 *Phys. Rev. Lett.* **91** 262001; Bauer G *et al.* (CDF) *Preprint* hep-ex/0312021; Abazov V M *et al.* (D0) *Preprint* hep-ex/0405004; Aubert B *et al.* (BaBar) *Preprint* hep-ex/0406022.
[39] Krokovny P *et al.* (Belle) 2003 *Phys. Rev. Lett.* **91** 262002; Abe K *et al.* (Belle) 2004 *Phys. Rev. Lett.* **92** 012002; Aubert B *et al.* (BaBar) 2004 *Phys. Rev.* **D69** 031101.
[40] Quigg C 2005 *Nucl. Phys. Proc. Suppl.* **142** 87 (*Preprint* hep-ph/0407124).
[41] Pennington M R 1997 *Nucl. Phys.* **A623** 189 (*Preprint* hep-ph/9612417); Pennington M R 2002 *Czech. J. Phys.* **52** B28.
[42] Nambu Y 1960 *Phys. Rev. Lett.* **4** 380; Nambu Y and Jona-Lasinio G 1961 *Phys. Rev.* **122** 345, **124** 246; Gell-Mann M and Levy M 1960 *Nuovo Cim.* **16** 705.
[43] Goldstone J 1961 *Nuovo Cim.* **19** 154.
[44] Stern J 1995 *Proc of Workshop on Physics and Detectors for DA$\Phi$NE* (eds. Baldini R *et al.*) (INFN, Frascati) pp. 609-622 (*Preprint* hep-ph/9510318); Stern J 1998 *Preprint* hep-ph/9801282.

[45] Martin A and Cheung F 1970 *Analyticity Properties and Bounds of the Scattering Amplitudes* (Gordon and Breach).
[46] Jin YS and Martin A 1964 *Phys. Rev.* **135** B1369.
[47] Adler S L 1965 *Phys. Rev.* **137** B1022, **139** B1638.
[48] Weinberg S 1966 *Phys. Rev. Lett.* **17** 616; Weinberg S 1979 *Physica* **96A** 327.
[49] Gasser J and Leutwyler H 1984 *Ann. Phys. (NY)* **158** 142; Gasser J and Leutwyler H 1985 *Nucl. Phys.* **B250** 465.
[50] Stern J, Sazdjian H and Fuchs N H 1993 *Phys. Rev.* **D47** 3814.
[51] Knecht M and Stern J 1995 *The Second DAΦNE Physics Handbook* (ed. Maiani L, Pancheri G and Paver N) (INFN, Frascati) pp. 169-190; Knecht M, Moussallam B and Stern J 1995 *The Second DAΦNE Physics Handbook* (ed. Maiani L, Pancheri G and Paver N) (INFN, Frascati) pp. 221-236.
[52] Ochs W 1973 Univ. of Munich thesis; Hyams B *et al.* 1973 *Nucl. Phys.* **B64** 134.
[53] Barrelet E 1972 *Nuovo Cim.* **8A** 331.
[54] Hoogland W *et al.* 1974 *Nucl. Phys.* **B69** 266; Hoogland W *et al.* 1977 *Nucl. Phys.* **B126** 109.
[55] Chew G F, Mandelstam S and Noyes H P 1960 *Phys. Rev.* **119** 478.
[56] Kawarabayashi K and Suzuki M 1966 *Phys. Rev. Lett.* **16** 255; Riazuddin and Fayyazuddin 1966 *Phys. Rev.* **147** 1071.
[57] Olsson M G 1967 *Phys. Rev.* **162** 1338.
[58] Pennington M R 1973 *Zeros in $\pi\pi$ scattering*, Proc. of the International Conference on $\pi\pi$ scattering and related topics, Tallahassee, Florida, A.I.P Conference Proceedings **13** pp. 89-116.
[59] Gomez F *et al.* (DIRAC) 2001 *Nucl. Phys. Proc. Suppl.* **96** 259; Gianotti P *et al.* (DIRAC) 2000 *Acta Phys. Polon. B* **31** 2571; Adeva B *et al.* (DIRAC) *J. Phys. G* **G30** 1929.
[60] Cabibbo N and Maksymowicz A 1965 *Phys. Rev.* **137** B438.
[61] Watson K M 1952 *Phys. Rev.* **88** 1163.
[62] Beier E W *et al.* 1972 *Phys. Rev. Lett.* **29** 511; Beier E W *et al.* 1973 *Phys. Rev. Lett.* **30** 399.
[63] Rosselet L *et al.* 1977 *Phys. Rev.* **D15** (1977) 374.
[64] Pislak S *et al.* (BNL-E865) 2001 *Phys. Rev. Lett.* **87** 221801 (*Preprint* hep-ex/0106071).
[65] Roy S M 1971 *Phys. Lett.* **B36** 353.
[66] Colangelo G, Gasser J and Leutwyler H 2001 *Nucl. Phys. B* **603** 125 (*Preprint* hep-ph/0103088); Ananthanarayan B, Colangelo G, Gasser J and Leutwyler H 2001 *Phys. Rept.* **353** 207 (*Preprint* hep-ph/0005297).
[67] Basdevant J L, Froggatt C D and Petersen J L 1974 *Nucl. Phys.* **B72** 413; Pennington M R and Protopopescu S D 1973 *Phys. Rev.* **D7** 1429, 2591.
[68] Colangelo G, Gasser J and Leutwyler H 2001 *Phys. Rev. Lett.* **86** 5008 (*Preprint* hep-ph/0103063).
[69] Gell-Mann M, Oakes R J and Renner B 1968 *Phys. Rev.* **175** 2195.
[70] Miransky V A 1985 *Nuovo Cim.* **90A** 149; Miransky V A 1994 *Dynamical Symmetry Breaking in Quantum Field Theories* (World Scientific) and references therein.
[71] Davies C T H *et al.* (HPQCD) 2004 *Phys. Rev. Lett.* **92** 022001 (*Preprint* hep-lat/0304004); Lepage P and Davies C 2004 *Int. J. Mod. Phys.* **A19** 877.
[72] Dyson F J 1949 *Phys. Rev.* **75** 1736; Schwinger J 1951 *Proc. Nat. Acad. Sc.* **37** 452 and 455.
[73] Bjorken J D and Drell S D 1965 *Relativistic Quantum Fields* (McGraw-Hill).
[74] Itzykson C and Zuber JB 1980 *Quantum Field Theory* (McGraw-Hill).
[75] Maskawa T and Nakajima H 1974 *Prog. Theor. Phys.* **52** 1326; Fukuda R and Kugo T 1976 *Nucl. Phys.* **B117** 250; Holdom B 1985 *Phys. Lett.* **150B** 301; Cohen A and Georgi H 1989 *Nucl. Phys.* **B314** 7; Leung C N, Love S T and Bardeen W 1989 *Nucl. Phys.* **B323** 493; and many more.
[76] Ward J C 1950 *Phys. Rev.* **78** 182.
[77] Green H S 1953 *Proc. Phys. Soc. (London)* **A66** 873; Takahashi Y 1957 *Nuovo Cim.* **6** 371.
[78] Atkinson D and Fry M 1979 *Nucl. Phys.* **B156** 301.
[79] Kızılersü A, Sizer T and Williams AG 2002 *Phys. Rev.* **D65** 085020; Kızılersü A, Schreiber A W, Sizer T and Williams AG 2002 *Nucl. Phys. Proc. Suppl.* **109A** 173.
[80] Ball J C and Chiu T W 1980 *Phys. Rev.* **D22** 2542.
[81] Bernstein J 1968 *Elementary Particles and their Currents* (W H Freeman) p. 64.
[82] Kızılersü A, Reenders M and Pennington M R 1995 *Phys. Rev.* **D52** 1242.
[83] Curtis D C and Pennington M R 1990 *Phys. Rev.* **D42** 4165.
[84] Curtis D C and Pennington M R 1992 *Phys. Rev.* **D46** 2663.
[85] Kızılersü A and Pennington M R 2005 (in preparation).
[86] Landau L D and Khalatnikov I M 1956 *Sov. Phys. JETP* **2** 69; Fradkin E S 1956 *Sov. Phys. JETP* **2** 361; Zumino B 1960 *J. Math Phys.* **1** 1.
[87] Bashir A and Raya A 2002 *Phys. Rev.* **D66** 105005 (*Preprint* hep-ph/0206277).

[88] Kondo K 1991 *Nucl. Phys.* **B351** 259; Holdom B 1996 *Phys. Rev.* **D54** 1068, *Prog. Theor. Phys. Suppl.* **123** 71.
[89] Richardson J L 1979 *Phys. Lett.* **B82** 272.
[90] Slavnov A A 1972 *Theor. Math. Phys.* **10** 99; Taylor J C 1971 *Nucl. Phys.* **B33** 436.
[91] Baker M, Ball J S and Zachariasen F 1981 *Nucl. Phys.* **B186** 531, 560.
[92] West G B 1982 *Phys. Lett.* **B115** 468; West G B 1983 *Phys. Rev.* **D27** 1878.
[93] Mandelstam S 1979 *Phys. Rev.* **D20** 3223.
[94] Bar-Gadda U 1980 *Nucl. Phys.* **B163** 312.
[95] Brown N and Pennington M R 1988 *Phys. Lett.* **B202** 257 (erratum-*ibid.* **B205** 596), *Phys. Rev.* **D38** 2266; Brown N and Pennington M R 1989 *Phys. Rev.* **D39** 2723.
[96] Pagels H 1977 *Phys.Rev.* **D15** 2991.
[97] Maris P and Roberts C D 1997 *Phys. Rev.* **C56** 3369.
[98] Langfeld K *et al.* 2003 *Phys. Rev.* **C67** 065206.
[99] Büttner K and Pennington M R 1995 *Phys. Rev.* **D52** 5220.
[100] Alkofer R and von Smekal L 2000 *Nucl. Phys.* **A680** 133 (*Preprint* hep-ph/0004141); Alkofer R and von Smekal L 2001 *Phys. Rept.* **353** 281 (*Preprint* hep-ph/0007355); Fischer C S and Alkofer R 2003 *Phys. Rev.* **D67** 094020 (*Preprint* hep-ph/0301094).
[101] Bonnet F D, Bowman P O, Leinweber D B and Williams A G 2000 *Phys. Rev.* **D62** 051501 (*Preprint* hep-lat/0002020).
[102] Watson, P 1999 *Preprint* hep-ph/9901454.
[103] Aston D *et al.* (LASS) 1988 *Nucl. Phys.* **B296** 493.
[104] Zou B S (1996) *Preprint* hep-ph/9611235, Talk presented at *34th Course of International School of Subnuclear Physics*, Erice, Sicily, July 1996; Anisovich V V, Bugg D V, Sarantsev A V and Zou, B S 1994 *Phys. Rev.* **D50** 972; Bugg D V, Zou B S and Sarantsev A V 1996 *Nucl. Phys.* **B471** 59.
[105] Au K-L, Morgan D and Pennington M R 1987 *Phys. Rev.* **D35** 1633.
[106] Anisovich V V and Sarantsev A V 1996 *Phys. Lett.* **B382** 429.
[107] van Beveren E *et al.* 1986 *Z. Phys.* **C30** 615.
[108] Tornqvist N 1975 *Z. Phys.* **C68** 647.
[109] Boglione M and Pennington M R 1997 *Phys. Rev. Lett.* **79** 1633.
[110] Weinstein J and Isgur N 1982 *Phys. Rev. Lett.* **48** 659; Weinstein J and Isgur N 1983 *Phys. Rev.* **D27** 588.
[111] Moussallam B 2000 *Eur. Phys. J.* **C14** 111 (*Preprint* hep-ph/9909292), *JHEP* **0008** 005 (*Preprint* hep-ph/0005245).
[112] Tandy P C 2003 *Prog. Part. Nucl. Phys.* **50** 305 (*Preprint* nucl-th/0301040).
[113] Maris P and Roberts C D 2003 *Int. J. Mod. Phys.* E **12** 297 (*Preprint* nucl-th/0301049).
[114] Bender A, Detmold W, Roberts CD and Thomas AW 2002 *Phys. Rev.* **C65** 065203 (*Preprint*: nucl-th/0202082).
[115] Sanz-Cillero J J, Donoghue J F and Ross A 2004 *Phys. Lett.* **B579** 86 (*Preprint* hep-ph/0305181); Young R D, Leinweber D B, Thomas A W and Wright S V 2002 *Phys. Rev.* **D66** 094507 (*Preprint* hep-lat/0205017); Young R D, Leinweber D B and Thomas A W 2004 *Nucl. Phys. Proc. Suppl.* **128** 227 (*Preprint* hep-lat/0311038).
[116] Aoki S *et al.* (CP-PACS) 1999 *Phys. Rev.* **D60** 114508 (*Preprint* hep-lat/9902018); Ali Khan A *et al.* (CP-PACS) 2002 *Phys. Rev.* **D65** 054505 (*Preprint* hep-lat/0105015).
[117] Maris P and Tandy P C 2000 *Phys. Rev.* **C62** 055204 (*Preprint* nucl-th/0005015); Jarecke D, Maris P and Tandy P C 2003 *Phys. Rev.* **C67** 035202 (*Preprint* nucl-th/0208019); Watson P, Cassing W and Tandy PC 2004 *Few Body Syst.* **35** 129 (*Preprint* hep-ph/0406340).
[118] Alkofer R, Watson P and Weigel H 2002 *Phys. Rev.* **D65** 094026 (*Preprint* hep-ph/0202053).
[119] Watson P, Benhaddou K and Pennington M R *in preparation*; Pennington M R 2005 *Nucl. Phys. Proc. Suppl.* **B141** 1 (*Preprint* hep-ph/0409156).
[120] Bhagwat M S *et al.* 2004 *Phys. Rev.* **C70** 035205.